\pdfoutput=1
\documentclass[11pt,twoside,a4paper,cmspaper,final,collab]{cms-tdr}
\def\svnVersion{898d546}\def\svnDate{2024/08/27}\def\cmsCernNoTag{CERN-EP-2023-075}\def\cmsCernDate{\today}\def\cmsMessage{Published in the Journal of High Energy Physics as \href{http://dx.doi.org/10.1007/JHEP08(2024)148}{\doi{10.1007/JHEP08(2024)148}.}}
\begin{document}\cmsNoteHeader{HIN-21-017}

\newcommand{\PbPb}{\ensuremath{\text{PbPb}}\xspace}
\newcommand{\pPb}{\ensuremath{\Pp\text{Pb}}\xspace}
\newcommand{\AMPT} {\textsc{ampt}\xspace}
\newcommand{\nch}   {\ensuremath{N_\text{ch}}\xspace}

\cmsNoteHeader{HIN-21-017}

\title{Multiplicity and transverse momentum dependence of charge-balance functions in \texorpdfstring{\pPb}{pPb} and \texorpdfstring{\PbPb}{PbPb} collisions at LHC energies}

\date{\today}

\abstract{
Measurements of the charge-dependent two-particle angular correlation function in proton-lead (\pPb) collisions at a nucleon-nucleon center-of-mass energy of $\sqrtsNN = 8.16\TeV$ and lead-lead (\PbPb) collisions at $\sqrtsNN = 5.02\TeV$ are reported. The \pPb and \PbPb data sets correspond to integrated luminosities of 186\nbinv and 0.607\nbinv, respectively, and were collected using the CMS detector at the CERN LHC. The charge-dependent correlations are characterized by balance functions of same- and opposite-sign particle pairs. The balance functions, which contain information about the creation time of charged particle pairs and the development of collectivity, are studied as functions of relative pseudorapidity ($\Delta \eta$) and relative azimuthal angle ($\Delta \phi$), for various multiplicity and transverse momentum (\pt) intervals. A multiplicity dependence of the balance function is observed in $\Delta \eta$ and $\Delta \phi$ for both systems. The width of the balance functions decreases towards high-multiplicity collisions in the momentum region $<$2\GeV, for \pPb and \PbPb results. Integrals of the balance functions are presented in both systems, and a mild dependence of the charge-balancing fractions on multiplicity is observed. No multiplicity dependence is observed at higher transverse momentum. The data are compared with \HYDJET, \HIJING, and \AMPT generator predictions, none of which capture completely the multiplicity dependence seen in the data. The comparison of results with different center-of-mass energies suggests that the balance functions become narrower at higher energies, which is consistent with the idea of delayed hadronization and the effect of radial flow.
}

\hypersetup{%
pdfauthor={CMS Collaboration},%
pdftitle={Multiplicity and transverse momentum dependence of charge balance function in pPb and PbPb collisions},%
pdfsubject={CMS},%
pdfkeywords={CMS, heavy ion, balance function, charge conservation, radial flow}} 

\maketitle 

\section{Introduction}
\label{sec:intro}

Ultrarelativistic heavy-ion collisions provide a means to investigate the properties of the quark-gluon plasma (QGP)~\cite{ARSENE20051,BACK200528,ADAMS2005102,ADCOX2005184,M_ller_2012,Armesto_2016}. This state of matter is formed in the first few moments ($\sim$3 $\times 10^{-24}$ seconds) of such collisions, and is characterized by large energy density compressed into a small volume. Two-particle angular correlations are used as a tool to study the properties of the system created in high-energy collisions~\cite{qgpcms_2pc,cmshin12015,PhysRevC.76.024914,Dusling_2016,gqp_hin11001,hin11006}. These correlations are usually measured as functions of $\Delta \eta$  and $\Delta \phi$, which denote the relative angle in pseudorapidity ($\eta$) and azimuthal angle ($\phi$), respectively. Many physical phenomena manifest themselves in these correlations: the collective behavior of the medium can be apparent in the long-range longitudinal structure at small $\Delta \phi$ angles~\cite{gqp_hin11001,hin11006,cmsppridge,PhysRevC.76.047901}, the jet-related correlations can be observed as a peak at small relative $\Delta \eta, \Delta \phi$ angles together with a broad $\Delta \eta$ structure at $\Delta \phi \sim \pi$, while correlations in relative momentum caused by resonance decays or quantum statistics, such as Bose--Einstein correlations, will appear at small relative angles only.

Recent theoretical studies suggest that the QGP evolves during a high-energy heavy ion collision, producing quarks in two waves~\cite{scott_chemqgp}. The first wave occurs during the first 5--10\unit{fm}$/c$ of the collision, when gluons thermalize into the QGP, followed by a isentropic expansion and hadronization, where most quark production occurs. The motivation behind the study in PbPb collisions stems from the expectation to differentiate between the early and late production of charges within the collision dynamics~\cite{ALICE:2021hjb}.

Charged-particle production is subject to local charge conservation, which ensures that for each created charge there is always an opposite balancing partner~\cite{scott_chemqgp,PhysRevLett.85.2689}. The electric charge balance function represents the probability that a charge $+q$ will see its balancing charge $-q$ within a limited range of $\Delta\phi$ and $\Delta\eta$~\cite{PhysRevLett.85.2689}. The width of the balance function represents a powerful tool to study the dynamics of particle production~\cite{Khachatryan:2016odn,qgpcms_2pc,PhysRevC.76.024914,cmshin12015}. Specifically, the width of the balance function is expected to be narrower when the particles are produced at a later stage of the system evolution. Conversely, a wider distribution would correspond to charge creation earlier in the evolution. Additionally, collective medium expansion, specifically the radial flow, may also affect the observed width of the correlated distributions. The azimuthal width of the balance function depends on the strength of the radial flow, while its longitudinal spread is related to the longitudinal momentum as $\Delta \eta$ $\sqrt{\smash[b]{m_\mathrm{0}^{2}  + \pt^{2}}}$, where \pt is the transverse momentum and  $m_\mathrm{0}$ is the particle mass~\cite{icpaqgp}. Therefore, radial flow can contribute to the narrowing of the balance functions for more central collisions.

Additionally, balance functions may provide a sensitive probe to study the hadronization of jets in proton-proton (pp) collisions~\cite{ACCDHW:1979mjc}. They can be used as a tool for studying the chemistry of the quark-gluon plasma~\cite{scott_chemqgp,ALICE:2021hjb}, the collision dynamics~\cite{Voloshin:2003ud,BOZEK2005247}, the hydrochemistry of particle formation~\cite{Pratt:2021xvg,Pratt:2018ebf}, and the balancing particle production~\cite{Jeon:2001ue,Pratt:2015jsa}. Moreover, the balance functions binned in the relative azimuthal angle, $\Delta\phi$, can effectively determine the diffusivity of light quarks~\cite{Pratt:2021xvg,Pratt:2018ebf}. Thus, charge balance functions provide some of the most compelling evidence for forming a state of matter at chemical equilibrium with sufficient number of light quarks produced in the early stages of the collision~\cite{Pratt:2019pnd}. Balance function integrals relate to net-charge fluctuations, which are crucial to understanding the transition from hadronic matter to the deconfined state and estimating QGP susceptibilities~\cite{Jeon:2001ue,Pruneau:2019baa,Pratt:2012dz,Pratt:2021xvg}. Finally, the balance functions are also valuable for confirming the chiral magnetic effect. The latter predicts an electric charge separation along the direction of the magnetic field, which can be experimentally observed as a charge-dependent correlation in the momentum space~\cite{Ye:2018jwq}.

The STAR Collaboration has performed measurements of the balance function in various collision systems, including AuAu, dAu, and pp collisions~\cite{rhicbf}. In AuAu collisions at $\sqrtsNN = 200\GeV$, for particles of $\abs{\eta} < 1.0$, the balance function was found to have a strong centrality dependence in both $\Delta \eta$ and $\Delta \phi$. A similar measurement covering the range $\abs{\eta} < 0.8$ was reported by the ALICE Collaboration at the CERN LHC~\cite{ccubf}. These measurements demonstrate that charge separation at kinetic freeze-out is sensitive to the details of the hadronization dynamics. However, more quantitative comparisons are required between experimental measurements and theoretical predictions in balance function studies to fully understand the underlying physics. Extending the acceptance to cover more of the produced particle pairs could reveal additional details of the mechanism(s) driving the particle correlations.

In this paper the charge-balance function is measured over a wide coverage of $\abs{\eta} < 2.4$ by exploiting the large acceptance of the CMS detector~\cite{CMS:2008xjf}. Results are presented as a function of charged-particle multiplicity and \pt in proton-lead (\pPb) and lead-lead (\PbPb) collisions at $\sqrtsNN = 8.16\TeV$ and 5.02\TeV, respectively. A comparison of the \PbPb and \pPb collisions can provide insight into the origin of long-range correlations observed in high-multiplicity \pPb collisions~\cite{cmshin12015}. This paper is organized as follows. The CMS detector is briefly discussed in Section~\ref{sec:CMS detector}. Section~\ref{sec:dataanslysis} describes the data samples and selection criteria. Section~\ref{sec:Ana} specifies the analysis procedure. Section~\ref{sec:sysuncert} reports on the various sources of systematic uncertainty. Section~\ref{sec:results} discusses the balance function results in both \pPb and \PbPb collisions, and comparison with models. Section~\ref{beamenergydep} presents the energy dependence of charge balance functions and comparisons with the previous lower $\sqrtsNN$ measurements. Finally, Section~\ref{sec:Summary} summarizes the findings. Tabulated results are provided in the HEPData record for this analysis~\cite{HEPData}.

\section{The CMS detector}
\label{sec:CMS detector}
The central feature of the CMS apparatus is a superconducting solenoid of 6\unit{m} internal diameter, providing a magnetic field of 3.8\unit{T}. Within the solenoid volume there is a silicon pixel and strip tracker, a lead tungstate crystal electromagnetic calorimeter (ECAL), and a brass and scintillator hadron calorimeter (HCAL), each composed of a barrel and two endcap sections. The silicon tracker consists of 1440 silicon pixel and 15,148 silicon strip detector modules (Phase-0). In 2017, an additional layer was added in both the barrel and endcap regions of the pixel detector and the number of silicon pixel modules increased to 1856 (Phase-1). The tracker detector measured the charged particles within the range $\abs{\eta} < 3.0$, and provides track resolutions of typically 1.5\% in \pt and 25--90 (20--75) \mum in the transverse impact parameter~\cite{CMS:2014pgm,DP-2020-049} in Phase-0 (-1) of pixel detector for nonisolated particles of $1 < \pt < 10\GeV$~\cite{Phase1Pixel}. The forward hadron (HF) calorimeter uses steel as an absorber and quartz fibers as the sensitive material. The two halves of the HF are located 11.2\unit{m} from the interaction region, one on each end, and together they provide coverage in the range $3.0 < \abs{\eta} < 5.2$. The HF calorimeters are subdivided into ``towers'' with $\Delta \eta \times \Delta \phi = 0.175 \times 0.175$, and energy deposited in a tower is treated as a detected hadron in this analysis. They also serve as luminosity monitors. A more detailed description of the CMS detector, together with a definition of the coordinate system used and the relevant kinematic variables can be found in Ref.~\cite{CMS:2008xjf}.

\section{Data samples and event selections}
\label{sec:dataanslysis}
The analysis presented in this paper is based on \PbPb collisions at $\sqrtsNN= 5.02\TeV$ collected by the CMS experiment in 2018. Approximately $4.27 \times 10^9$ \PbPb events were used, corresponding to an integrated luminosity 0.607\nbinv~\cite{CMS:LUM-18-001,CMS:LUM-17-003}. The data samples were collected by the CMS experiment with a two-tiered trigger system. The first level trigger (L1) consists of custom hardware processors and, uses information from the calorimeters and muon detectors to select events at a rate of around 100\unit{kHz} within a fixed latency of about 4\mus~\cite{CMS:2020cmk}. The second level or high-level trigger (HLT) consists of a farm of processors running a version of the full event reconstruction software optimized for fast processing, and reduces the event rate to around 1\unit{kHz} before data storage~\cite{CMS:2016ngn}. The MB events are triggered by requiring signals above the readout threshold of 3 \GeV in each of the HF calorimeters\cite{CMS:2016ngn}. Further selections are applied offline to reject events from background processes (beam-gas interactions and nonhadronic collisions), as discussed in Ref.~\cite{strangehadron}. In the offline analysis, events are required to have at least one interaction vertex, based on two or more reconstructed tracks, with a distance of less than 15 cm from the center of the nominal interaction point along the beam axis, $z_\text{vtx}$. The primary vertex is taken to be the vertex corresponding to the highest track multiplicity in the event, evaluated using tracking information alone, as described in Section 9.4.1 of Ref.~\cite{CMS-TDR-15-02}. In the final analysis, the \PbPb collision events are required to have at least two calorimeter towers in each HF detector with energy deposits of more than 4 \GeV per tower. These criteria select $(99 \pm 2)$\% of inelastic hadronic \PbPb collisions. Finding values higher than 100\% reflects the possible presence of ultra-peripheral (nonhadronic) collisions in the selected event sample.\\\

The \pPb data were recorded in 2016 and approximately $1.37 \times 10^9$ events were used, corresponding to an integrated luminosity of 186\nbinv~\cite{CMS:LUM-17-002,CMS:LUM-17-003}. The beam energies were $6.5\TeV$ for protons and $2.56\TeV$ per nucleon for lead nuclei, resulting in $\sqrtsNN= 8.16\TeV$. The nucleon-nucleon center of mass in the \pPb collisions is not at rest with respect to the laboratory frame because of the energy difference between the colliding particles. Massless particles emitted at $\eta_\text{cm} = 0$ in the nucleon-nucleon center-of-mass frame will be detected at $\eta = -0.465$ (clockwise proton beam) or $0.465$ (counterclockwise proton beam) in the laboratory frame. To select high-multiplicity \pPb collisions, a dedicated high-multiplicity trigger was implemented~\cite{strangehadron}. At L1, the \pPb events were triggered by requiring at least one track with $\pt > 0.4\GeV$  in the pixel tracker during a \pPb bunch crossing and at least one tower in one of the two HF detectors having an energy above 1 \GeV.  In addition, the total number of ECAL\text{+}HCAL towers with the transverse energies above a threshold of 0.5\GeV is required to exceed 120 (ECAL) and 150 (HCAL). The events that pass the L1 trigger are subsequently processed at the HLT. \\

Track reconstruction is performed online as part of the HLT with the same reconstruction algorithm used offline~\cite{CMS:2016ngn}. The number of tracks with $\abs{\eta} < 2.4$ and $\pt > 0.4\GeV$ (denoted as the primary tracks, i.e. originated at the primary vertex and satisfying the high-purity criteria ~\cite{CMS:2014pgm}), and a distance of closest approach of less than 0.12 cm to the primary vertex, is determined for each event~\cite{symmcum}. The primary tracks are used to perform the analysis, and to define event categories based on the charged-particle multiplicity ($N_\text{trk}^\text{offline}$). The multiplicity classification (120--150, 150--185, 185--250, $\geq$250)  in this analysis is identical to that used in Ref.~\cite{2013213}, where more details are provided, including a table relating $N_\text{trk}^\text{offline}$ to the fraction of minimum bias triggered events. When measuring the charge-balance function in \pPb collisions, the same event may contain multiple independent interactions (pileup), which constitutes a background for the analysis of high-multiplicity events. The average number of collisions per event in \pPb data varied between 0.10--0.25, and is negligible in \PbPb collisions. A similar procedure to that described in~\cite{symmcum} is used for identifying and rejecting events with pileup, which is based on the number of tracks associated with each reconstructed vertex and the distance between the vertices.

\section{Analysis methods}
\label{sec:Ana}
Charged particle tracks are selected if the significance of the longitudinal ($\rd_{z}$) and transverse ($\rd_{xy}$) distance from the beam axis satisfies $\abs{\rd_{z}}/\sigma_{z} < 3$ and $\abs{\rd_{xy}}/\sigma_{xy} < 3$, where $\sigma_{z}$ and $\sigma_{xy}$ are the measurement uncertainties. The relative uncertainty in \pt, $\sigma_{\pt}/\pt$, must be less than 10\%. To ensure high tracking efficiency and to reduce the rate of misreconstructed tracks, particles are selected within $\abs{\eta} < 2.4$. For this analysis, we have applied a minimum \pt cutoff value of 0.4 (0.5) \GeV for \pPb (\PbPb) collisions. Simulation studies based on \HYDJET (version 1.8) ~\cite{Hydjet}, \AMPT (version 1.1)~\cite{ampt} in \PbPb and \HIJING (version 1.3)~\cite{HIJING,HIJING2} in \pPb are used to estimate the geometrical acceptance and efficiency for the primary track reconstruction as well as the rate of misreconstructed tracks. Each reconstructed track is weighted by the inverse of the correction factor, $f_\mathrm{c} = {AE}/(1-F)$, as a function of pseudorapidity and transverse momentum. The weight factor accounts for the detector acceptance $A(\eta,\pt)$, reconstruction efficiency $E(\eta,\pt)$, and the fraction of misreconstructed tracks $F(\eta,\pt)$.
The acceptance is defined as the probability that a charged particle generates enough hits in the tracker to be reconstructed by the track-finding algorithm, while the efficiency is defined as the likelihood that these hits will be used to reconstruct a track with parameters representative of the original particle. A detailed analysis of tracking performance based on Monte Carlo (MC) simulations and collision data can be found in Ref~\cite{CMS:2010mua}. Simulated MC events show that the combined geometrical acceptance and reconstruction efficiency for the primary tracks is about 60\% for the 0--5\% most central PbPb collisions over the full acceptance of $\abs{\eta} < 2.4$ and 65\% for $\abs{\eta} < 1.0$. The fraction of misreconstructed tracks is within 1--2\% for $\abs{\eta} < 1.0$ and 14\% for $\abs{\eta} < 2.4$. The contribution due to the secondary tracks coming from the beampipe and the detector material is also considered in this analysis and this found to be less than 0.01\% or negligible. We note that the requirements on the distance from the primary vertex imposed on the selected tracks, as described above, are effective in removing the $\Pep\Pem$ contamination in the opposite-sign particle correlations. This contamination stems from photon conversions in the detector material and is suppressed effectively, thanks to the high resolution of inner pixel detector layers and their proximity to the beam pipe responsible for the majority of this type of background. Track splitting and merging can impact pairs of tracks in close spatial and momentum proximity, affecting two-particle correlations. We studied the potential impact of these effects on the physics quantities of interest by comparing the nominal results with those constructed by removing all charged-particle pairs with momentum separation below a given threshold (minimal allowed separation of 0.1, 0.15, and 0.2 \GeV were considered) along with the cos($\theta)< 0.9999$ requirement, which yielded negligible difference in the final measurements. In \PbPb collisions, additional selections are applied to reject the misreconstructed tracks: the number of hits in the silicon tracker is required to be larger than 11 and the $\chi^{2}$ per degree of freedom per layer of the silicon detector must be less than 0.18. Systematic uncertainty calculations related to the track selection variations are discussed in the next Section. The MC simulations of the CMS detector response are based on \GEANTfour~\cite{GEANT4}. The \PbPb collision centrality is defined as a fraction of the inelastic hadronic cross section, with 0\% corresponding to the full overlap of the two colliding nuclei. The event centrality is determined offline and is based on the total transverse energy measured in the HF calorimeters, using the methodology described in Ref.~\cite{CMS:2011iwn}. The value of \nch (charged-particle multiplicity) is corrected for the tracking efficiency and misidentification rate in both systems. For the \nch calculation, a minimum \pt threshold of 0.5 (0.4) \GeV is applied for \PbPb (\pPb) collisions. The centrality binning for \PbPb and multiplicity binning for \pPb collisions used for this measurement are listed in Table~\ref{fig:nchtable}. Table~\ref{fig:nchtable} also presents values of the corrected average charged-particle multiplicity $\langle \nch \rangle$ within $\abs{\eta} < 2.4$ for different centrality bins, and multiplicities in \PbPb and \pPb collisions~\cite{2013213}.

\begin{table}[htb]
\topcaption{Corrected average \nch ($\langle \nch \rangle$) values, calculated for different multiplicities in \PbPb collisions at $\sqrtsNN = 5.02\TeV$ and in \pPb collisions at 8.16\TeV.}
\label{fig:nchtable}
\centering
\begin{tabular}{@{\;\;\;\;}c@{\;\;\;}c@{\;\;\;}c@{\;\;\;}c@{\;\;\;}c@{\;\;\;}c@{}c@{\qquad}c@{\;\;\;}c@{\;\;\;}c@{\;\;\;}c@{}}
\hline
\multicolumn{2}{c}{\PbPb} & & \multicolumn{2}{c}{\pPb} \\
 \cline{1-2} \cline{4-5}
Centrality {(\%)} & {$\langle \nch \rangle$ } && {$N_\mathrm{trk}^{\mathrm{offline}}$} & {$\langle \nch \rangle$}\\

\hline
        0--10    & 3770 $\pm 189$    && 0--40     &    24 $\pm 1$\\
        10--20   & 2540 $\pm 127$   && 40--80    &    73  $\pm 3$\\
        20--30   & 1678 $\pm 84$   && 80--120   &    118  $\pm 5$\\
        30--40   & 1057 $\pm 53$   && 120--150  &    165  $\pm 7$\\
        40--50   & 620  $\pm 31$   && 150--165  &    196  $\pm 8$\\
        50--60   & 328  $\pm 16$   && 165--185  &    214  $\pm 9$\\
        60--70   & 160  $\pm 8$   && 185--200  &    236   $\pm 9$\\
        70--80   & 65   $\pm 3$   && 200--225  &    254   $\pm 10$\\
        & && 225--250 & 285  $\pm 11$\\
        & && 250--270 & 314  $\pm 13$\\
        & && 270--300 & 342  $\pm 14$\\

\hline
\end{tabular}
\end{table}

The differential correlation function is constructed using the standard CMS method~\cite{2013213, qgpcms_2pc, strangehadron,cmsppridge,gqp_hin11001, hin11006, cmshin12015}. In each event, every ``trigger'' particle within a specified \pt interval is matched with other ``associated'' particles within a corresponding \pt interval. The trigger and associated particles may be selected from the same or different \pt intervals~\cite{jet1,jet2,ALICE:2018jco}. The trigger particles are defined, for each track multiplicity class, as charged particles originating from the primary vertex (PV) within a given \pt ranges and $\abs{\eta} < 2.4$. There can be more than one trigger particle in the event, the total number of trigger particles is denoted as $N_{\text{trig}}$. The signal distribution $S(\Delta \eta,\Delta \phi)$ is constructed by using pairs of particles within the same event per trigger particle ~\cite{qgpcms_2pc},
\begin{equation} \label{eqn_sig}
S(\Delta \eta,\Delta \phi) = \frac{1}{N_\text{trig}}\frac{\rd^{2}N^\text{same}}{{\rd\Delta \eta} \ {\rd\Delta \phi}},
\end{equation}
where $N^\text{same}$ is the number of pairs in ($\Delta \eta$, $\Delta \phi$)  bin where $\Delta \eta$ and $\Delta \phi$ are the relative angular variables between the particles of the pairs. The so-called mixed event distribution $M(\Delta \eta, \Delta \phi)$ is constructed using the mixed event technique~\cite{symmcum} by pairing the trigger particles in each event with the associated particles from 10 different random events within the same 2 cm wide $z_\text{vtx}$ range and from the same track multiplicity class, as shown in Table~\ref{fig:nchtable}:
\begin{equation} \label{eqn_bkg}
 M(\Delta \eta,\Delta \phi) = \frac{1}{N_\text{trig}}\frac{\rd^{2}N^\text{mix}}{\rd\Delta \eta \ \rd\Delta \phi},
\end{equation}
where $N^\text{mix}$ is the number of mixed event pairs in a given ($\Delta \eta,\Delta \phi$) bin. The correlation function is constructed using the normalized signal and mixed event distributions:
\begin{equation} \label{eqn_corr}
 C_{2}(\Delta \eta,\Delta \phi)= M(0,0) \frac{S(\Delta \eta,\Delta \phi)}{M(\Delta \eta,\Delta \phi)},
\end{equation}
where $M(0,0)$ represents the mixed-event associated yield for both particles of the pair going in approximately the same direction (with a bin width of 0.3 in $\Delta\eta$ and $\pi/6$ in $\Delta\phi$). The $M(0,0)$ bin has the highest pair-acceptance, as for a given particle passing the analysis selection criteria, the conditional probability for the second particle to be accepted as well is highest in the closest spatial proximity to the first one. Therefore, the ratio $M(0,0)/M(\Delta \eta,\Delta \phi)$ is the pair-acceptance correction factor used to derive the corrected per-trigger-particle associated yield distribution~\cite{CMS:2014und}. The signal and mixed-event distributions are first calculated for each event, and then averaged over all the events within the same track multiplicity class, for each \pt bin. The correlation function is denoted by $C_{2}(\Delta \eta,\Delta \phi)$ in terms of the relative $\Delta \eta$ and $\Delta \phi$ variables. Using the positively and negatively charged particles, we construct four different charge combinations, which can be written as $C_{2}(+,-),C_{2}(+,+),C_{2}(-,+), C_{2}(-,-)$. The functions $C_{2}(+,+)$ and $C_{2}(-,-)$ are called SS correlations, and the other two are called OS correlations. The SS correlations are affected by the Hanbury--Brown--Twiss effect~\cite{PhysRevC.78.054906,hbt}, by Coulomb repulsion, and by a contribution from minijet production~\cite{PhysRevD.84.112004}. The OS correlations contain a minijet component~\cite{PhysRevD.84.112004}, an attractive Coulomb contribution~\cite{PhysRevC.78.054906}, and correlations due to the decay of resonances. The OS and SS correlations exhibit long-range rapidity correlations, called ''ridge-like'' correlations. The balance function combines same-sign subtractions to isolate the opposing charge statistically~\cite{PhysRevLett.85.2689,scott_chemqgp,ALICE:2018jco}. The balance function $B(\Delta \eta,\Delta \phi)$ is defined as
\begin{equation} \label{eqn_balfun}
B(\Delta \eta,\Delta \phi)=\frac{1}{2}[C_{2}(+,-)+C_{2}(-,+)- C_{2}(+,+)-C_{2}(-,-)].
\end{equation}
\section{Systematic uncertainties}
\label{sec:sysuncert}
Systematic uncertainties are calculated by varying the event and track selections for both \PbPb and \pPb collisions events. The balance function is calculated for three ranges of z-vertex of PV: $\abs{v_{z}} < 3$ cm, $-15 < v_{z} < -3$ cm; and $3 < v_{z} < 15$ cm. Similarly, the track quality requirements are varied, by changing $\abs{\rd_{z}}/\sigma_{z}$ and $\abs{\rd_{xy}}/\sigma_{xy}$ from 2 to 5, $\sigma_{\pt}/\pt$ from 0.05--0.10, and the $\chi^{2}$ per layer from 0.15--0.18. Moreover, the centrality calibration is varied to estimate the related systematic uncertainties in the width of the balance function for \PbPb collisions. Finally, the impact of pileup in \pPb collisions is estimated by varying the pileup selection of events in the analysis by changing the required separation between reconstructed vertices and their numbers of associated tracks. The systematic uncertainties for each source are estimated from the difference between the nominal and varied results. The maximum variation is taken as the final systematic uncertainty for each source, and the total systematic uncertainty is evaluated by adding all the sources in quadrature. In \PbPb simulations a discrepancy between $\Delta \phi$ balance functions obtained for particle level information and from reconstructed particle tracks was observed. This discrepancy is related to a reduced track finding efficiency for close-by low-\pt ($<$2\GeV) tracks in central \PbPb collisions. A residual correction is a ratio of generated with reconstructed tracks, was obtained from MC simulations, where three models (\HYDJET, \HIJING and \AMPT) provided consistent correction functions for the range $0.3 <\abs{\Delta\eta}< 1.0$. The difference between corrected and uncorrected data was used as an conservative estimate of the corresponding systematic uncertainty. The maximum uncertainty was found to be 13.5\% in the $\langle \abs{\Delta\phi}\rangle$ comparison of the balance function discussed in Table~\ref{fig:nchtab}. 

\begin{table}[ht]
\topcaption{Summary of systematic uncertainties calculated in $\langle \abs{\Delta \eta} \rangle$ and $\langle \abs{\Delta\phi} \rangle$ for \PbPb collisions at {$\sqrtsNN = 5.02\TeV$} and \pPb collisions at 8.16\TeV.}
\centering
\begin{tabular}{@{\;\;\;\;}c@{\;\;\;}c@{\;\;\;}c@{\;\;\;}c@{\;\;\;}c@{\;\;\;}c@{}c@{\qquad}c@{\;\;\;}c@{\;\;\;}c@{\;\;\;}c@{}}
\hline
\multirow{2}{*}{Uncertainty source} &\multicolumn{2}{c}{\PbPb } & & \multicolumn{2}{c}{\pPb}\\
 \cline{2-3} \cline{5-6}
& $\langle \abs{\Delta \eta} \rangle$ & $\langle \abs{\Delta\phi} \rangle$ && ${\langle \abs{\Delta \eta} \rangle}$ & ${\langle \abs{\Delta \phi} \rangle}$\\

\hline

       Vertex selection            & 0.005 & 0.009  && 0.016 & 0.012\\
       Centrality calibration      & 0.005 & 0.005  &&  ---  &  --- \\
       Pileup selection            &  ---  & ---    && 0.002 & 0.001\\
       Track quality requirements  & 0.004 & 0.012  && 0.017 & 0.004\\
       Tracking efficiency         & 0.001 & 0.005  && 0.001 & 0.003\\
       MC closure test             & 0.002 & 0.062  && 0.001 & 0.001\\
\hline
\end{tabular}
\label{fig:sys1}
\end{table}

Table~\ref{fig:sys1} lists the maximum absolute systematic uncertainties calculated for both collision systems in terms of $\langle \abs{\Delta \eta} \rangle$ and $\langle \abs{\Delta\phi} \rangle$. The systematic uncertainties in the amplitudes, widths, and integrals of the balance functions, due to the track and vertex selections, are estimated by varying these selections. For this analysis, we applied a minimum \pt requirement (0.4 \GeV for \pPb and 0.5 \GeV for \PbPb collisions) because of the inefficiency in the low-\pt tracking. 
\begin{table}[htb]
\topcaption{Summary of percentage systematic uncertainties calculated in $\langle \abs{\Delta \eta} \rangle$ and $\langle \abs{\Delta\phi} \rangle$ for \PbPb collisions at {$\sqrtsNN = 5.02\TeV$} and \pPb collisions at 8.16\TeV.}
\centering
\begin{tabular}{@{\;\;\;\;}c@{\;\;\;}c@{\;\;\;}c@{\;\;\;}c@{\;\;\;}c@{\;\;\;}c@{}c@{\qquad}c@{\;\;\;}c@{\;\;\;}c@{\;\;\;}c@{}}
\hline
\multirow{2}{*}{Uncertainty source} &\multicolumn{2}{c}{\PbPb (\%)} & & \multicolumn{2}{c}{\pPb (\%)}\\
 \cline{2-3} \cline{5-6}
& $\langle \abs{\Delta \eta} \rangle$ & $\langle \abs{\Delta\phi} \rangle$ && ${\langle \abs{\Delta \eta} \rangle}$ & ${\langle \abs{\Delta \phi} \rangle}$\\

\hline

       Vertex selection  & 0.8   & 1.3 &&   3.2 & 0.7\\
       Centrality calibration   & 0.8 & 0.8  && --- & ---\\
       Pileup selection   &  --- & ---  && 0.4 & 0.1\\
       Track quality requirements   & 0.7 & 3.5  &&2.7 & 2.8\\
       Tracking efficiency      & 1.2 & 1.0     &&1.0 & 3.0\\
       MC closure test    & 0.5 & 13.5   &&1.0 & 2.0\\
\hline
\end{tabular}
\label{fig:nchtab}
\end{table}

This measurement is also extended to higher values of \pt  ($2 < p_\text{T,asso} < 3 < p_\text{T,trig} < 4\GeV$, $3 < p_\text{T,asso} < 8 < p_\text{T,trig} < 15\GeV$). The \pt of the trigger particle is denoted by $p_\text{T, trig}$, whereas that of the associated particle is denoted by $p_\text{T, asso}$. The systematic uncertainty values from each source, in all multiplicity classes and $\pt<2\GeV$, are summarized in Table~\ref{fig:nchtab} for the two systems. The maximum systematic uncertainties in the width of the $\langle \abs{\Delta \eta}\rangle$ and $\langle \abs{\Delta\phi} \rangle$ are measured to be 6.0\% for \PbPb collisions and 3.0\% for \pPb collisions for intermediate values of transverse momentum, $2 < p_\text{T,asso} < 3 < p_\text{T,trig} < 4\GeV$. The maximum systematic uncertainties in $\langle \abs{\Delta \eta}\rangle$ and $\langle \abs{\Delta\phi} \rangle$ are found to be 4\% in \PbPb collisions; for \pPb collisions, 4\% and 6\% are found for the lower ($2 < p_\text{T,asso} < 3 < p_\text{T,trig} < 4\GeV$) and the higher ($3 < p_\text{T,asso} < 8 < p_\text{T,trig} < 15\GeV$) values of \pt, respectively. In addition, the systematic uncertainties are calculated for the integral of the balance function. The highest variation for \PbPb collisions is 3\%, whereas the maximum difference for \pPb collisions is 5\%.

\section{Results}
\label{sec:results}
\begin{figure}[thb]
  \centering
    \includegraphics[width=1.0\textwidth]{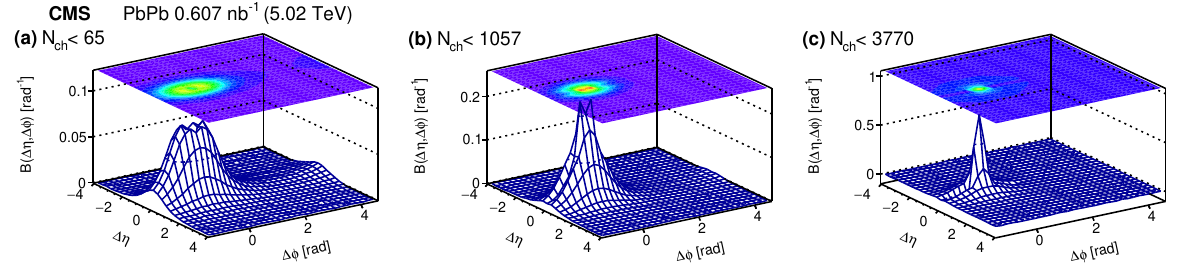}
    \\[1ex]
    \includegraphics[width=1.0\textwidth]{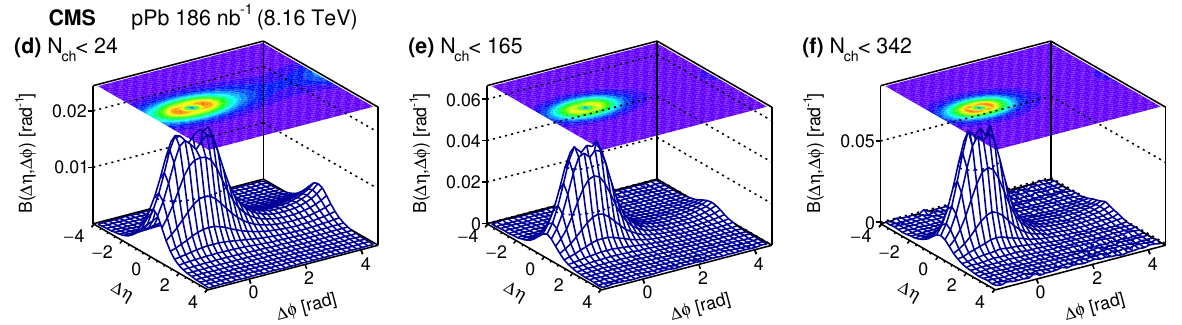}
  \caption{The balance function is shown in terms of $\Delta \eta$ and $\Delta \phi$ in \PbPb collisions at $\sqrtsNN = 5.02\TeV$ (upper panels) and for \pPb collisions at 8.16\TeV (lower panels). From left to right, the results are shown for the lowest to highest multiplicity classes in \PbPb and \pPb collisions. The trigger and associated particles in \PbPb (\pPb) collisions satisfy the condition $0.5 (0.4) < p_\text{T,asso} < p_\text{T,trig} < 2.0\GeV$.}
  \label{fig:2dbfpPb}
\end{figure}

The balance functions for nonidentified charged particles are presented as a function of $\Delta \eta$ and $\Delta \phi$ in different multiplicity classes and \pt ranges for both collision systems in Fig.~\ref{fig:2dbfpPb}.

The upper panels in Fig.~\ref{fig:2dbfpPb} show the centrality dependence of the charge-balance function in \PbPb collisions. The magnitude of the balance function changes with multiplicity, with higher values corresponding to collisions with higher multiplicity. A narrower balance function distribution is observed in central \PbPb collisions. This is consistent with particle production at later times in the collision process for the larger system created in more central collisions, leading to a smaller separation in $\Delta\eta$ and $\Delta\phi$~\cite{rhicbf}.

The lower panels in Fig.~\ref{fig:2dbfpPb} represent the multiplicity dependence of the balance function in \pPb collisions. The balance function is observed to also become narrower in $\Delta \eta$ and $\Delta \phi$ with increasing multiplicity. A similar depletion structure around $(\Delta\eta,\Delta\phi) = (0,0)$ is also seen in mid-central to peripheral \PbPb events, as shown in upper panels of Fig.~\ref{fig:2dbfpPb}  and previously in Ref.~\cite{alicebf}. This type of structure is more pronounced in \pPb collisions in the smaller range of multiplicities. One possible mechanism that could create such a structure in both collision systems is the charge-dependent short-range correlations, such as Coulomb attraction or repulsion, or quantum statistical correlations~\cite{deplection}.

\begin{figure}[thb]
    \centering
    \includegraphics[width=0.49\textwidth]{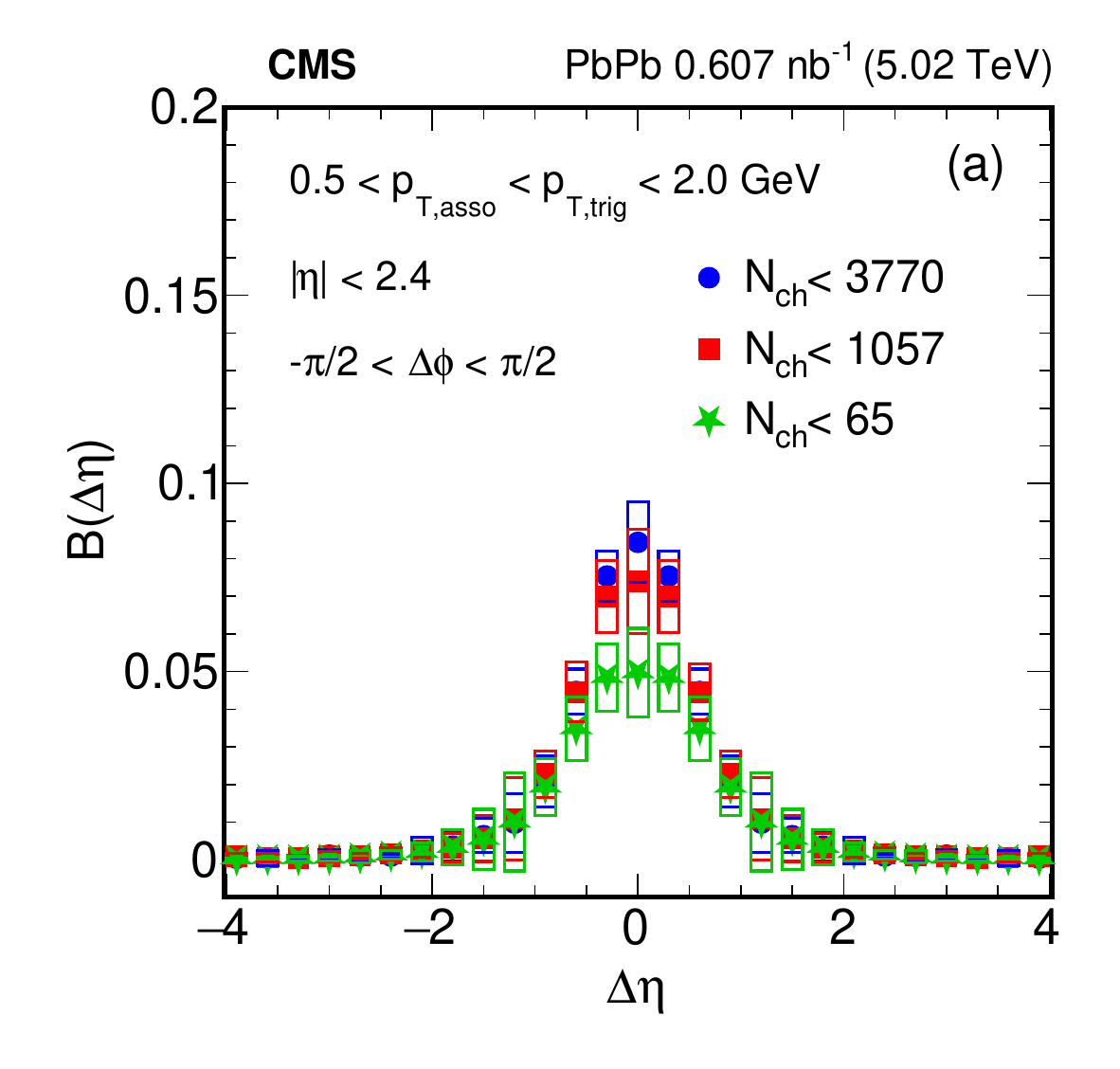}
    \includegraphics[width=0.49\textwidth]{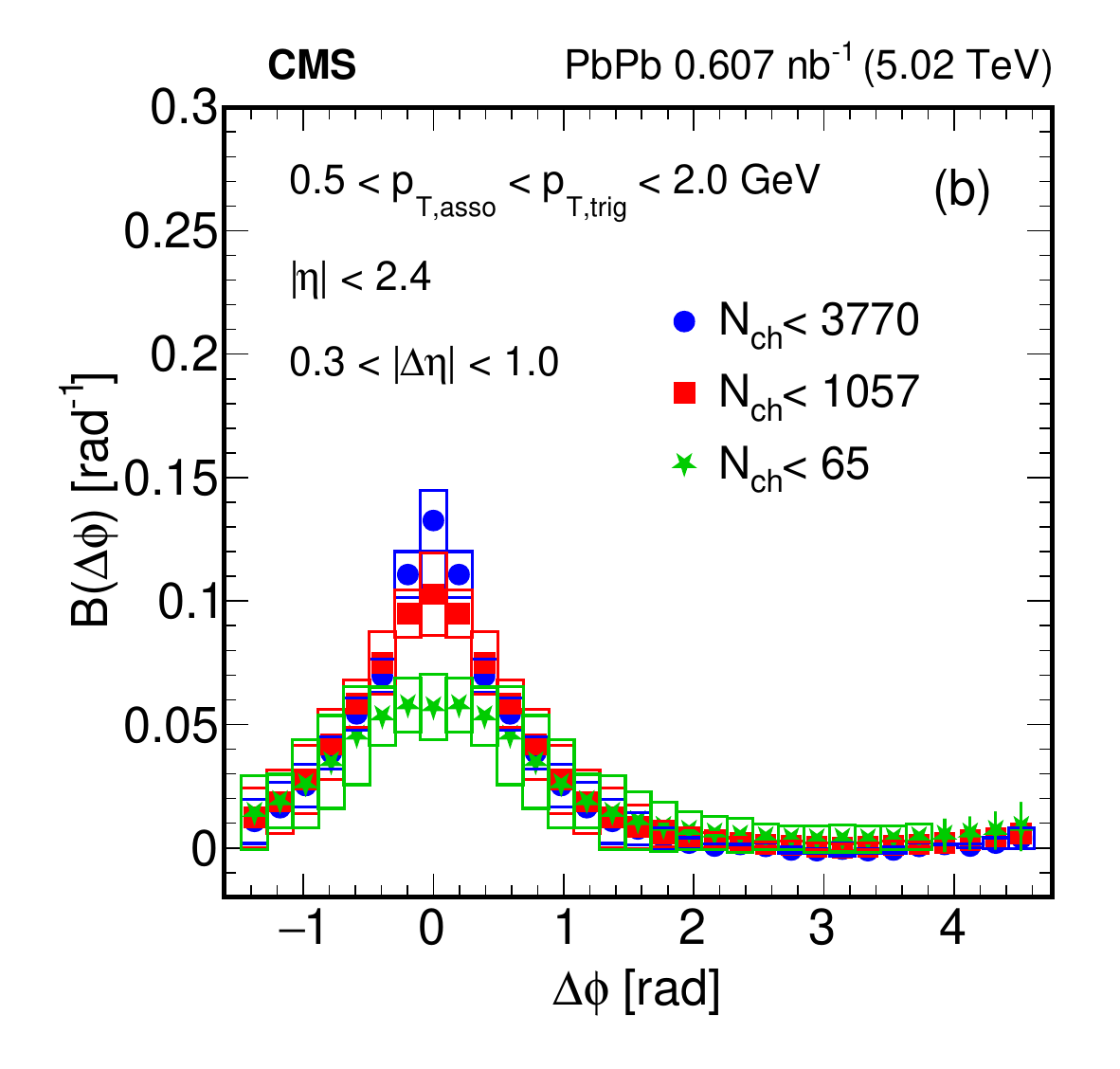}
    \includegraphics[width=0.49\textwidth]{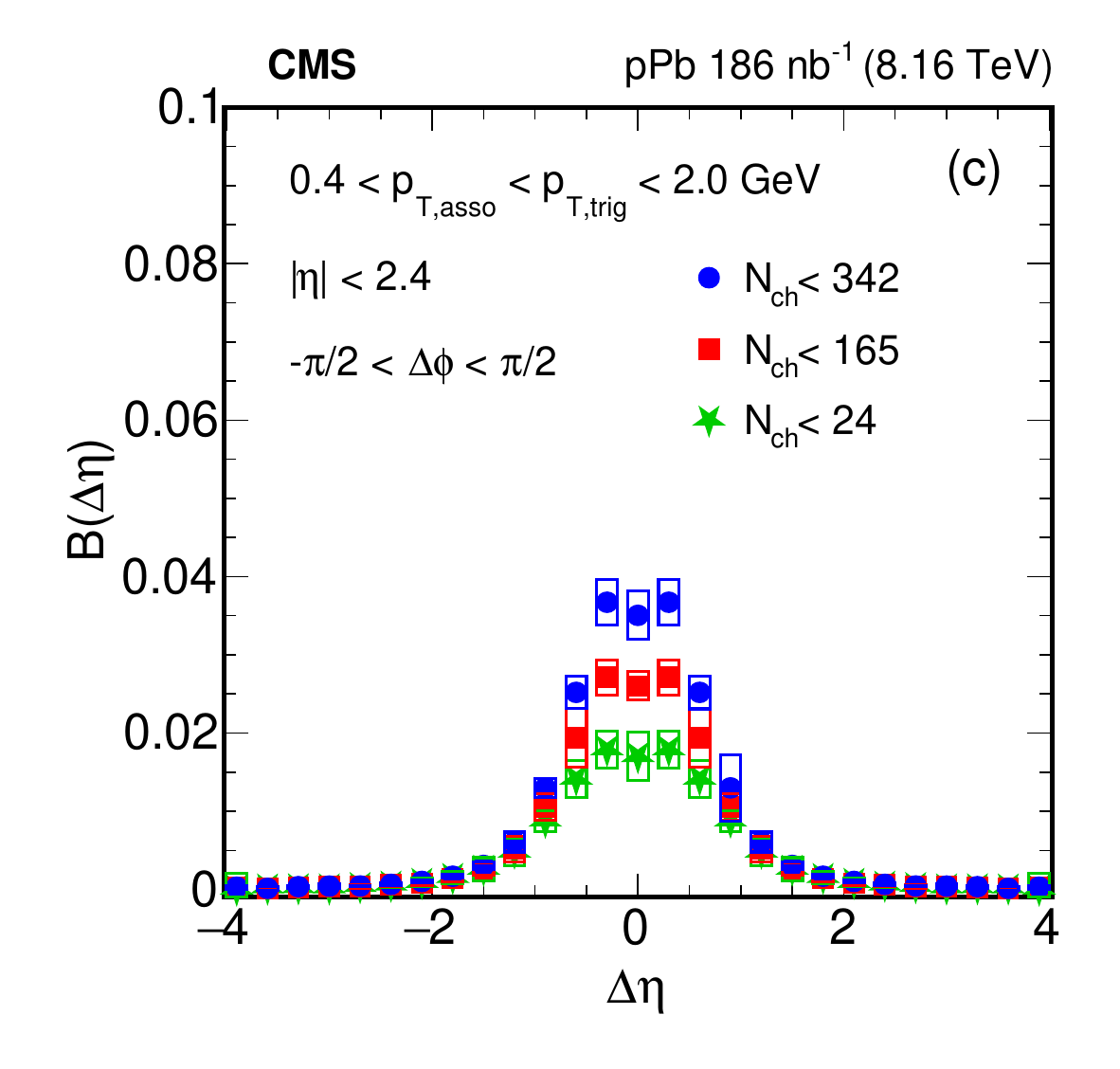}
    \includegraphics[width=0.49\textwidth]{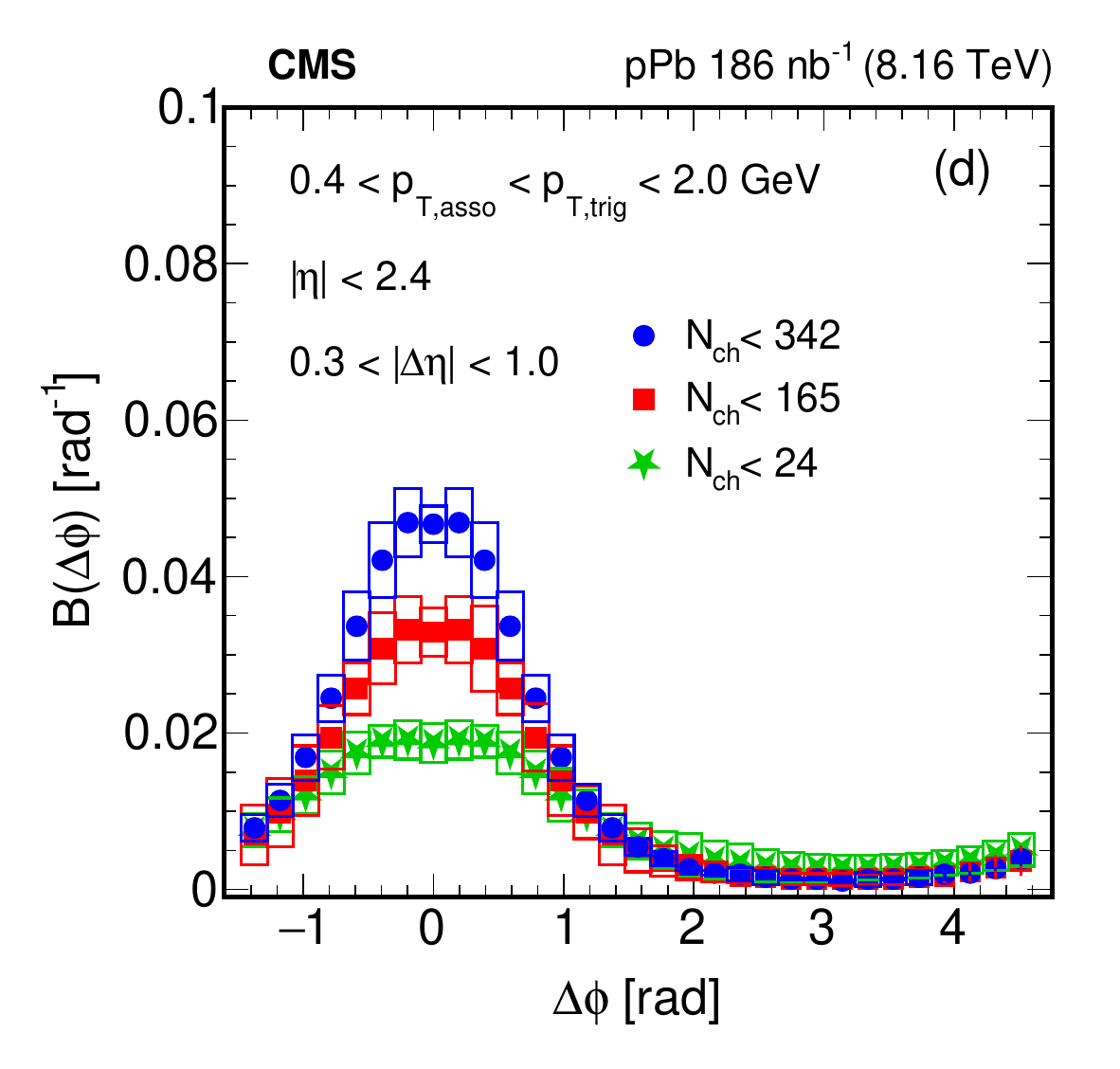}
  \caption{The projection of the balance function is presented in the upper panel for \PbPb (lower panel for \pPb) collisions as a function of $\Delta\eta$ (left column) and in $\Delta\phi$ (right column). The statistical uncertainties of the data points are smaller than the marker size and rectangular boxes indicate the systematic uncertainties.}
\label{fig:ondp}
\end{figure}

Figure~\ref{fig:ondp} shows 1D projections, derived for $\Delta\eta$ ($\abs{\Delta \phi} < \pi/2$ range) and $\Delta \phi$ ($0.3 < \abs{\Delta \eta} < 1.0$ range). The balance function distributions show a strong multiplicity dependence in $\Delta\eta$ and $\Delta\phi$ on the near-side $\abs{\Delta \phi} < \pi/2 $, for both collision systems. As before, a narrower peak is observed in high-multiplicity \pPb collisions as compared to low-multiplicity ones.

\begin{figure}[thb]
    \centering
    \includegraphics[width=0.49\textwidth]{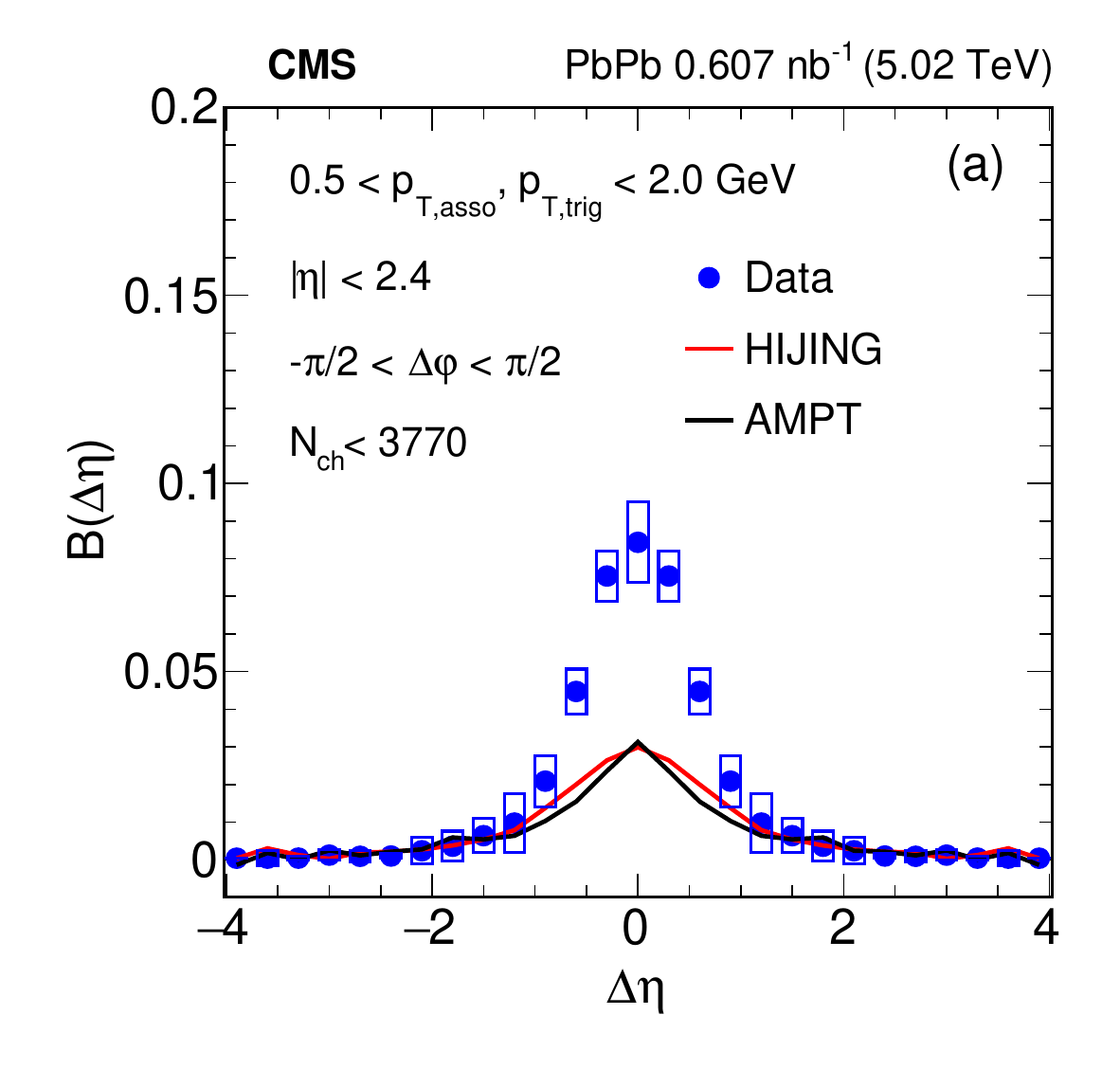}
    \includegraphics[width=0.49\textwidth]{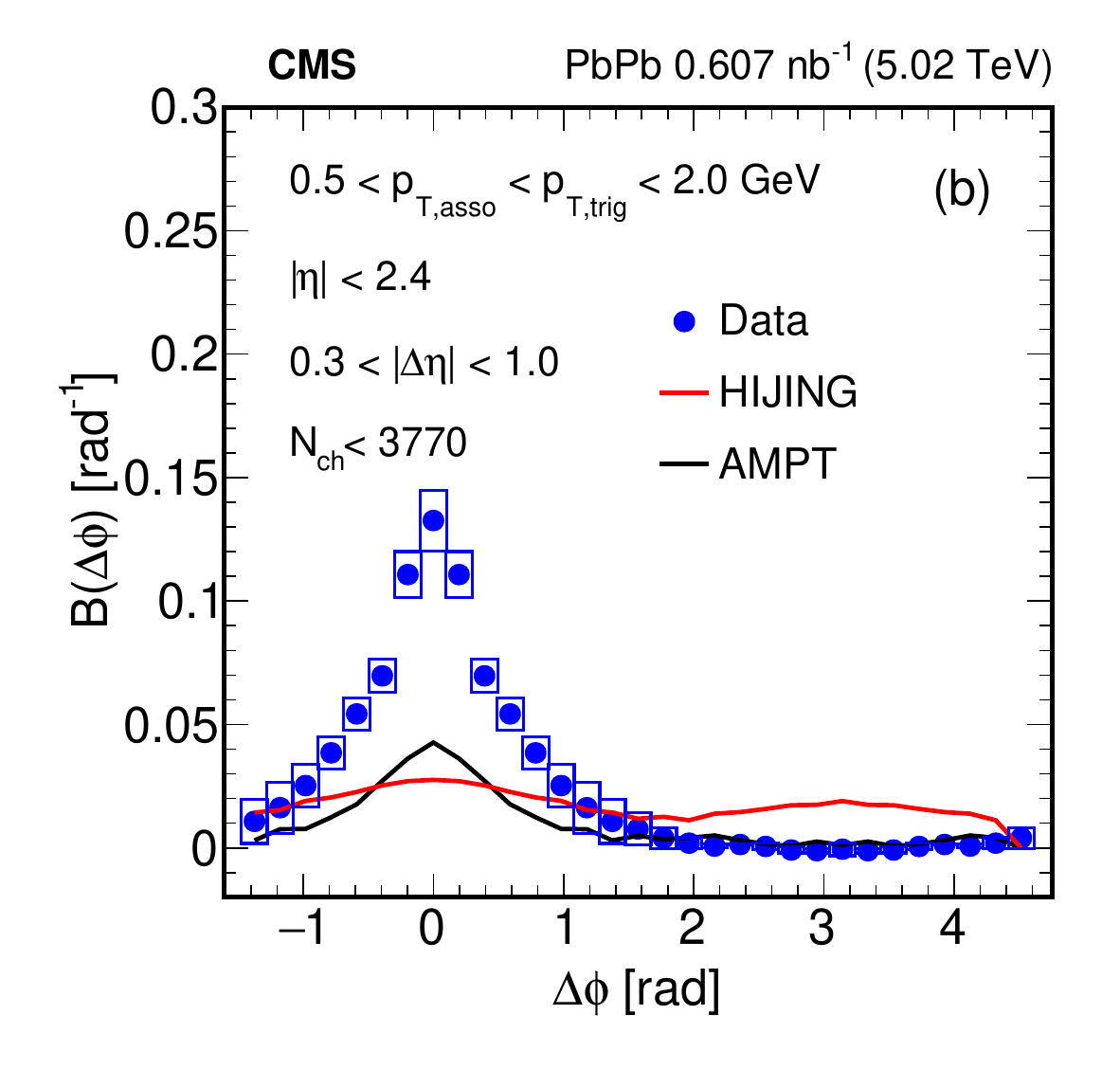}
    \includegraphics[width=0.49\textwidth]{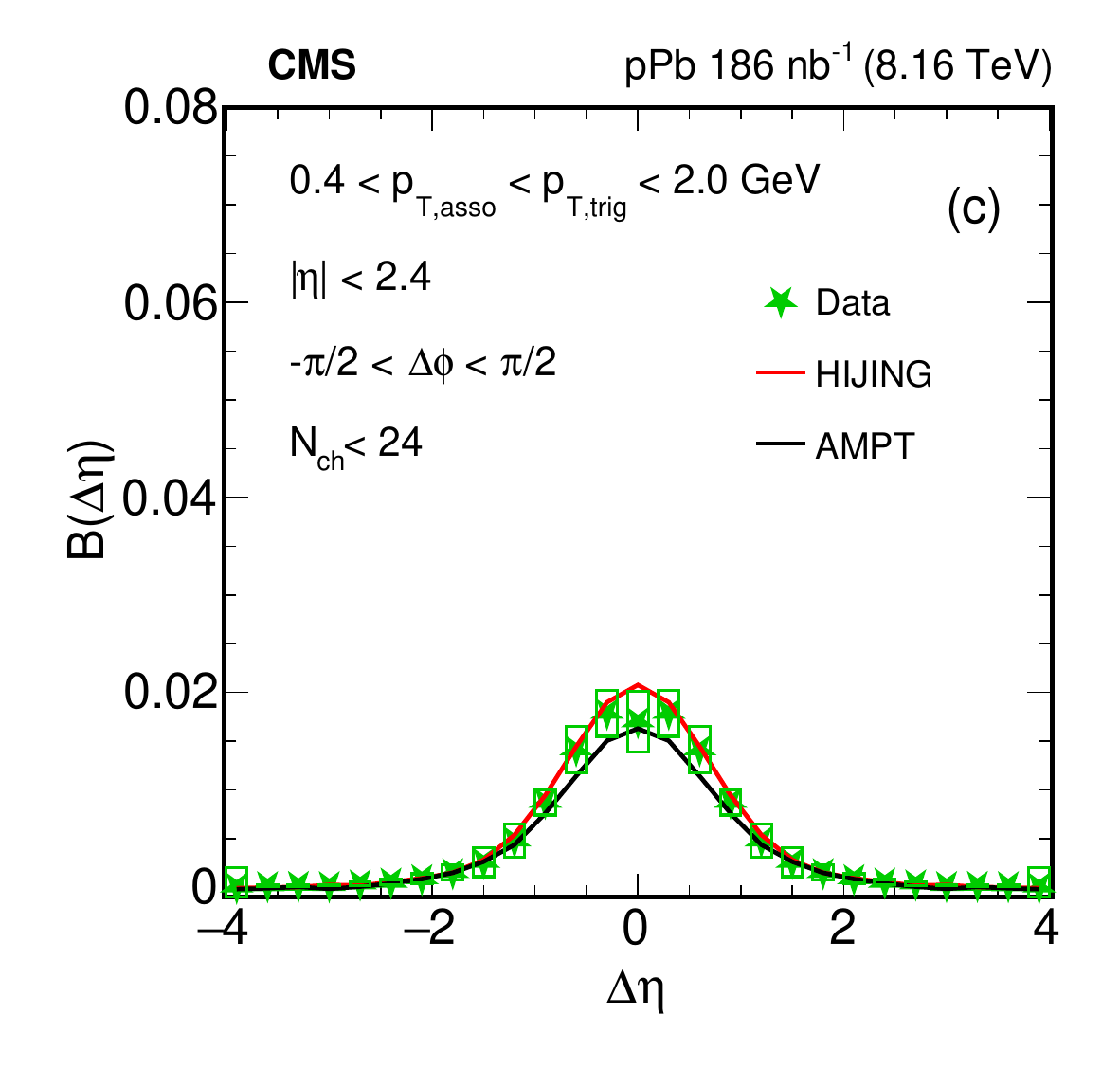}
    \includegraphics[width=0.49\textwidth]{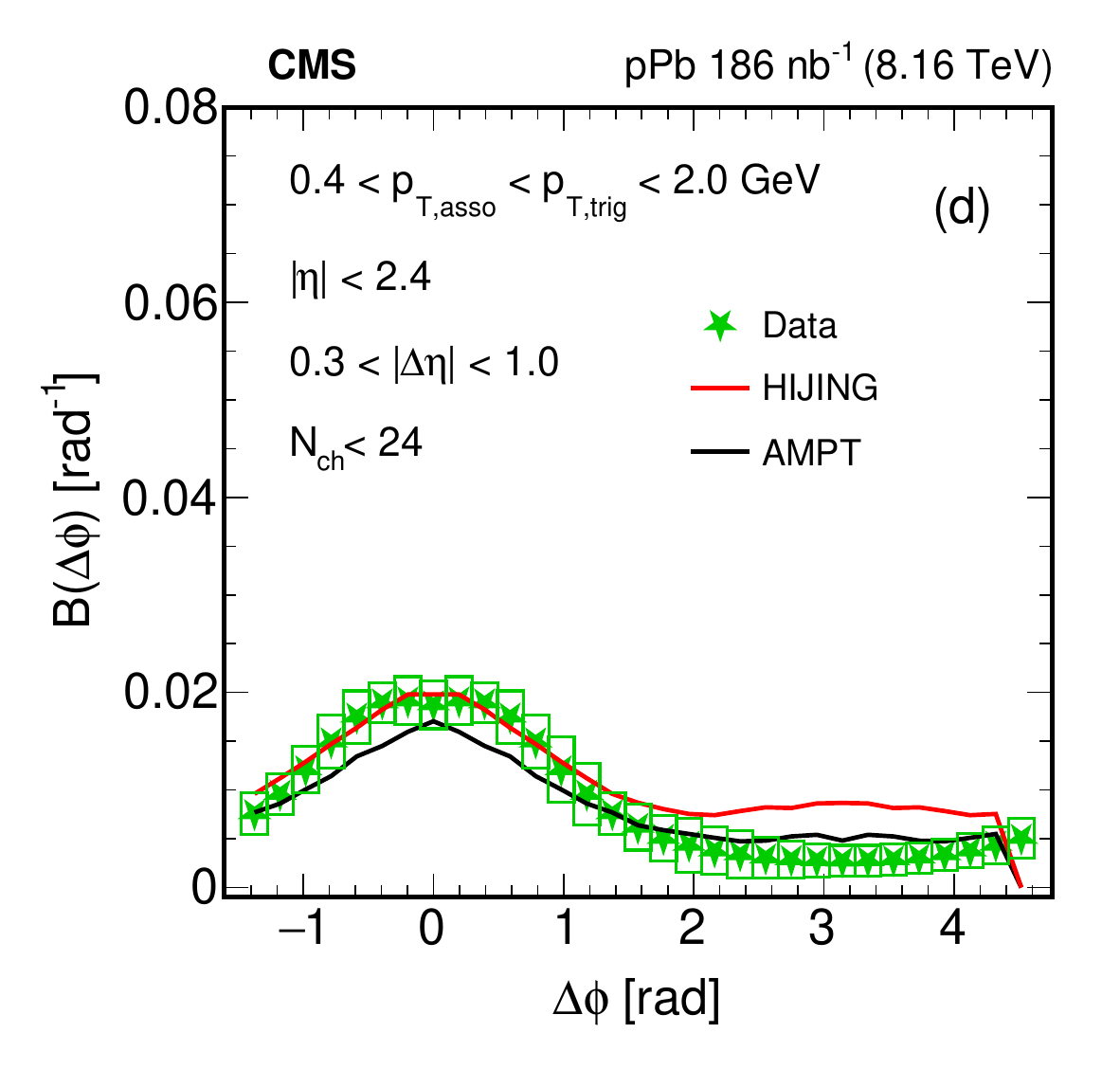}
  \caption{The comparison of the balance function with \AMPT and \HIJING event generators is presented in the upper panel ((a) and (b)) for \PbPb collisions and the lower panel ((c) and (d)) for \pPb collisions as a function of $\Delta\eta$ (left column) and in $\Delta\phi$ (right column). For \PbPb collisions, only the highest multiplicity ($\nch< 3770$) and for \pPb collisions, only the lowest multiplicity ($\nch< 24$) are shown.}
  \label{fig:pbpb_ppb_model}
\end{figure}

Figure \ref{fig:pbpb_ppb_model} presents the near-side projection of the balance functions in \PbPb and \pPb collisions, and its comparisons with different MC model calculations. Neither \AMPT nor \HIJING can fully explain the balance function projections in $\Delta\eta$ and $\Delta\phi$ in \PbPb collisions, as they both underestimate the balance function's magnitude, and anticipate far broader distributions. However, the one-dimensional projection is quantitatively in agreement with both the models in  $\Delta\eta$, while \AMPT slightly underestimates the \pPb data in $\Delta\phi$ comparisons.

\begin{figure}[thb]
    \centering
    \includegraphics[width=0.49\textwidth]{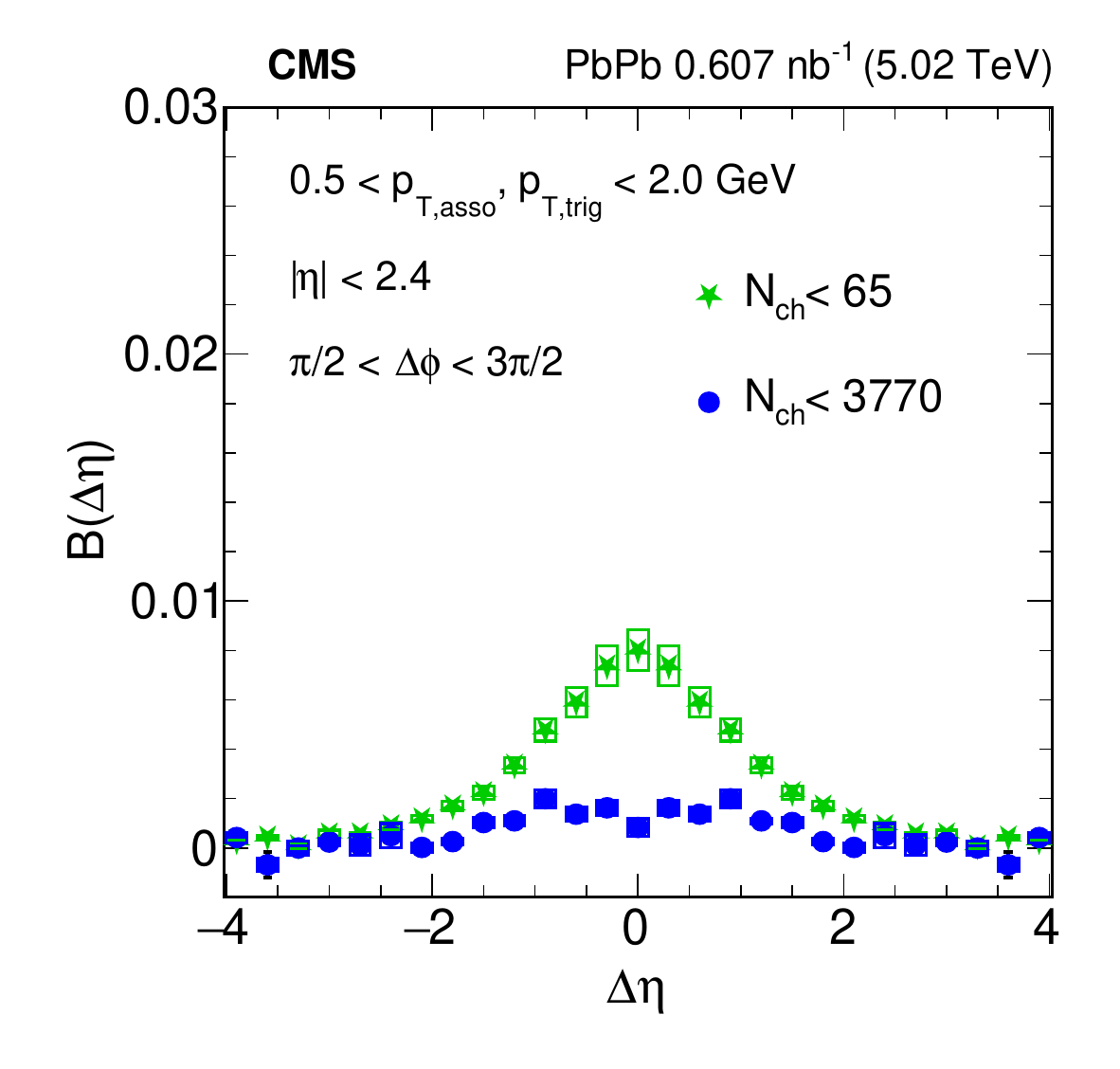}
    \includegraphics[width=0.49\textwidth]{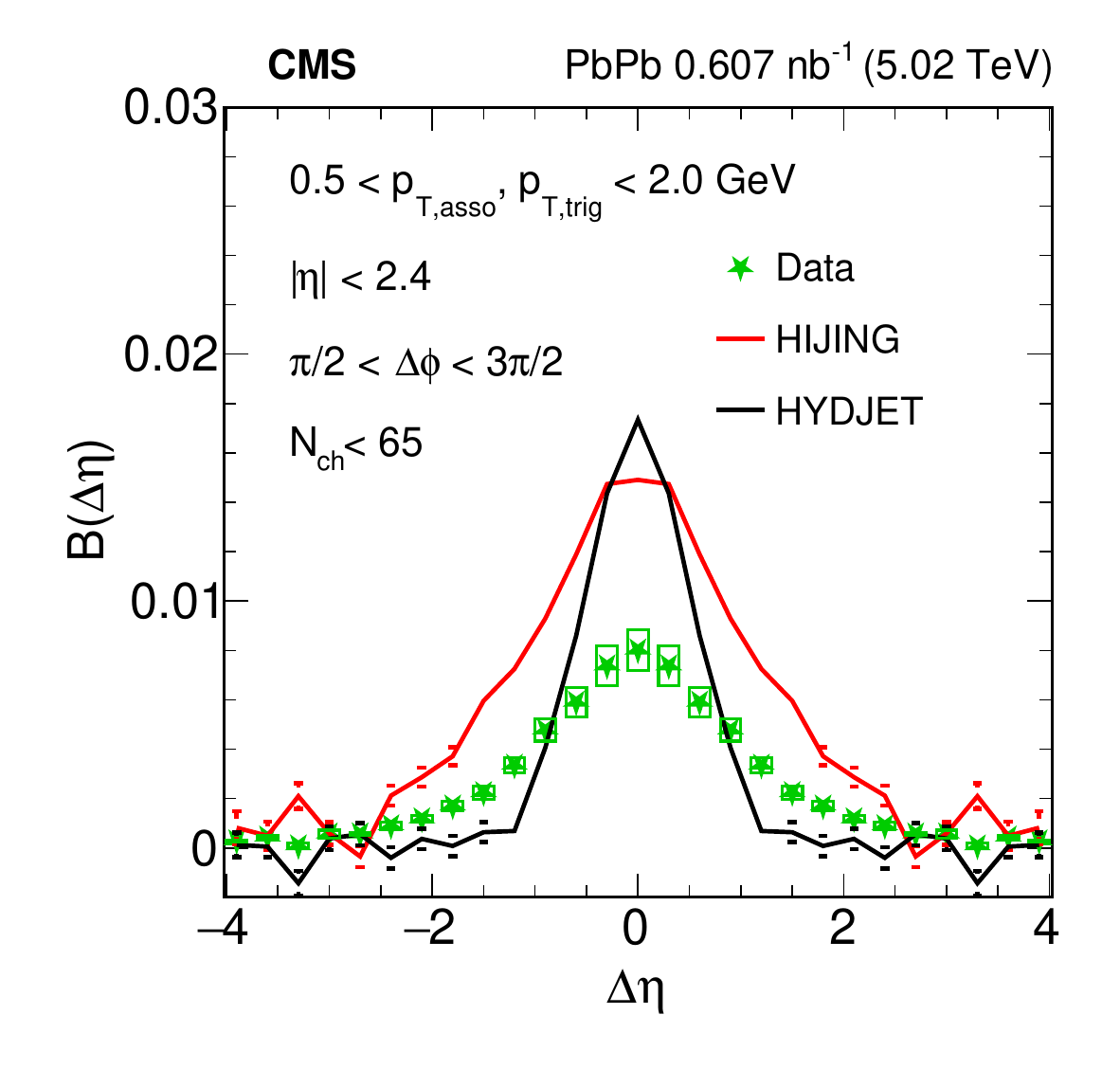}
    \includegraphics[width=0.49\textwidth]{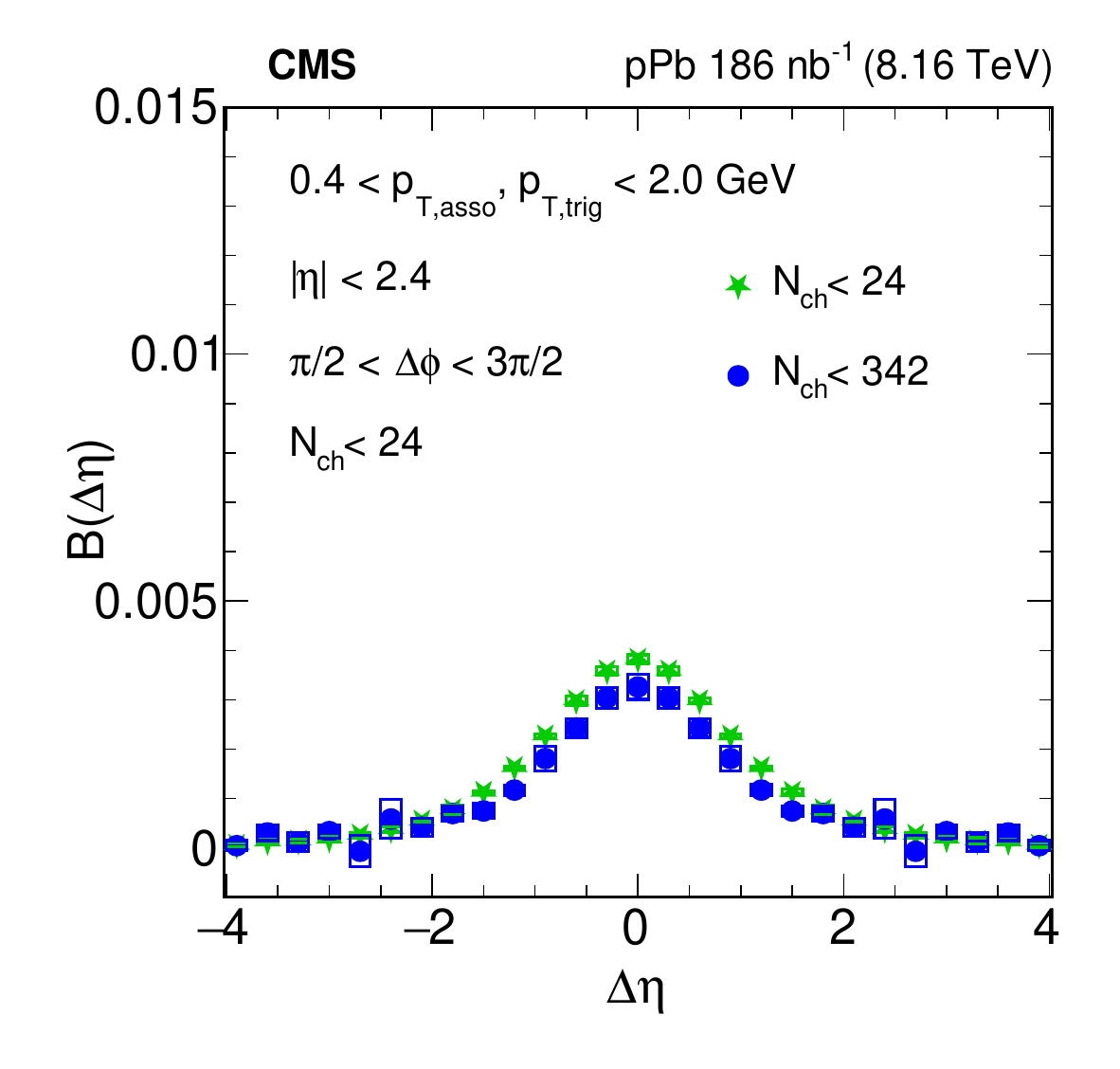}
    \includegraphics[width=0.49\textwidth]{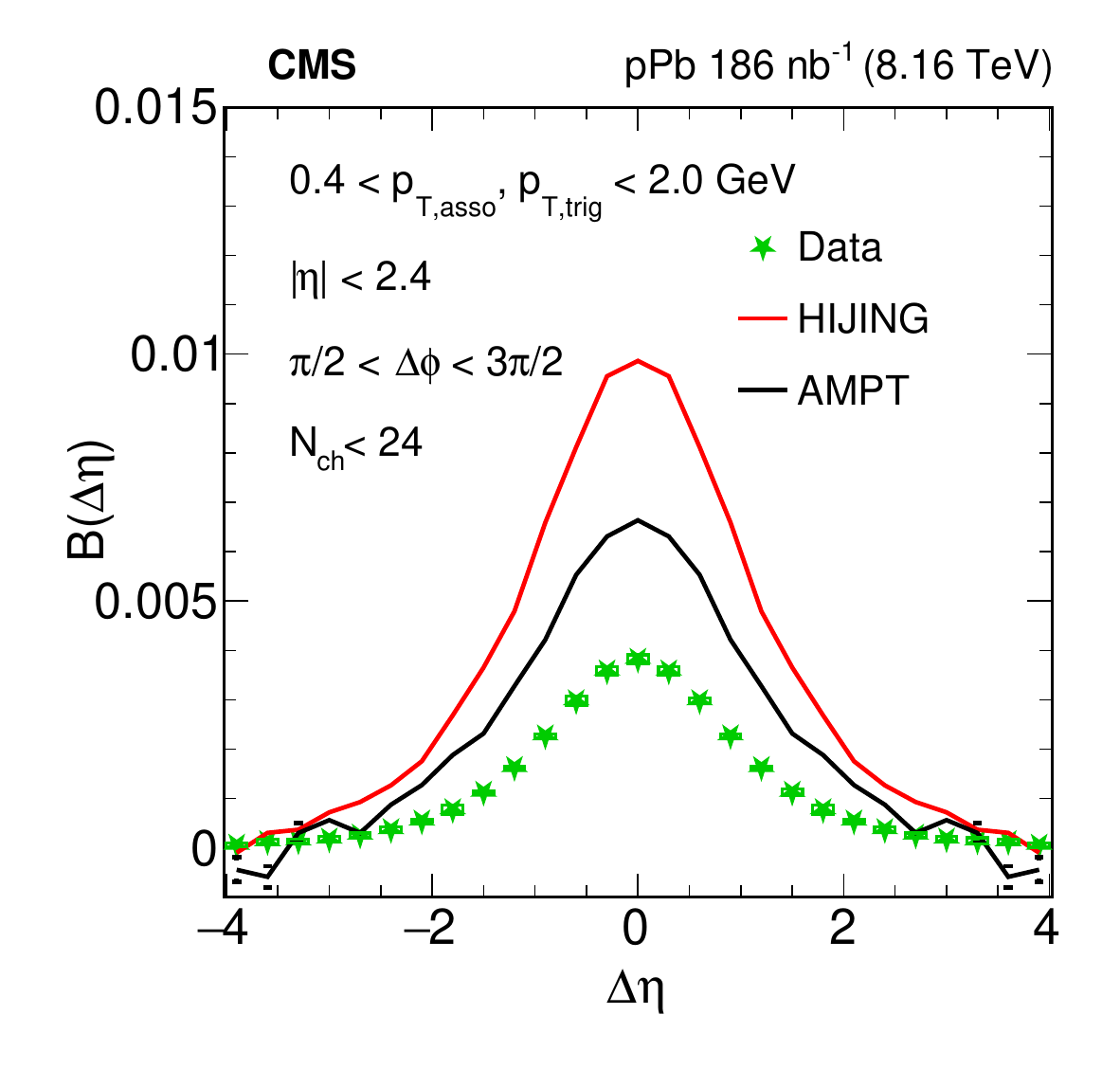}
  \caption{The away-side projection of the balance function for the charged-particles with $\pt< 2\GeV$ is shown for the \PbPb upper panel ((a) and (b)) and \pPb ((a) and (b)) collisions. The data results are compared with the different event generators with the lowest multiplicity in both the systems.}
  \label{fig:away_ppb_pbpb_model}
\end{figure}

Figure~\ref{fig:away_ppb_pbpb_model} presents the away-side ($-\pi/2 < \Delta\phi < 3\pi/2$) projection of the charge balance functions for $\pt< 2\GeV$ in both \pPb and \PbPb collisions. The away-side of the balance function demonstrates a distinct pattern, a larger magnitude of $B(\Delta\eta)$ is observed in the low-multiplicity events compared to their counterparts in high-multiplicity events. It is seen that none of these models from \AMPT, \HIJING, and \HYDJET are in quantitative agreement with the data point and exhibit a correlation peak on the away-side of a significantly higher magnitude.
\clearpage
\subsection{Balance function integral}
\begin{figure}[thb]
\centering
\includegraphics[width=0.49\textwidth]{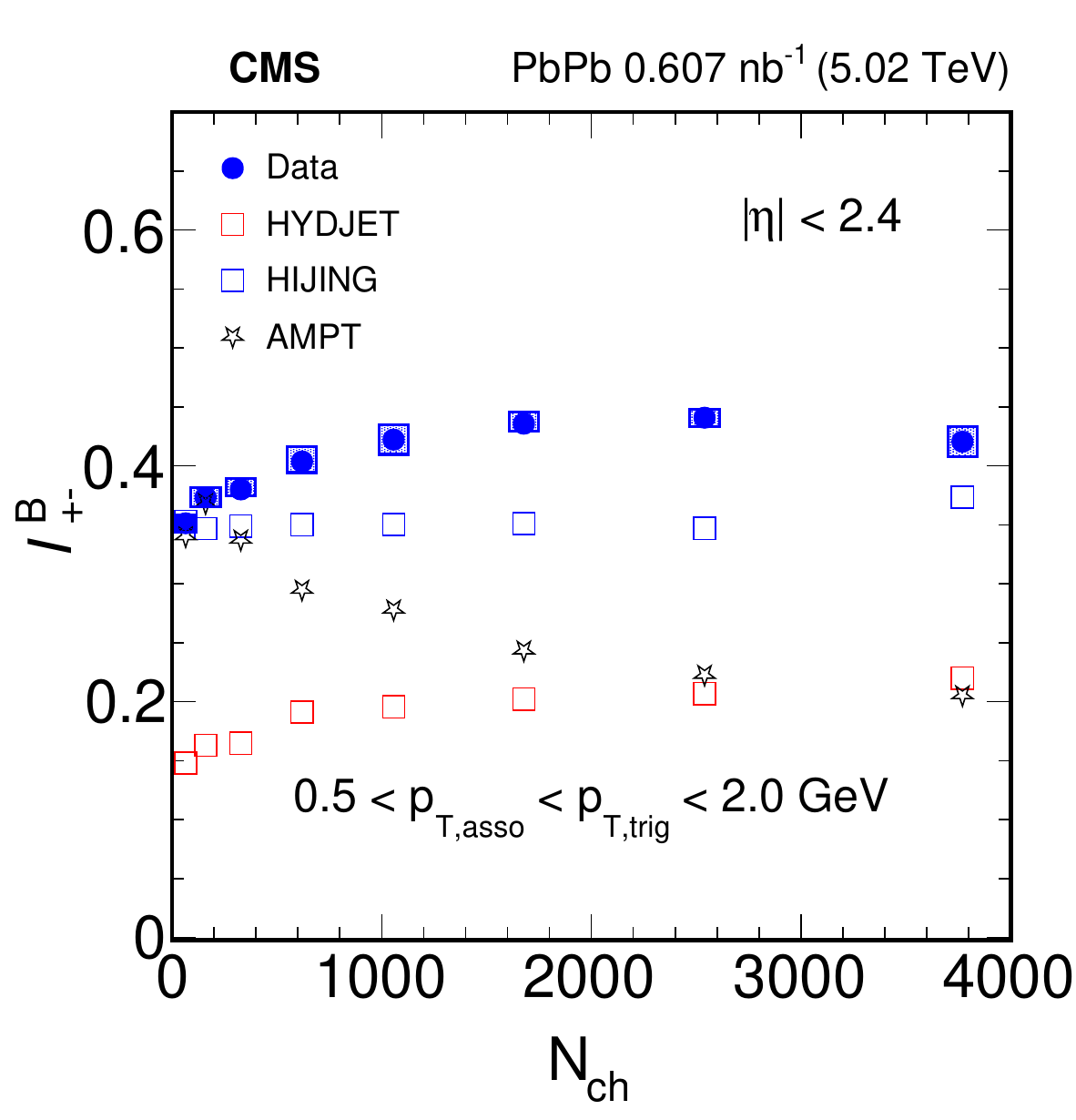}
\includegraphics[width=0.49\textwidth]{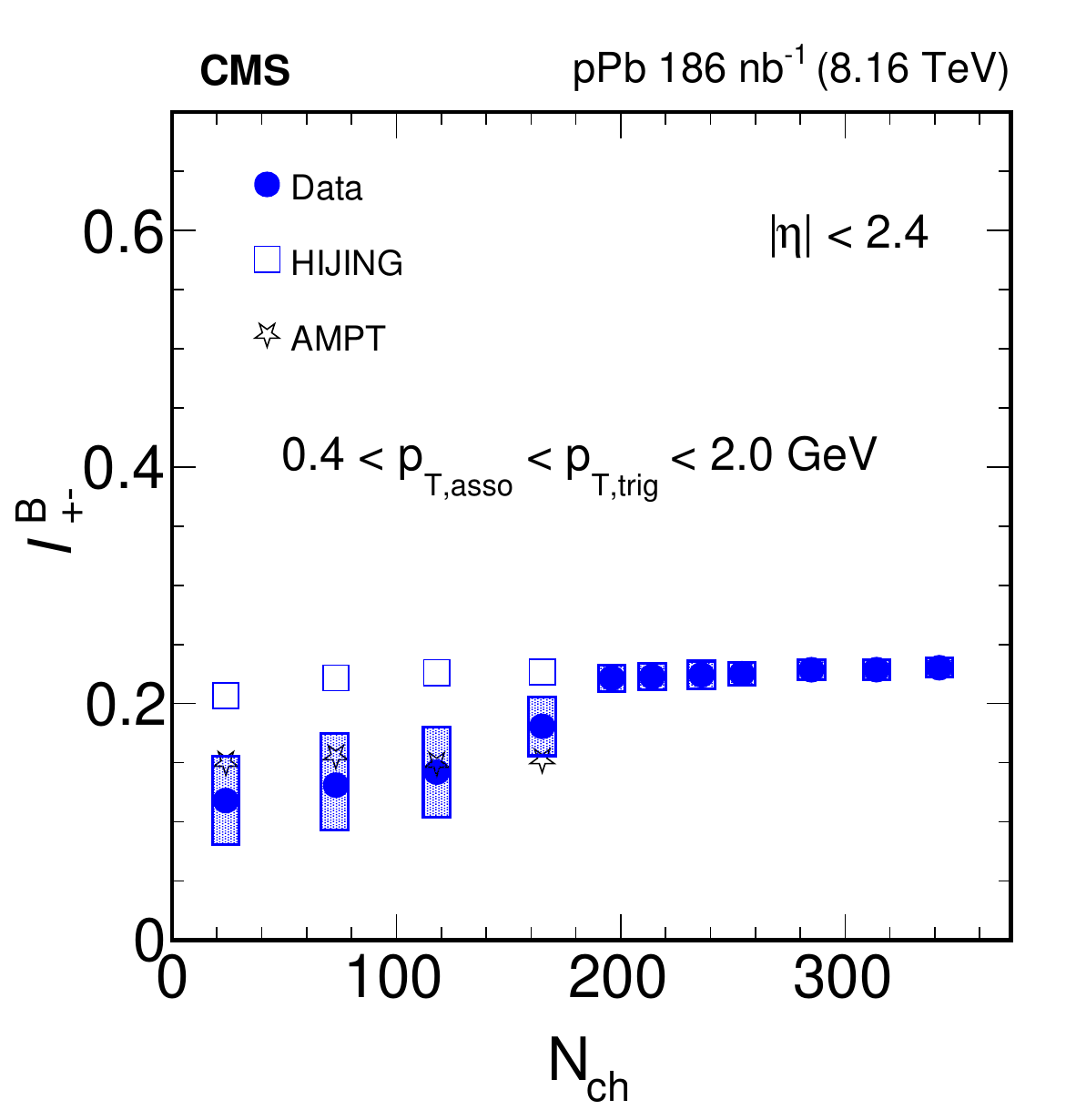}
\caption{The integral of the balance functions ($I_{+-}^{B}$) in \PbPb collisions at $\sqrtsNN = 5.02\TeV$ (left) and in \pPb collisions at $\sqrtsNN = 8.16\TeV$ (right). The balancing partners fall within the specified pseudorapidity range ($\abs{\eta} < 2.4$) and have transverse momenta within the range of $0.5 < p_\mathrm{T} < 2.0\GeV$ (for PbPb) and $0.4 < p_\mathrm{T} < 2.0\GeV$ (for \pPb). The statistical uncertainties in the data points are smaller than the marker size and rectangular boxes indicate the systematic uncertainties.}
\label{integralbf}
\end{figure}

The left and right plots in Fig.~\ref{integralbf} present the integral of the balance functions in terms of \nch, in \PbPb and \pPb collisions, respectively. By definition, the balance function is a conditional density. It is the likelihood that one event will occur under certain conditions while another possibility has occurred. In the ideal case~\cite{scott_chemqgp,Jeon:2001ue}, the integral of the balance function over the full phase space (i.e., all possible values of $\eta$, \pt, and $\phi$) is unity by construction, which means that the total charge in the collision is conserved. However, experimentally, because of the finite detector acceptance, the integral does not capture all the balancing partners due to the \pt selection made. The integral values of the balance function in \PbPb and \pPb collisions are determined to be 0.35--0.42 and 0.11--0.23, for $\pt < 2.0\GeV$, respectively. The integrals of the balance functions in the two systems show a notable difference. This may be due to the radial flow focusing pairs of both positive and negative particles into the same \pt range. In both these collisions, a minimal change in the integral with multiplicity classes is observed. Recently, ALICE reported the integral for nonidentified charged hadrons with $\abs{\eta}< 0.8$ and $0.2 < \pt < 2.0\GeV$~\cite{alicebf}. Comparison with the model calculations from \HIJING, \HYDJET, and \AMPT are also shown in Fig. 5. The \HIJING predictions show a weak dependence of the balance function integrals on the event multiplicity for both \PbPb and \pPb collision systems. The \HYDJET calculations, available only for PbPb interactions, show an increasing trend toward higher-multiplicity events, similar to that seen in the data but significantly underestimating the magnitude of the measured integral values. Calculations from the \AMPT model predict a decreasing trend towards the high multiplicity for \PbPb collisions but little to no dependence on \nch for \pPb collisions. However, the multiplicity range for the latter predictions is limited. We note that within the \nch overlap range, the \AMPT calculations agree with the measured balance function integrals from \pPb data within the uncertainties. A mild dependence of the integral of the balance function with collision centrality is observed, which could suggest the increase of multiplicity fluctuations for central events compared to peripheral events~\cite{STAR:2008szd}. The integral of the balance function over the full acceptance is related to the charge fluctuations~\cite{Jeon:2001ue}. Additionally, the integral of balance functions is sensitive to the hydrochemistry of the collisions~\cite{Pratt:2015jsa}, which is necessary to infer contributions to single-particle spectra from hadronic resonance decays based on models.

\subsection{Balance function width}
\label{widthestimation}
The balance function distribution width can be used to quantify how tightly the balancing partners are correlated and can be characterized by the averages $\langle \abs{\Delta \eta} \rangle$ and $\langle \abs{\Delta \phi} \rangle$, with $\langle \abs{\Delta \eta} \rangle$ given in Eq.~(\ref{eu_eqnon}),
\begin{equation} \label{eu_eqnon}
\langle |\Delta \eta| \rangle = \frac{\sum_{i} B(\Delta\eta_{i})  |\Delta\eta_{i}|}{\sum_{i} B(\Delta\eta_{i})},
\end{equation}
where $\sum_{i} B(\Delta\eta_{i})$ is the balance function value for each $\Delta\eta_{i}$ bin, with the sum running over all bins $i$. The absolute average value of the balance function distribution is estimated in $\Delta\eta$ and $\Delta\phi$. The width of the balance function in $\Delta\eta$ and $\Delta\phi$ decreases with increasing \nch, more significantly in the smaller \nch range. For this analysis, we have used the range $ \abs{\Delta\eta} < 3$ for the $\langle \abs{\Delta \eta} \rangle$  calculations, and $\abs{\Delta\phi} < 1.5$ for the $\langle \abs{\Delta \phi} \rangle$ calculations because of the probability of detecting both balancing charge-partners decreases with the increase of $\Delta \eta$ and $\Delta\phi$ windows. The balance function determined in a specific pseudorapidity window $B_{+-}(\Delta\eta|\eta_\text{max})$ can be connected to the balance function over an infinite interval under the assumption of a boost-invariant system (independent of rapidity)~\cite{EHS:2005lzp} according to the Eq.~(\ref{boostinveq}), 
\begin{equation}
\label{boostinveq}
B_{+-}(\Delta\eta|\eta_\mathrm{max}) = B_{+-}(\Delta\eta|\infty) \, (1-\frac{\Delta\eta}{\eta_\text{max}}).
\end{equation}
The factor $(1-\frac{\Delta\eta}{\eta_\text{max}})$ represents the probability that a particle's partner, separated by $\Delta\eta$, will fall within the finite rapidity window~\cite{Jeon:2001ue}.

\subsubsection{Balance function in low transverse momentum and comparison with models}
The results are compared with predictions from the \HYDJET (\PbPb collisions only)~\cite{Lokhtin:2005px}, \AMPT, and \HIJING MC event generators, by means of $p$-values~\cite{pval} from a $\chi^{2}$ test accounting for statistical uncertainties only. The \HYDJET is composed by a combination of the soft, hydro-type state, and the hard multi-jets. In case of \AMPT simulations, the string melting option is employed, with the generator parameters tuned to the available LHC experimental results. The \HIJING model includes multiple minijet production, nuclear shadowing of parton distribution functions, and mechanisms of jet interactions in dense matter.

\begin{figure}[thb]
    \centering
    \includegraphics[width=0.49\textwidth]{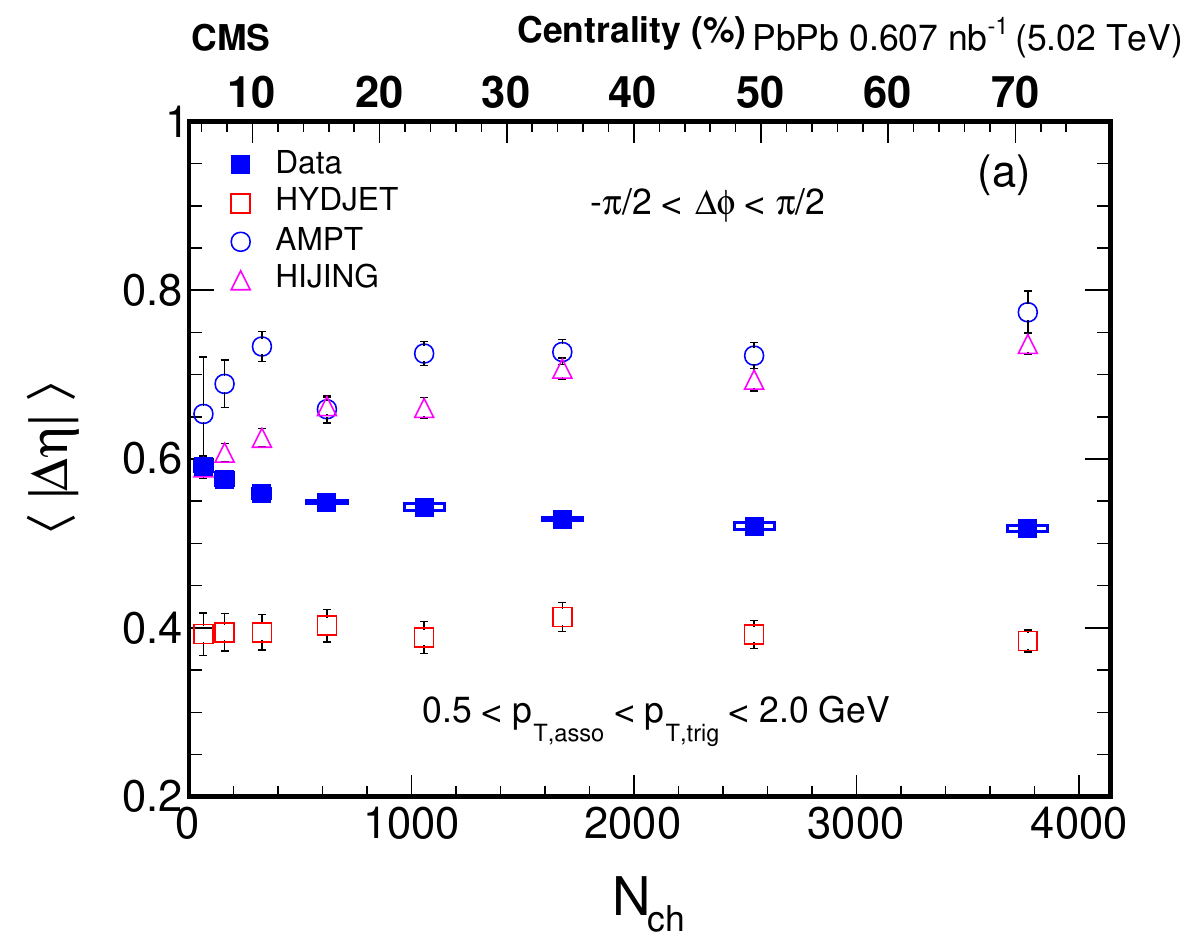}
    \includegraphics[width=0.49\textwidth]{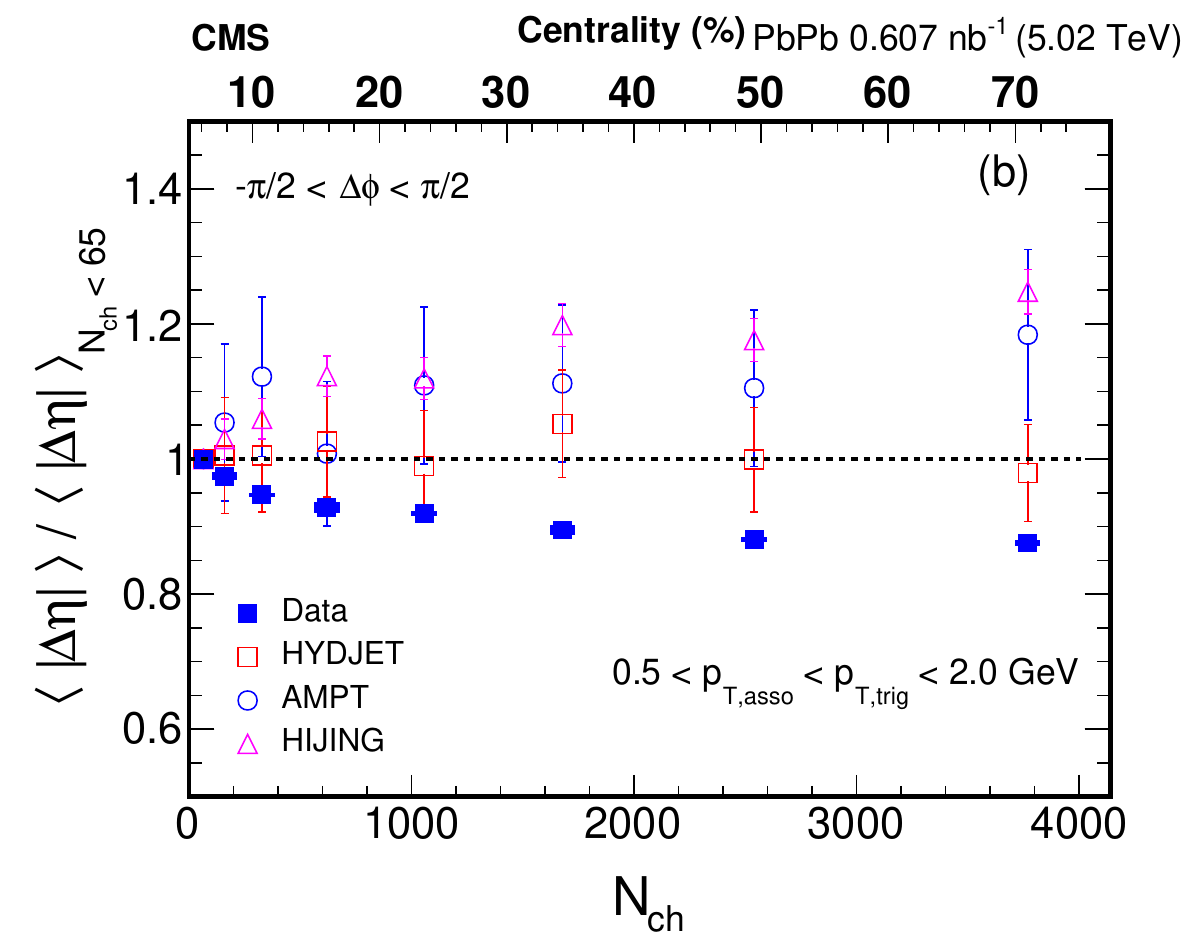}
    \includegraphics[width=0.49\textwidth]{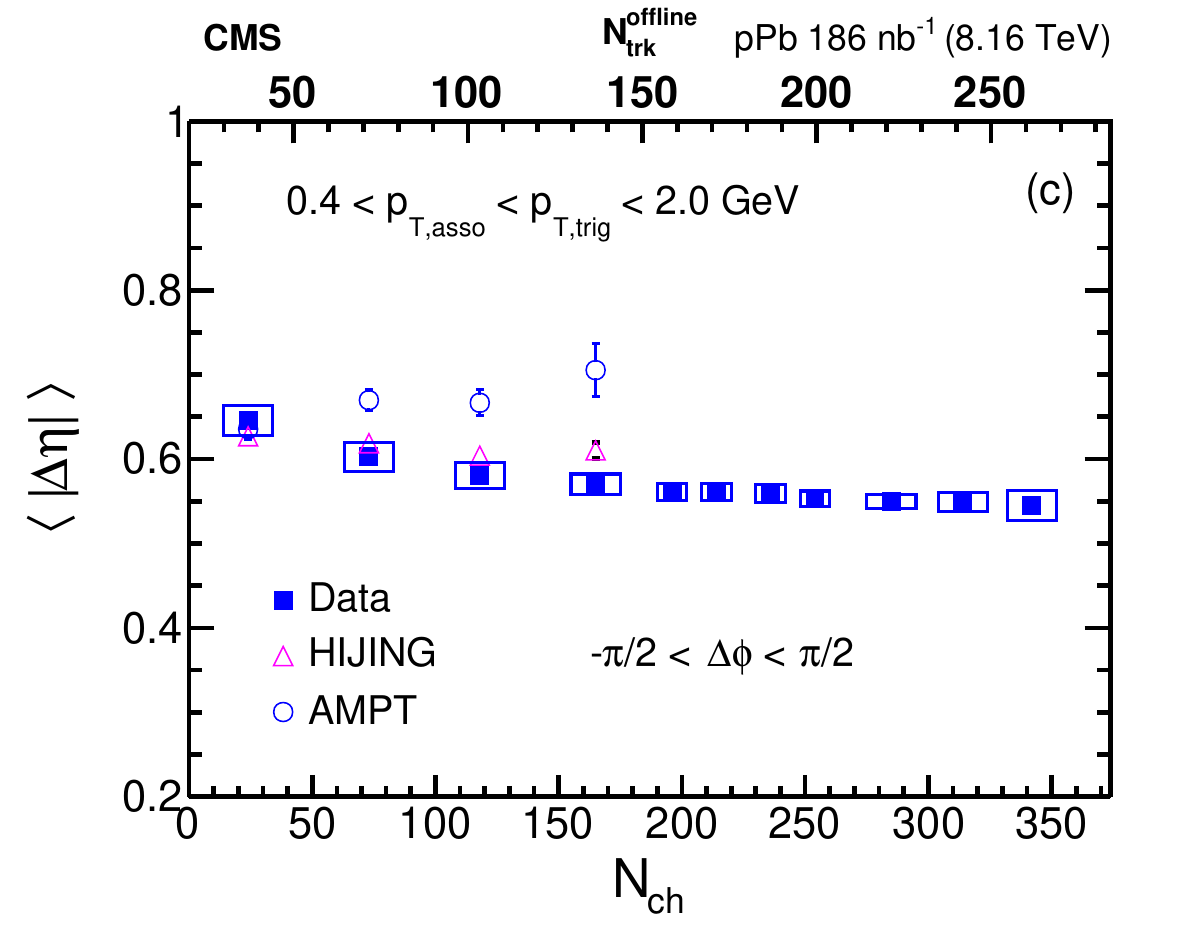}
    \includegraphics[width=0.49\textwidth]{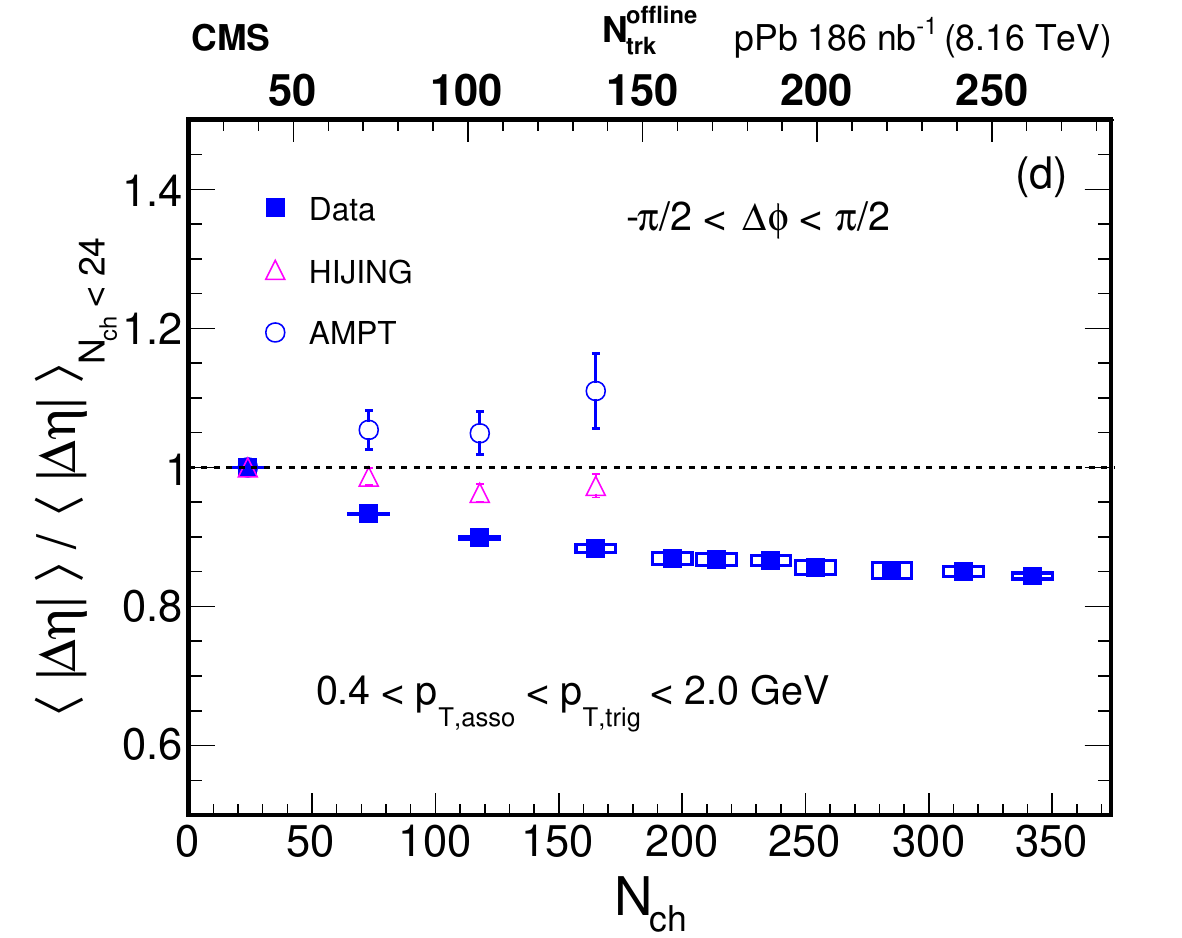}
    \caption{The width of the balance function in $\langle \abs{\Delta \eta} \rangle$ and the ratio of $\langle \abs{\Delta \eta} \rangle/ \langle \abs{\Delta \eta} \rangle_{\nch < 65}$ and  $\langle \abs{\Delta \eta} \rangle / \langle \abs{\Delta \eta} \rangle_{\nch < 24}$ are shown as functions of \nch for \PbPb collisions in $\sqrtsNN = 5.02\TeV$ (upper panels) and \pPb collisions in $\sqrtsNN = 8.16\TeV$ (lower panels), respectively. The statistical uncertainties of the data points are smaller than the marker size and rectangular boxes indicate the systematic uncertainties.}
\label{fig:ondbfun}
\end{figure}

Figure~\ref{fig:ondbfun} presents the experimental results of width values with \nch, showing a strong multiplicity dependence of the $\langle \abs{\Delta \eta} \rangle$ for both collision systems. In \HYDJET, $\langle \abs{\Delta \eta} \rangle$ does not show any significant dependence on \nch. In this model, local charge conservation for more peripheral events (smaller multiplicities) has more influence on the charge-balance function than for large multiplicities. Comparing \HIJING predictions with the \PbPb and \pPb data, no clear multiplicity dependence is seen in the model calculations. \HIJING does not explain the experimental data properly as the $p$-value  is smaller than 0.01. In addition, the magnitude of the balance function widths is larger in \HIJING than in the data because the collective flow effect is not present in the \HIJING model. \\

\begin{figure}[thb]
\centering
  \includegraphics[width=0.49\textwidth]{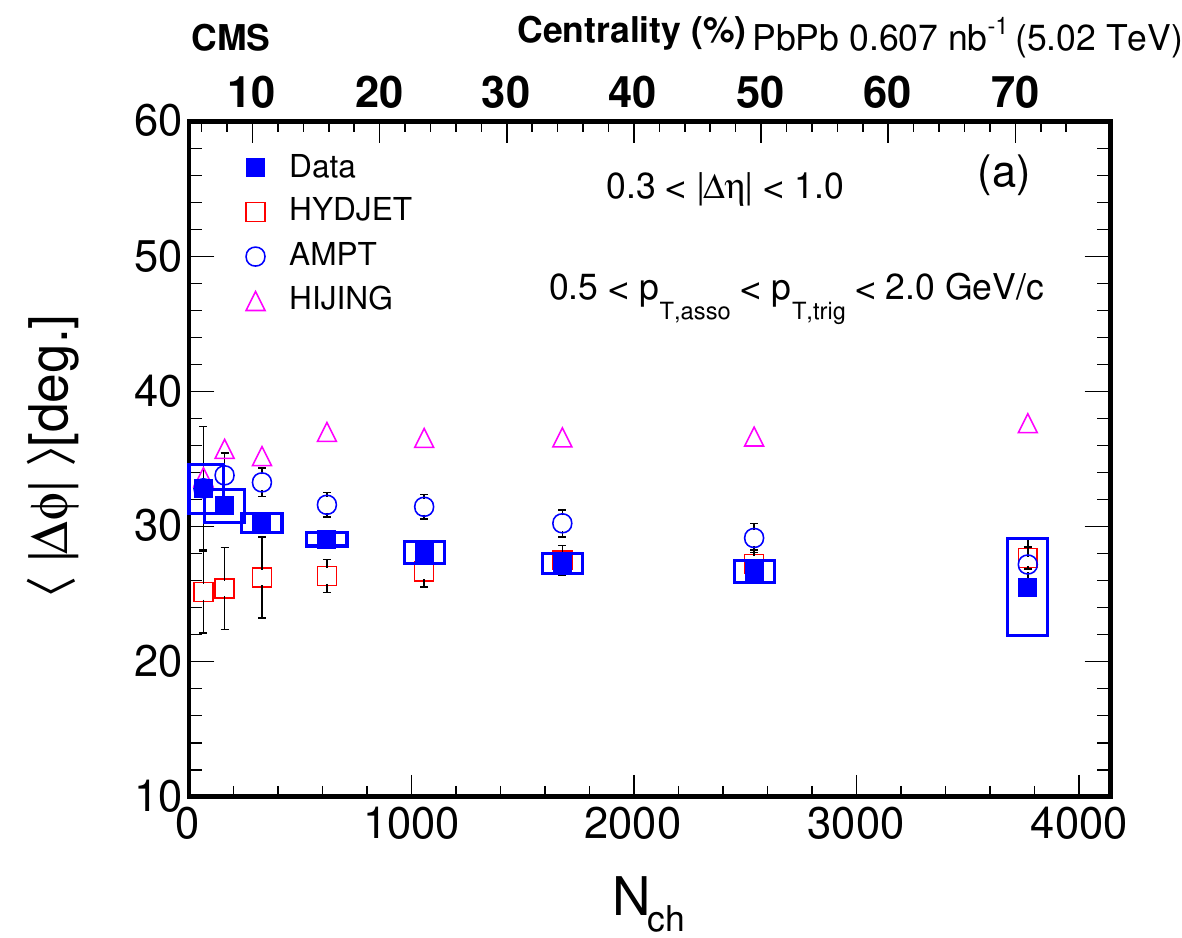}
  \includegraphics[width=0.49\textwidth]{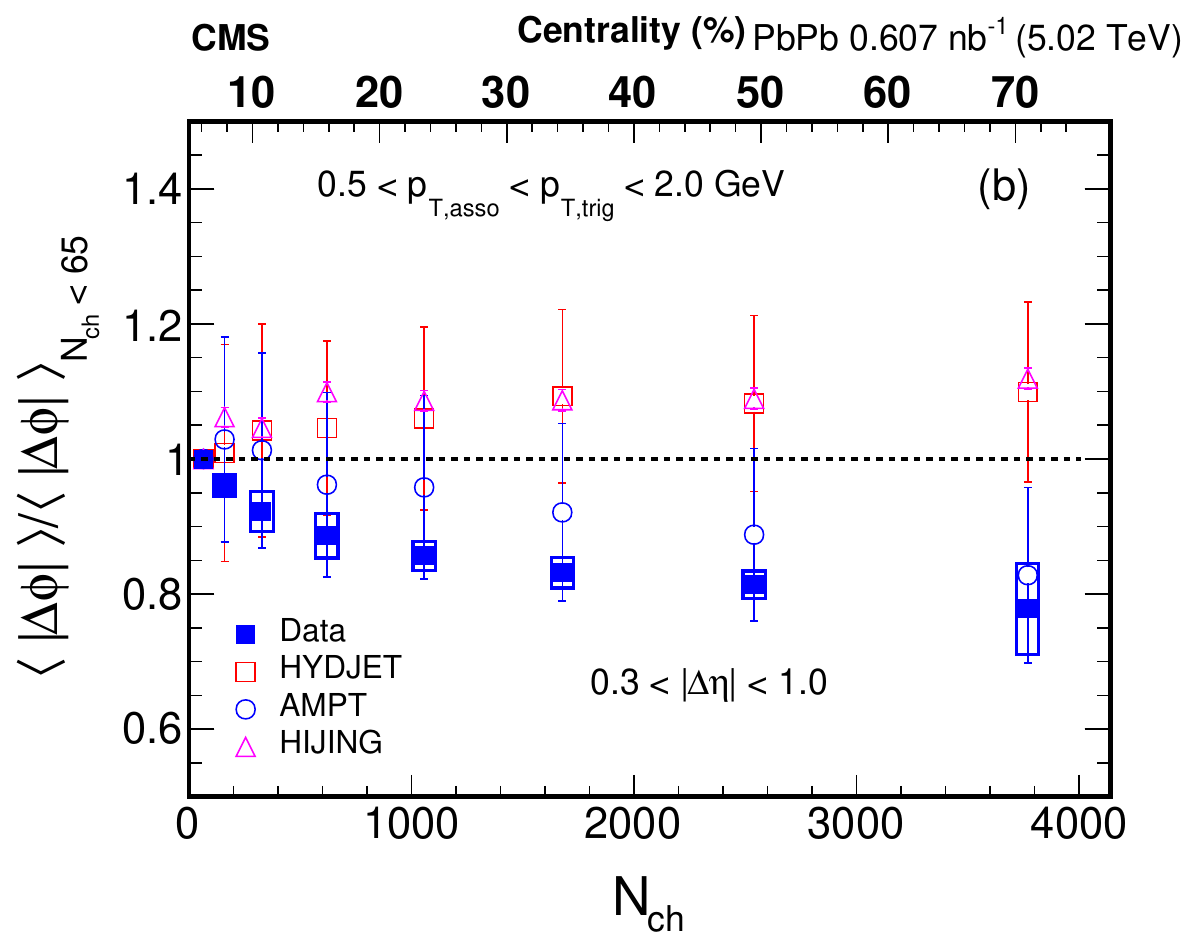}
  \includegraphics[width=0.49\textwidth]{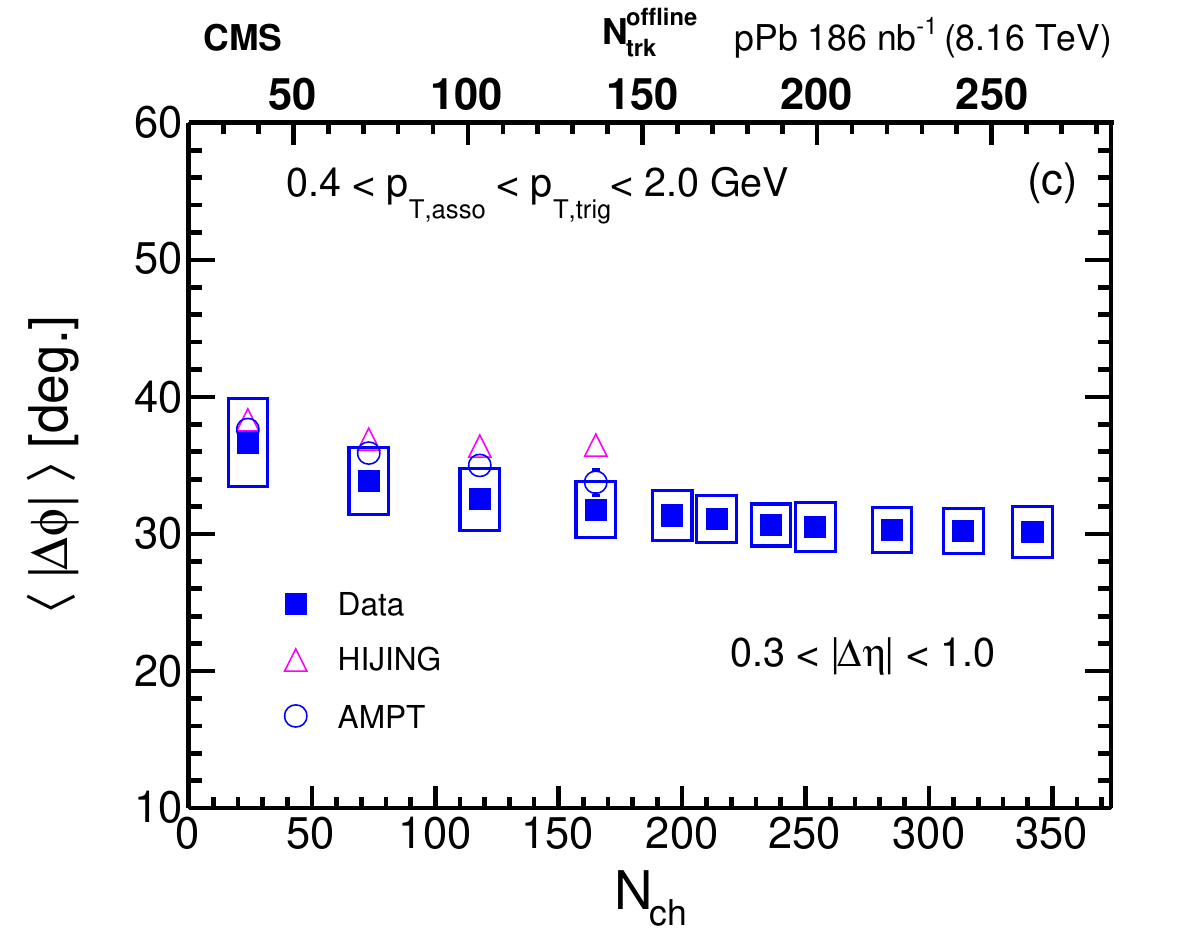}
  \includegraphics[width=0.49\textwidth]{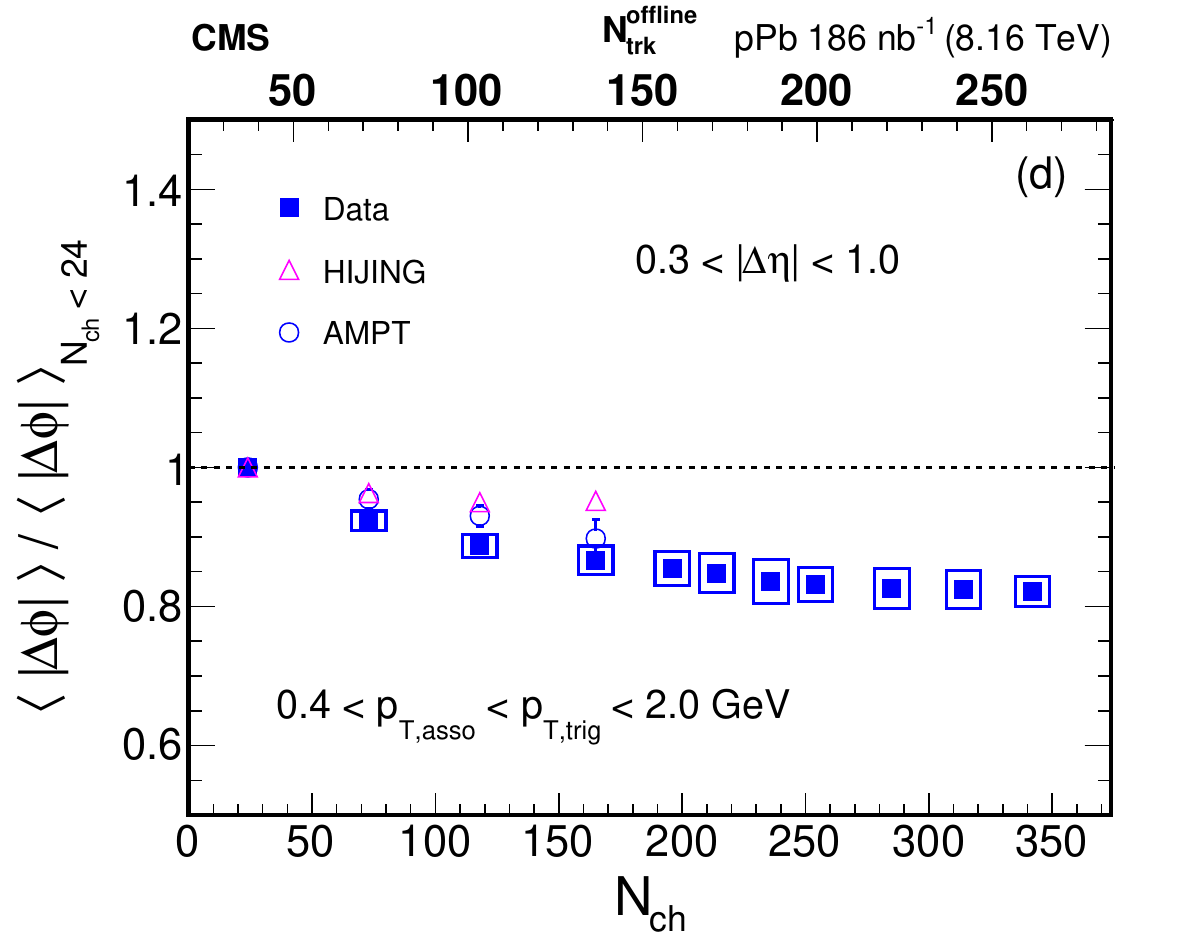}
  \caption{The width of the balance function in $\langle \abs{\Delta \phi} \rangle$ and the ratios of $\langle \abs{\Delta \phi} \rangle/\langle \abs{\Delta \phi} \rangle_{\nch < 65}$ and $\langle \abs{\Delta \phi} \rangle/ \langle \abs{\Delta \phi} \rangle_{\nch < 24}$ are shown as functions of \nch for \PbPb collisions in $\sqrtsNN = 5.02\TeV$ (upper panels) and \pPb collisions in $\sqrtsNN = 8.16\TeV$ (lower panels), respectively. The statistical uncertainties of the data points are smaller than the marker size and rectangular boxes indicate the systematic uncertainties.}
\label{fig:lowpTmeandphi}
\end{figure}

The data results are also compared with the \AMPT model, which includes the quark coalescence and the decay of resonances. When comparing the $\langle \abs{\Delta \eta} \rangle$  in both collision systems, \AMPT predicts larger $\langle \abs{\Delta \eta} \rangle$ than data ($p$-value of 0.01 in \pPb), and overall shows worse agreement than \HIJING ($p$-value of 0.01 in \pPb). We estimate the relative decrease of the width, which is expressed by the ratio of $\langle \abs{\Delta \eta} \rangle$ for each multiplicity class to the lowest multiplicity value, i.e., $\langle \abs{\Delta \eta}\rangle_{\nch < 65}$ (for \PbPb) and $\langle \abs{\Delta \eta}\rangle_{\nch < 24}$ (for \pPb) in order to compare the width in both collision systems within the same multiplicity range. However, due to the limitations in the model calculations, we are constrained within a specific range of multiplicity compared to the pPb data.

The right plots of Fig.~\ref{fig:ondbfun} present the normalized width in $\Delta \eta$, where the data results are compared with different models and this indicates the model prediction does not show significant multiplicity dependence. Our experimental findings, based on considering only the statistical uncertainties from the limited sample size, suggest that the relative change in \pPb collisions appears to similar to that in \PbPb collisions.

Figure~\ref{fig:lowpTmeandphi} presents the experimental findings for $\langle \abs{\Delta\phi} \rangle$ in \PbPb and \pPb collisions. A significant change in the balance function width is observed with multiplicity. Similarly, the data results are compared with the various MC predictions. The \HYDJET and \HIJING generators are not able to reproduce the trend of data results in the case of \PbPb collisions. A significant multiplicity dependence is shown in $\langle \abs{\Delta \phi} \rangle$ because of the radial flow effect in \AMPT, which acts over the balancing partners by preserving their initial-state correlations in $\Delta\phi$, in both systems. This trend is also reflected in Figs.~(\ref{fig:lowpTmeandphi}b) and (\ref{fig:lowpTmeandphi}d), where the relative decrease of the width in $\langle \abs{\Delta\phi} \rangle$ has a strong contribution from collective final state effects. The normalized value of $\langle \abs{\Delta\phi} \rangle$ in \pPb collisions has a similar ratio to \PbPb data. The \HIJING and \AMPT predictions ($p$-values of 0.01 and 0.02) are able to describe the decreasing trend of the \pPb data with \nch for small values of \nch, where the correlations are dominated by resonance decays. On the other hand, the two generators show little dependence on \nch for larger values of \nch in \pPb, whereas the data continues its decreasing trend, as demonstrated in Fig.~(\ref{fig:lowpTmeandphi}d).

\subsection{Transverse momentum dependence of balance functions}

\begin{figure}[thb]
\centering
    \includegraphics[width=0.49\textwidth]{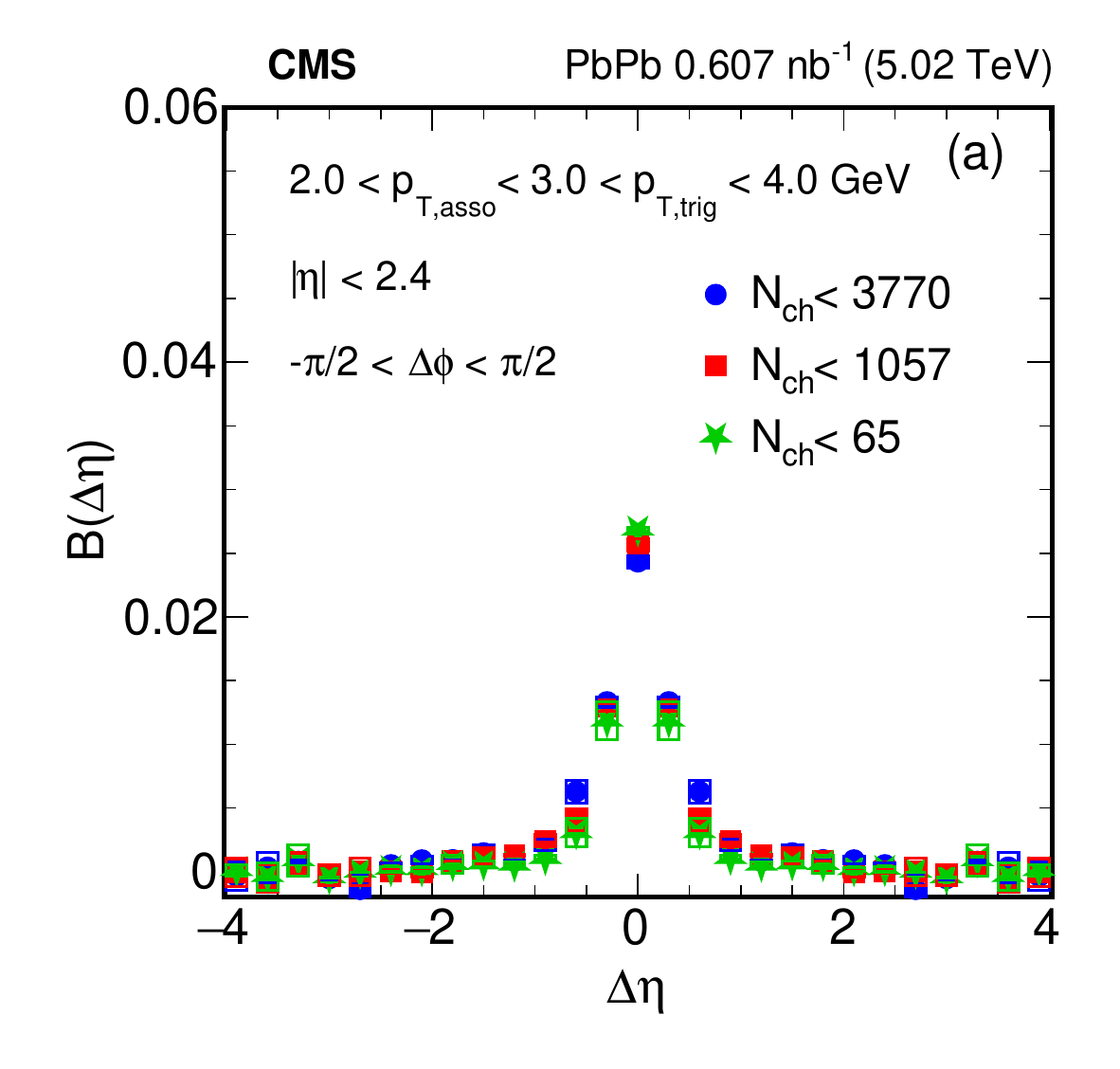}
    \includegraphics[width=0.49\textwidth]{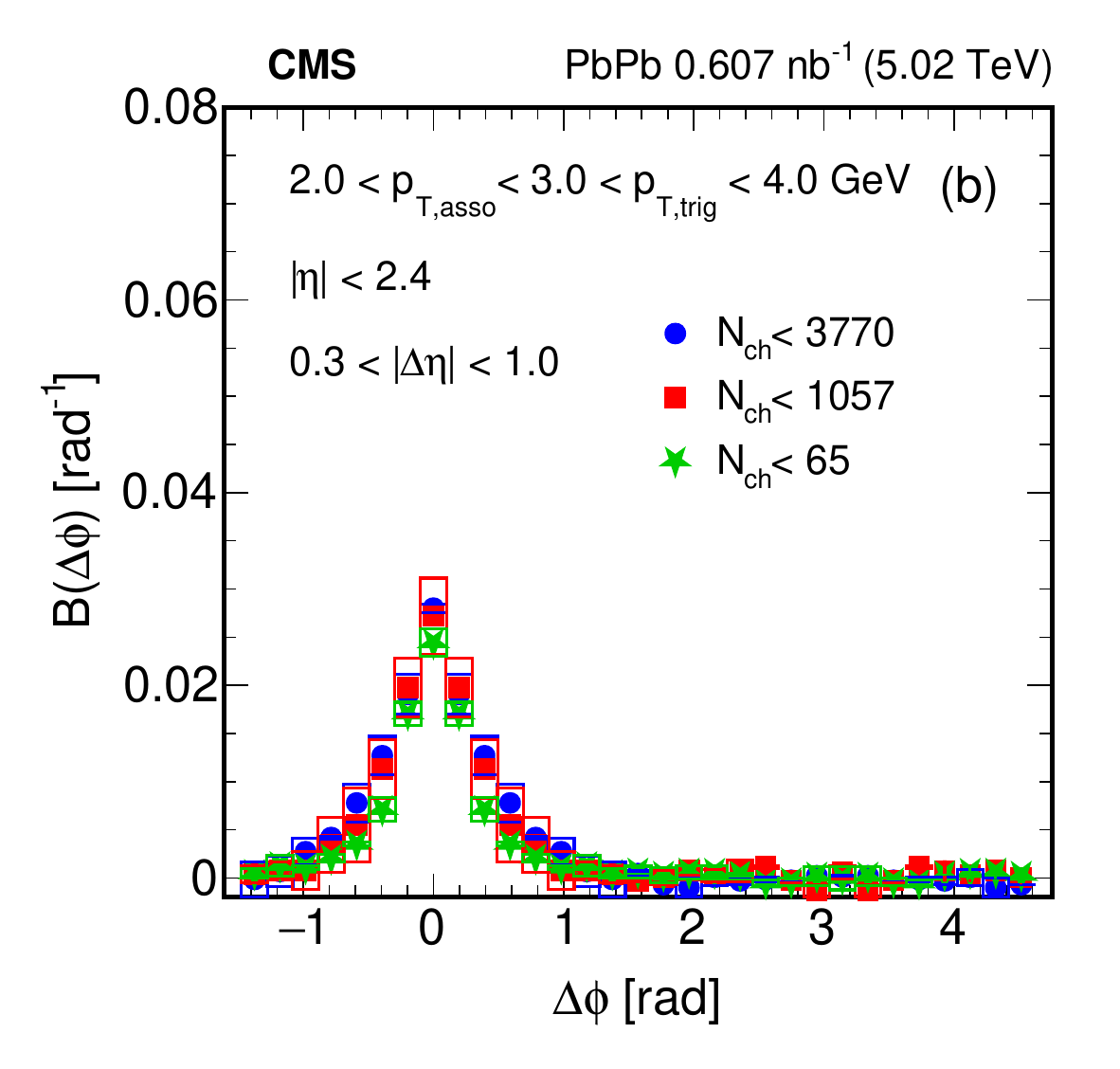}
    \includegraphics[width=0.49\textwidth]{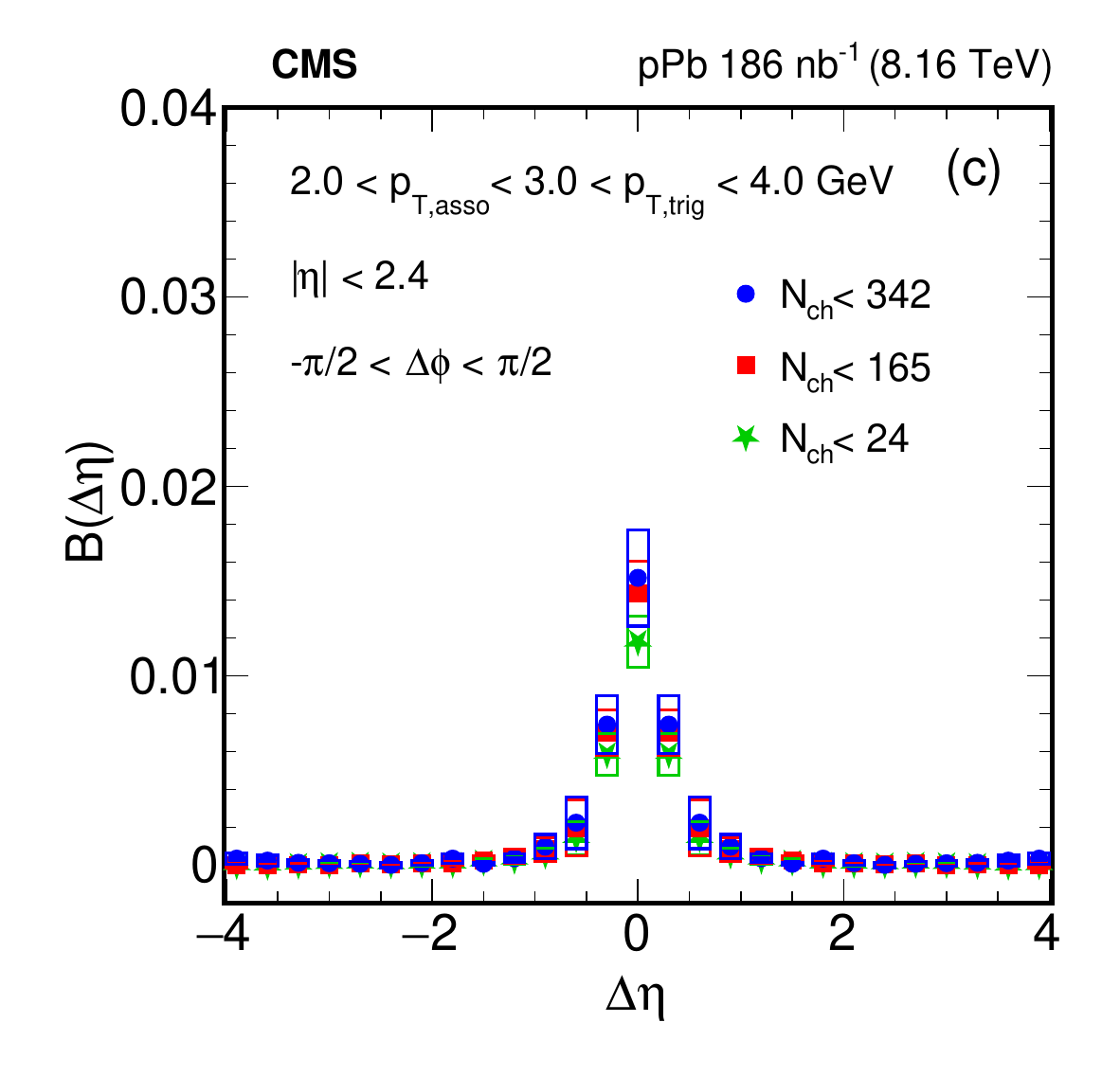}
    \includegraphics[width=0.49\textwidth]{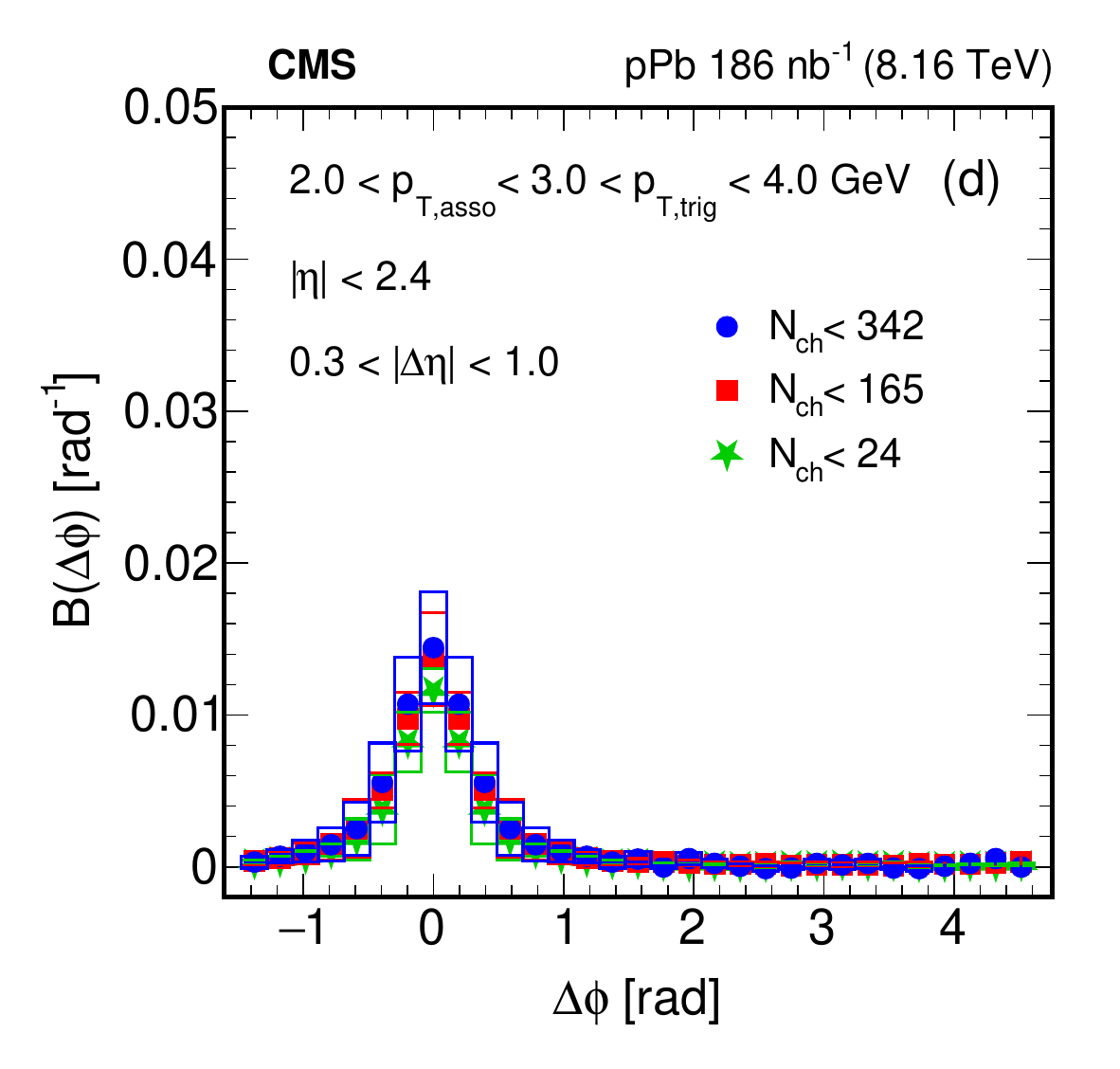}
\caption{The projection of the balance function is presented for \PbPb in $\sqrtsNN = 5.02\TeV$ (upper panels) and \pPb in $\sqrtsNN = 8.16\TeV$ (lower panels) collisions as a function of $\Delta \eta$ (left column) and $\Delta \phi$ (right column), for $2.0 < p_\text{T,asso} < 3.0 < p_\text{T,trig} < 4.0\GeV$ ranges. The 1D projection is derived for $\Delta \eta$ in near-side ($\abs{\Delta \phi} < \pi/2$) and $\Delta \phi$ ($0.3 < \abs{\Delta \eta} < 1.0)$ regions.}
\label{fig:itmpT}
\end{figure}

\begin{figure}[thb]
\centering
    \includegraphics[width=0.49\textwidth]{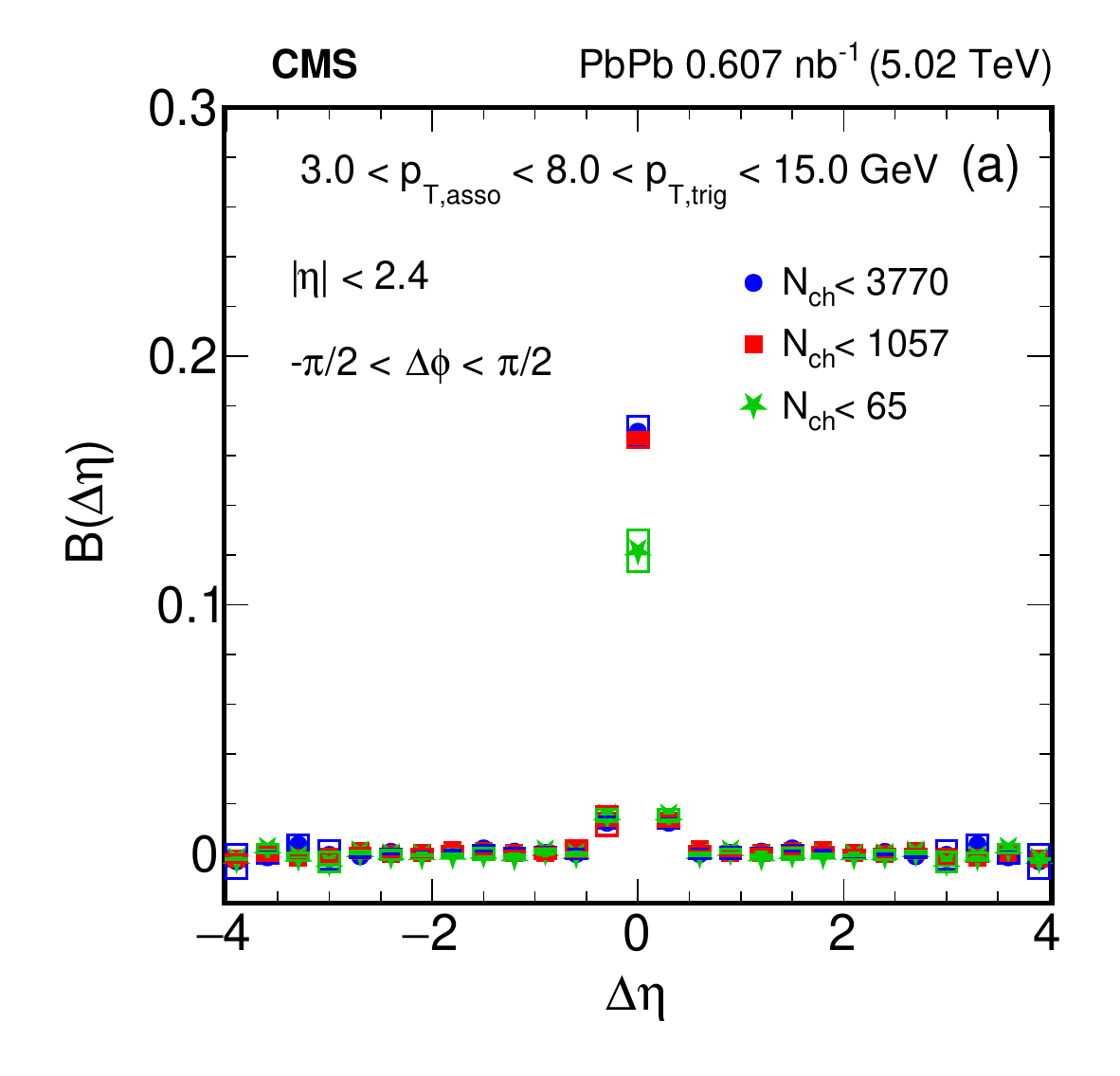}
    \includegraphics[width=0.49\textwidth]{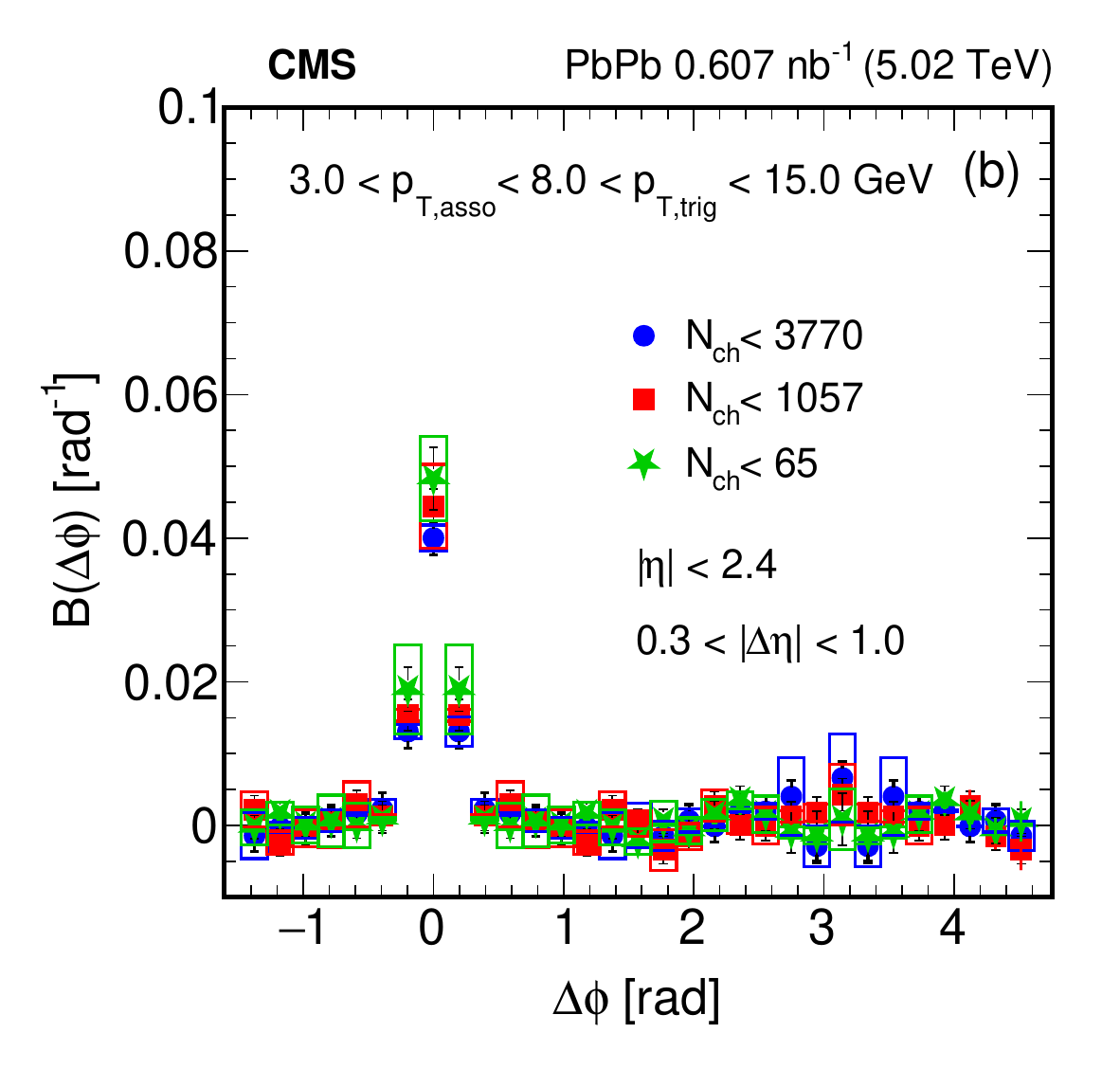}
    \includegraphics[width=0.49\textwidth]{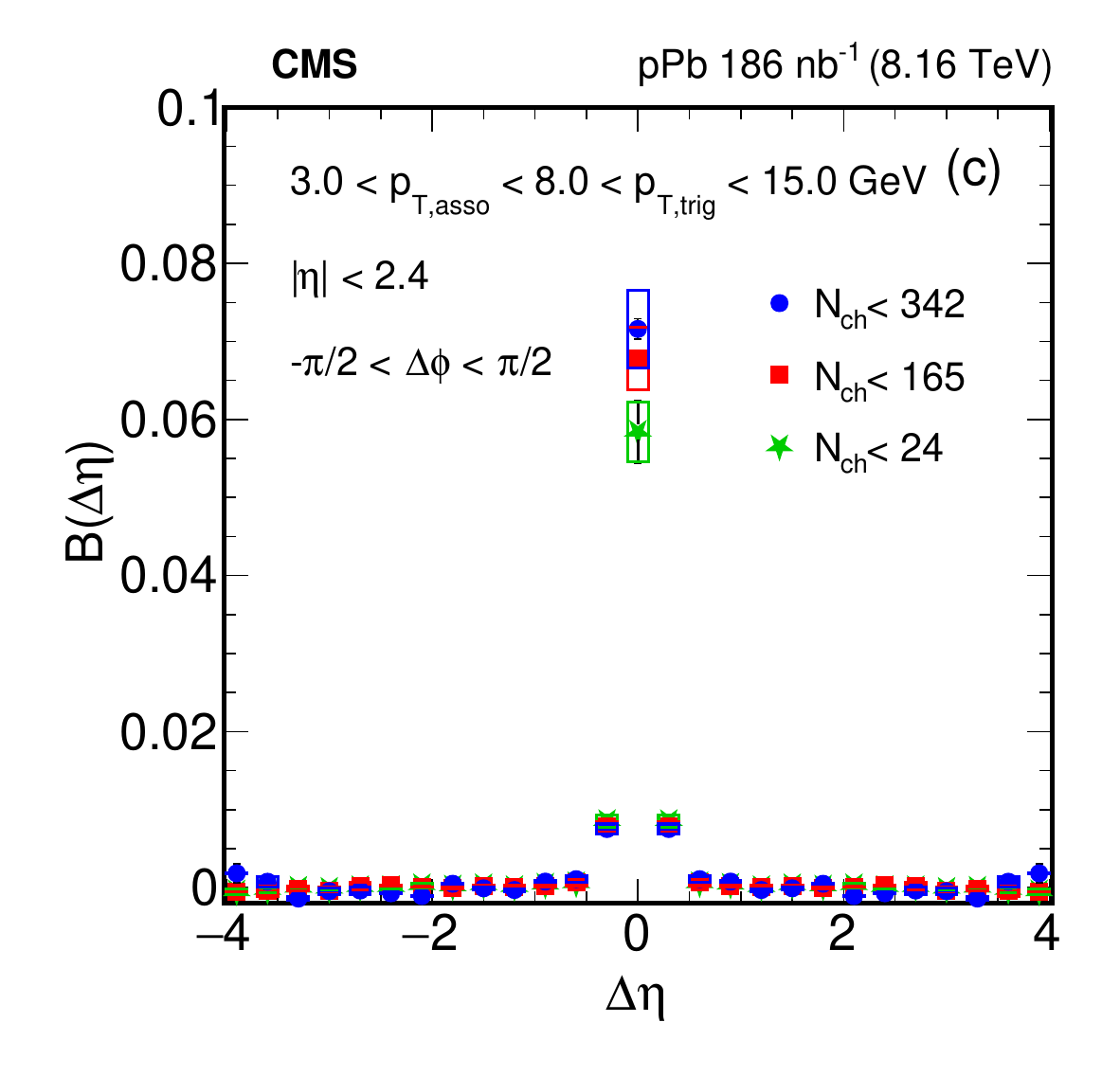}
    \includegraphics[width=0.49\textwidth]{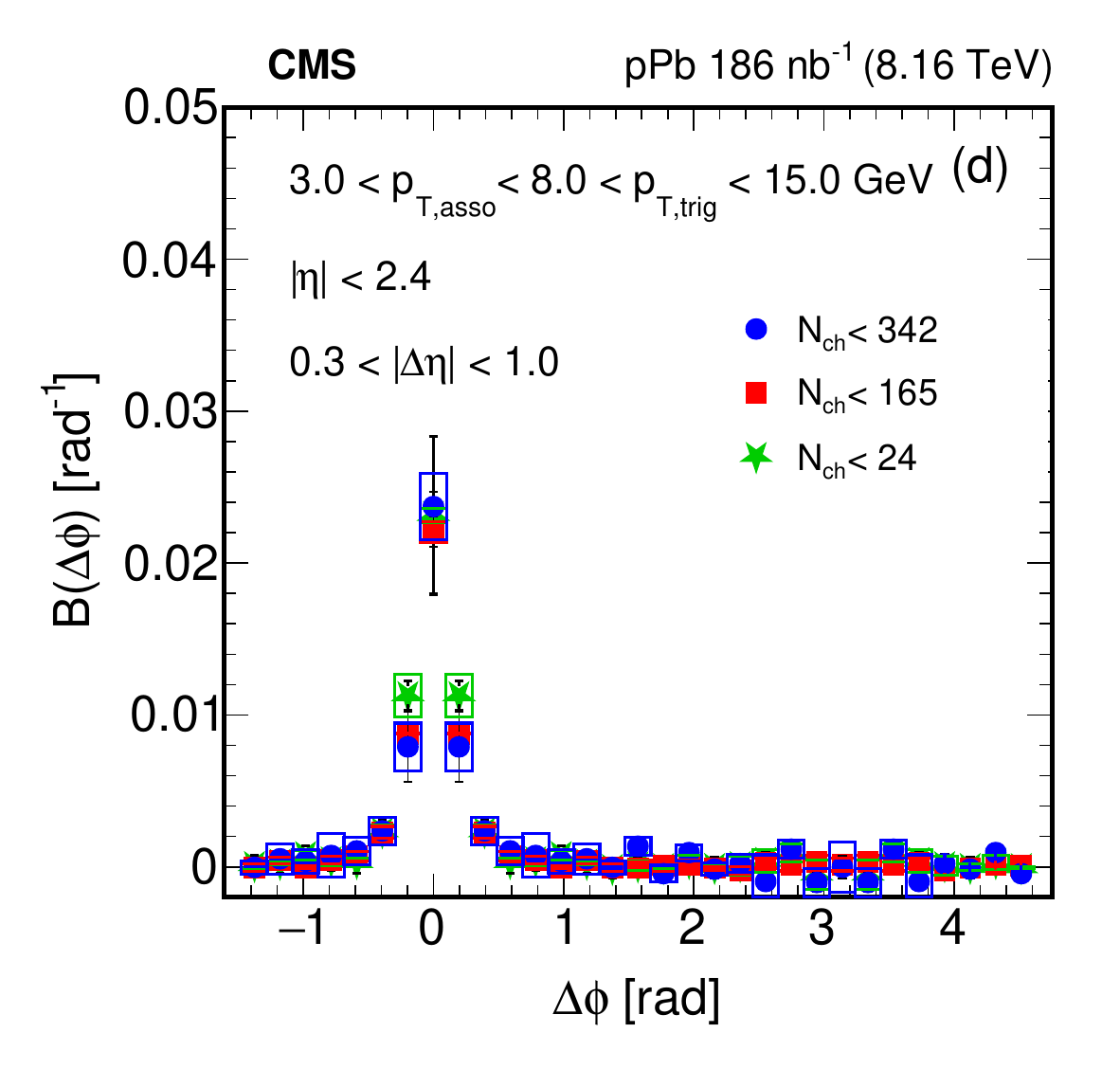}
  \caption{The projection of the balance function is presented for \PbPb in $\sqrtsNN = 5.02\TeV$ (upper panels) and \pPb in $\sqrtsNN = 8.16\TeV$ (lower panels) collisions as a function of $\Delta \eta$ (left column) and $\Delta \phi$ (right column), for $3.0 < p_\text{T,asso} < 8.0 < p_\text{T,trig} < 15.0\GeV$ ranges. The 1D projection is derived for $\Delta \eta$ in near-side ($\abs{\Delta \phi} < \pi/2$) and $\Delta \phi$ ($0.3 < \abs{\Delta \eta} < 1.0)$ regions.}
\label{fig:highpT1d}
\end{figure}

\begin{figure}[thb]
\centering
    \includegraphics[width=0.49\textwidth]{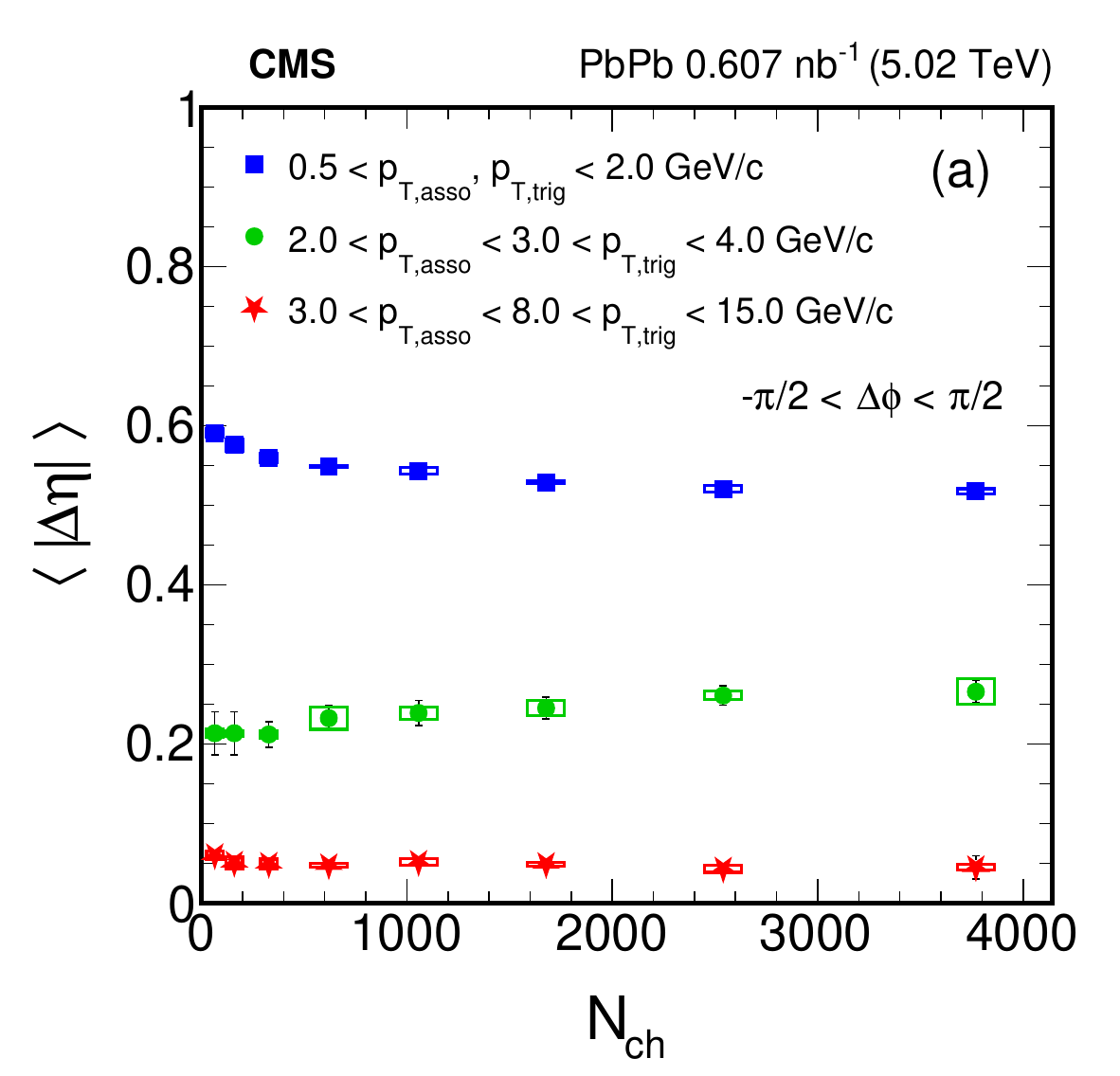}
    \includegraphics[width=0.49\textwidth]{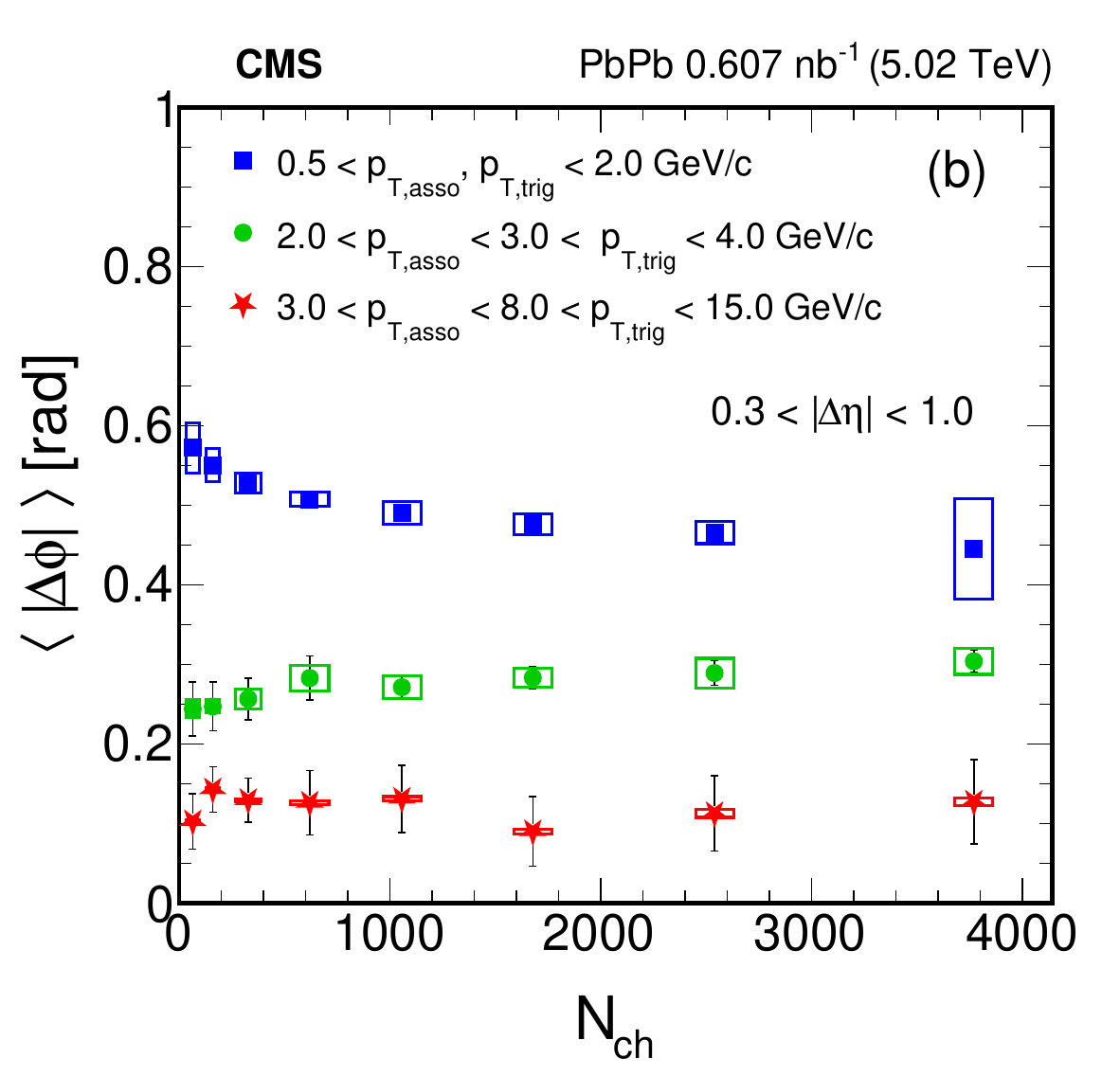}
    \includegraphics[width=0.49\textwidth]{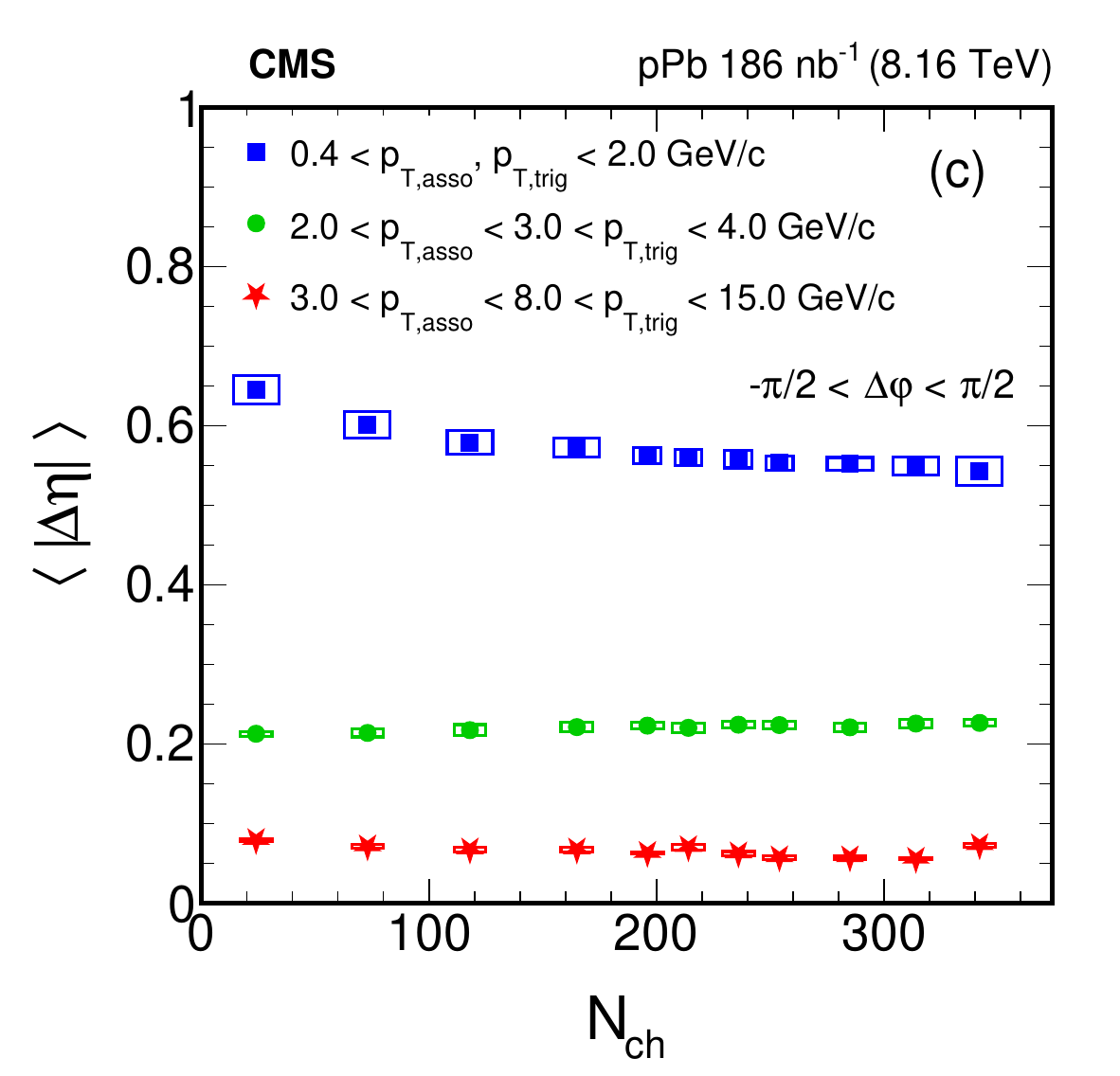}
    \includegraphics[width=0.49\textwidth]{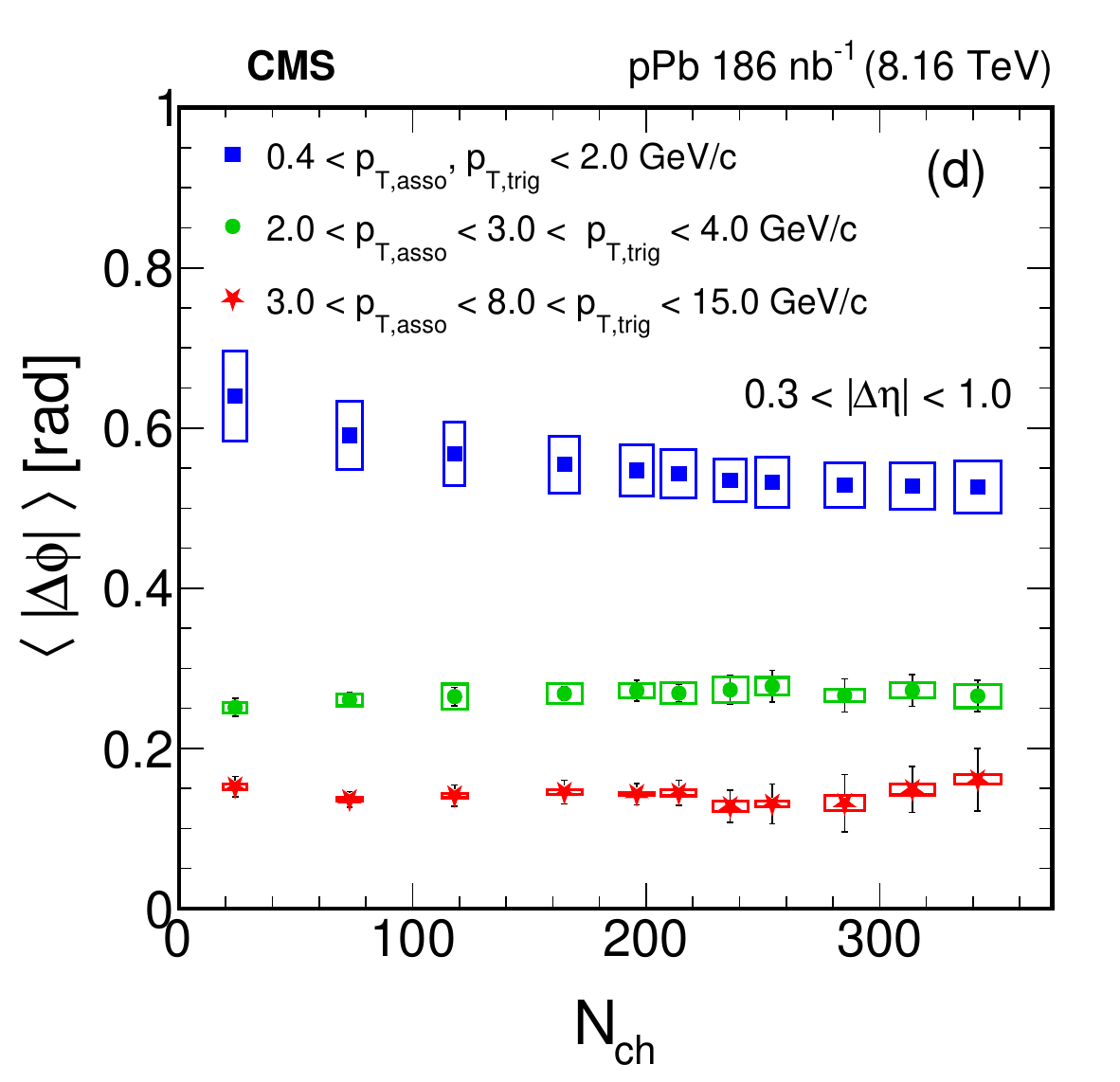}
  \caption{The width of the balance function in $\Delta \eta$ (left column) and $\Delta \phi$ (right column) is calculated for different \pt interval in \PbPb in $\sqrtsNN = 5.02\TeV$ (upper panels) and \pPb collisions in $\sqrtsNN = 8.16\TeV$ (lower panels). The vertical lines indicate the statistical uncertainties of the data points, and the rectangular boxes indicate the systematic uncertainties.}
\label{fig:highpT}
\end{figure}

This measurement is extended to higher values of the \pt ($>$2\GeV) to study if the narrowing or the widening of the balance function is constrained to the bulk particle production at low \pt ($\pt < 2\GeV$) or is also connected to hard process. Figures~\ref{fig:itmpT} and~\ref{fig:highpT1d} represent the 1D projections of the balance function in  $\Delta \eta$ and $\Delta \phi$ for the trigger and associated particles in the intermediate-\pt ($2 < p_\text{T,asso} < 3 < p_\text{T,trig} < 4\GeV$) and high-\pt ($3 < p_\text{T,asso} < 8 < p_\text{T,trig} < 15\GeV$) ranges. The upper panels show the plots for \PbPb collisions, and the lower panel is for \pPb collisions. It can be seen that they become narrower for increasing \pt, as compared with low-\pt results, and exhibit a smaller multiplicity dependence. The width of the balance function in $\Delta\eta$ is narrower in the high-\pt range than in the low- and intermediate-\pt ranges, which is consistent with the findings in $\Delta\phi$. This implies that the effects of  radial flow on the balance function is weaker at higher \pt, and the balance function at high \pt is more sensitive to other effects such as jet fragmentation and medium response~\cite{jetmed,jetpbpb}. 

The width of the balance functions in $\langle \abs{\Delta \eta} \rangle$ and $\langle \abs{\Delta \phi} \rangle$, for the different values of \pt, are presented in Fig.~\ref{fig:highpT} as a function of \nch, for both \PbPb and \pPb collisions. The narrowing of the balance function width in the low-\pt region is understood in a delayed hadronization picture, where the particles are produced at later stages of the evolution of the long-lived medium formed in these collisions. Also in comparison with higher \pt, the multiplicity dependence in low-\pt \PbPb collisions is attributed to the centrality dependence of the radial flow, which retains part of the initial correlations of the balancing partners. These results suggest that the balance function is a useful tool to investigate the interplay between soft and hard processes in heavy-ion collisions at different \pt ranges. Similarly, the multiplicity dependence in low-\pt \pPb collisions could be explained by collectivity. Collectivity in small collision systems is already suggested by the observation of long-range ridge correlations in \pPb collisions~\cite{cmshin12015,cmshinppridge,ATLASpPb}. The similarity of the balance functions in \pPb and \PbPb collisions suggests a similar origin of particle correlations in these two colliding systems.

\section{Beam energy dependence}
\label{beamenergydep}
\begin{figure}[thb]
\centering
    \includegraphics[width=0.49\textwidth]{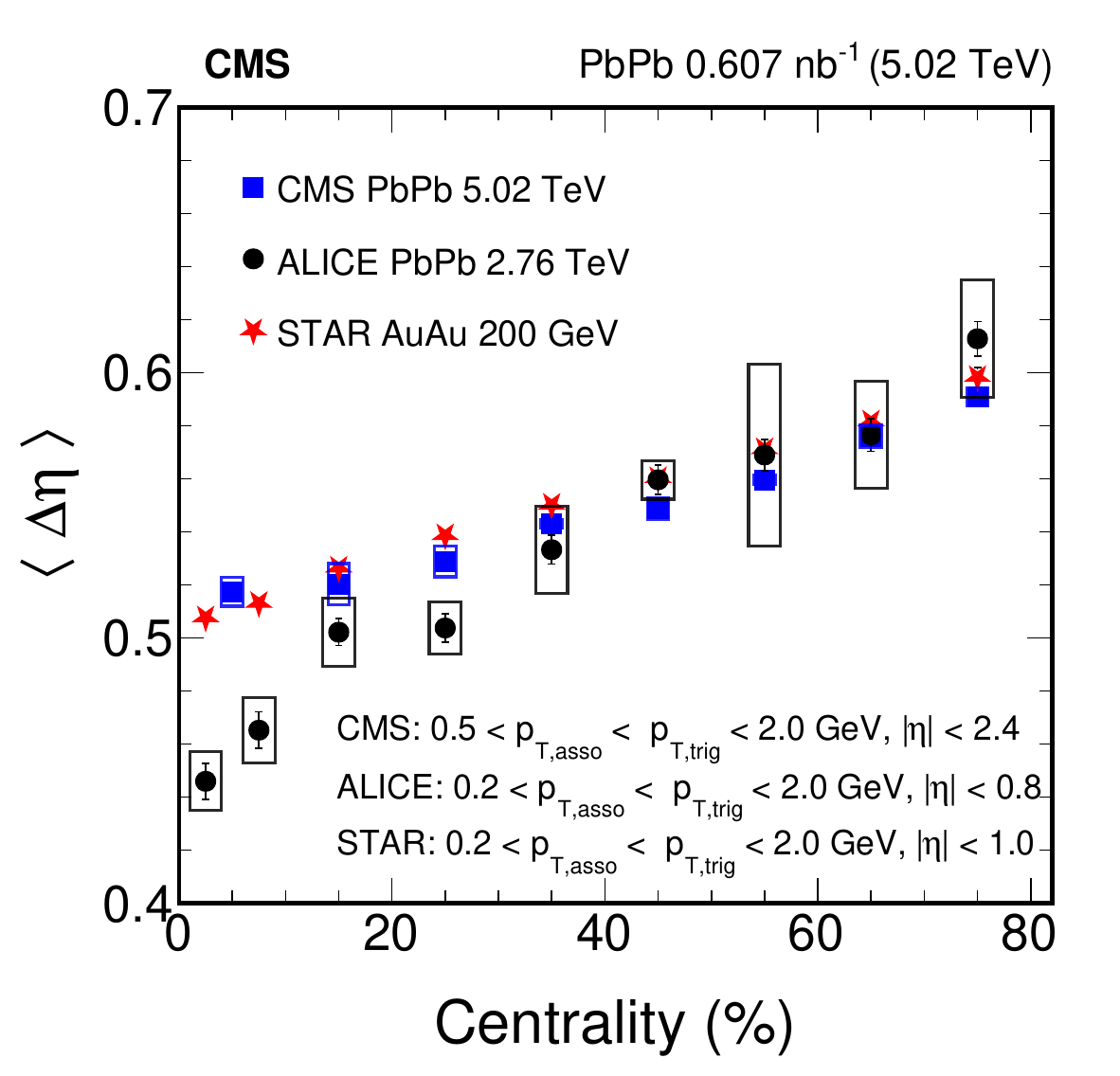}
    \includegraphics[width=0.49\textwidth]{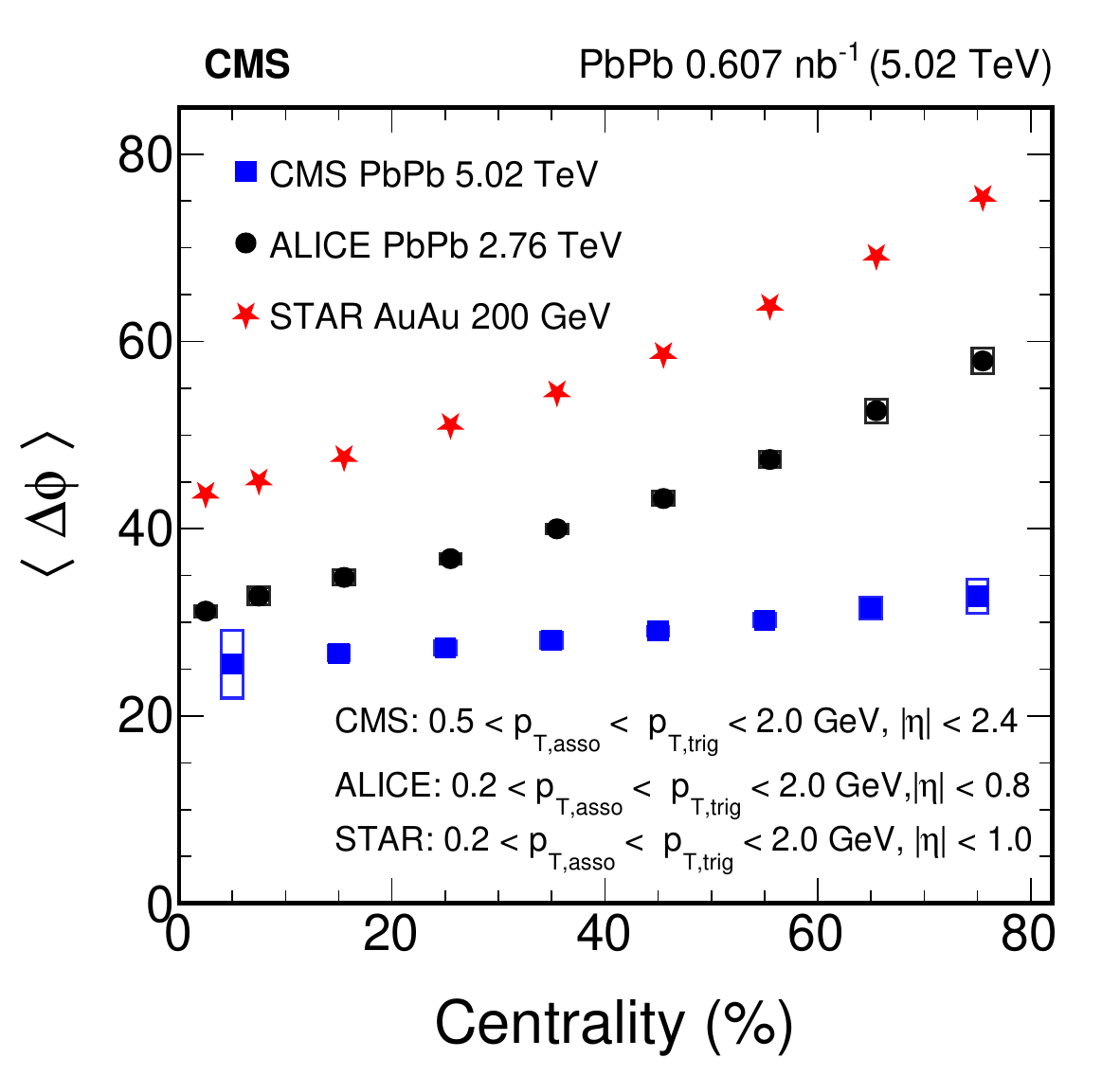}
    \includegraphics[width=0.49\textwidth]{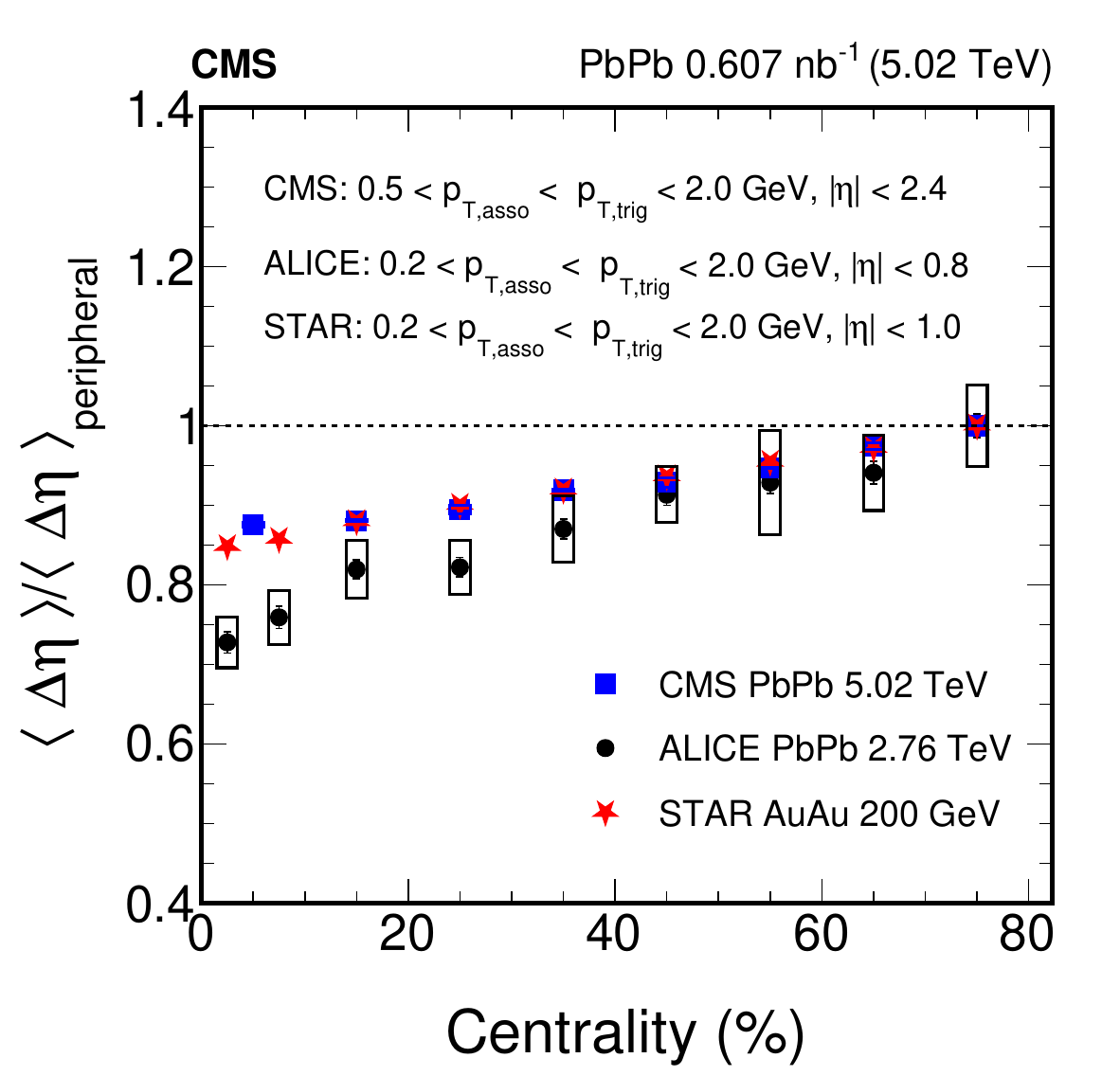}
    \includegraphics[width=0.49\textwidth]{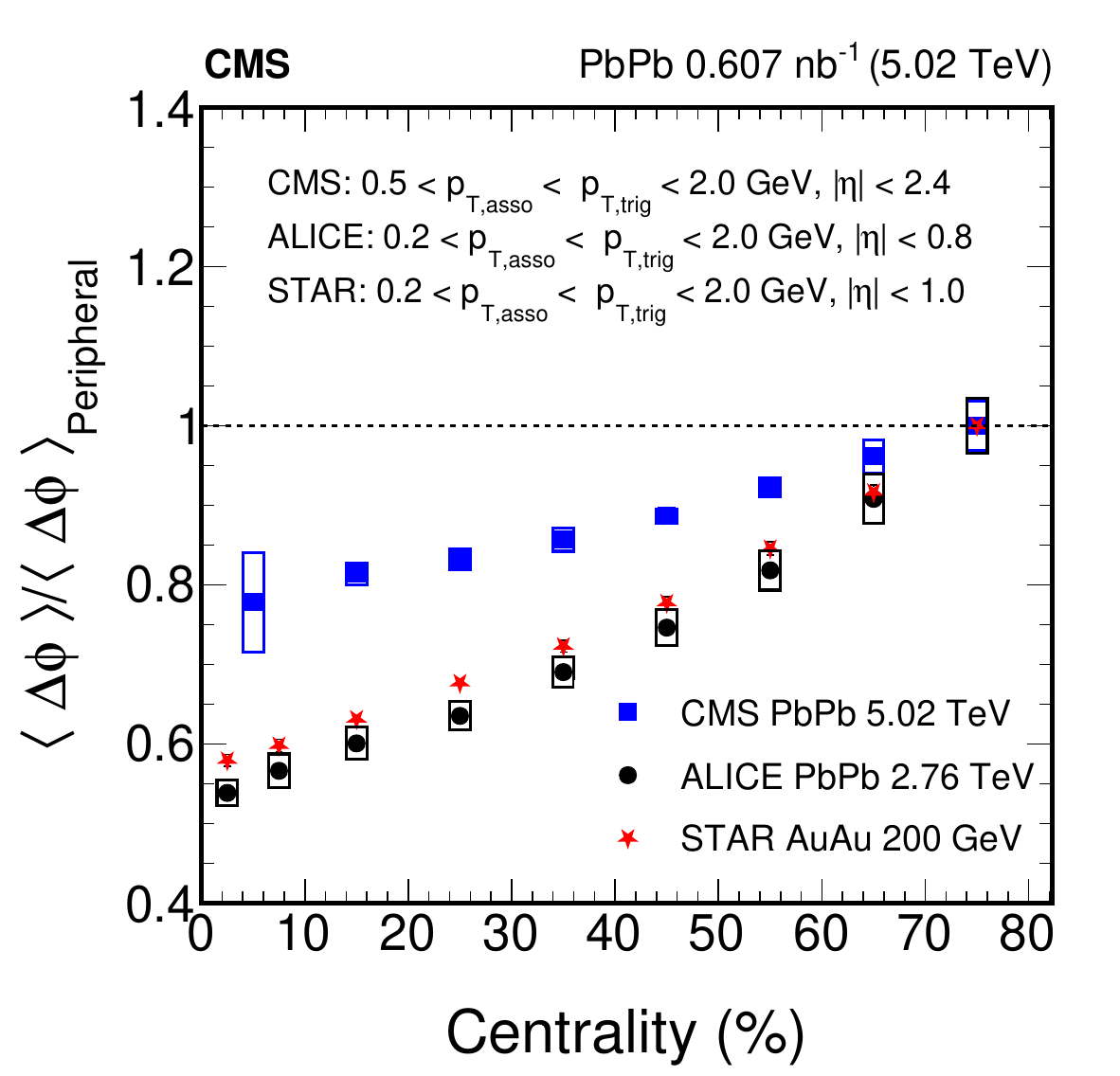}
  \caption{The centrality dependence of the balance function width (upper panels) and relative change in the width (lower panels) in $\Delta\eta$ and $\Delta\phi$. The CMS results are compared to the STAR results at $\sqrtsNN = 200\GeV$ in AuAu collisions and ALICE at $\sqrtsNN = 2.76\TeV$ in \PbPb collisions. The STAR and ALICE measured their results in $0.2 < \pt < 2.0\GeV$ ranges with a limited pseudorapidity coverage ($\abs{\eta} < 1.0$ for STAR and $\abs{\eta} < 0.8$ for ALICE). The vertical lines indicate the statistical uncertainties, and the rectangular boxes indicate the systematic uncertainties.}
\label{beamenndep}
\end{figure}

In the upper panel of the Fig.~\ref{beamenndep}, a comparison is presented of the balance function widths in $\langle \abs{\Delta\eta} \rangle$ and $\langle \abs{\Delta\phi} \rangle$ as a function of centrality. The STAR and ALICE Collaborations reported their results at $\sqrtsNN = 200\GeV$ in AuAu collisions and $\sqrtsNN = 2.76\TeV$ in \PbPb collisions, respectively, with a transverse momentum interval of $0.2 < \pt < 2.0\GeV$~\cite{ccubf,rhicbf}. The balance function width calculated for STAR does not include any systematic uncertainties. The data points from both the ALICE and STAR experiments are corrected for acceptance and detector effects to make a proper comparisons with the CMS results. For all three experiments, a significant centrality dependence is observed in $\langle \Delta\eta \rangle$ and $\langle \Delta\phi \rangle$. The balance function width is found to be narrower at the LHC energies than at RHIC, which is consistent with the idea that the system produced during the collisions has a larger radial flow due to the increase of the center-of-mass energy~\cite{hin11006,CMS:2015yux}. Additionally, the narrowing of the balance functions suggests that a longer duration of the QGP at the LHC can reduce the separation between charge pairs during their creation at hadronization. The lower panel of the Fig.~\ref{beamenndep} shows the relative decrease of the width in $\langle \Delta\eta \rangle$ and $\langle \Delta\phi \rangle$ from peripheral to central collisions. 

\clearpage
The relative change in the width in $\langle \Delta\eta \rangle$ is consistent in three different energies for RHIC and LHC within the uncertainties. This finding could shed light on the claim that the late-stage production of balancing partners was mainly responsible for narrowing the width in $\Delta\eta$. On the other hand, a significant difference is seen with respect to the STAR and ALICE results, when studying the relative width in $\Delta\phi$. This might be attributed to a stronger effect from radial flow on $\Delta\phi$ at higher energies and the different kinematic selections. Another factor to be considered at the LHC is a greater occurrence of more energetic jet-like structures resulting in particles emitted preferentially in cones with smaller $\Delta\eta$ and $\Delta\phi$ than at RHIC. Therefore, further theoretical and experimental studies are necessary to reconcile the late-stage creation of charges.

\section{Summary}
\label{sec:Summary}
This paper presents a measurement of the charge-balance function for nonidentified charged particles in proton-lead (\pPb) and lead-lead (\PbPb) collisions using the broad pseudorapidity coverage of the CMS detector. For both systems, the dependence of the balance function on relative pseudorapidity ($\Delta\eta$) and relative azimuthal angle $\Delta\phi$ of particle pairs is studied for different multiplicity classes and transverse momentum (\pt) ranges. It is observed that the width in both $\Delta\eta$ and $\Delta\phi$ decreases with charged particle multiplicity (\nch) in \pPb and \PbPb systems for $\pt < 2\GeV$. These results are consistent with the system possessing a large radial flow, with particle creation at a later stage of the collision, or both. A mild dependence of the integral of the charge balance functions with collision multiplicity is observed in \PbPb and \pPb collisions. The multiplicity dependence is weaker for higher \pt as compared with the $\pt < 2\GeV$ region, which implies that the balancing partners are strongly correlated. The data are compared with \HYDJET, \HIJING, and \AMPT generator predictions, none of which capture completely the multiplicity dependence seen in the data. The comparisons of \PbPb results with those at lower energies from STAR and ALICE show a similar dependence of the widths in $\Delta\eta$ and $\Delta\phi$ as a function of centrality. However, a significant difference is seen in the widths in $\Delta\phi$ as compared to both STAR and ALICE, suggesting a potential interplay from radial flow, different kinematic selection, and jet-like structures. By studying the charge-balance functions in both small and large systems, we have explored the evolution of particle production mechanisms and the transition from a small to a large system behavior. Further study of the balance function with identified particles can provide valuable insights into the hadronization process of the quark-gluon plasma (QGP) and a crucial benchmarks that constraints the theoretical models of hadron production and its transport in heavy-ion collisions.

\begin{acknowledgments}
  We congratulate our colleagues in the CERN accelerator departments for the excellent performance of the LHC and thank the technical and administrative staffs at CERN and at other CMS institutes for their contributions to the success of the CMS effort. In addition, we gratefully acknowledge the computing centers and personnel of the Worldwide LHC Computing Grid and other centers for delivering so effectively the computing infrastructure essential to our analyses. Finally, we acknowledge the enduring support for the construction and operation of the LHC, the CMS detector, and the supporting computing infrastructure provided by the following funding agencies: SC (Armenia), BMBWF and FWF (Austria); FNRS and FWO (Belgium); CNPq, CAPES, FAPERJ, FAPERGS, and FAPESP (Brazil); MES and BNSF (Bulgaria); CERN; CAS, MoST, and NSFC (China); MINCIENCIAS (Colombia); MSES and CSF (Croatia); RIF (Cyprus); SENESCYT (Ecuador); MoER, ERC PUT and ERDF (Estonia); Academy of Finland, MEC, and HIP (Finland); CEA and CNRS/IN2P3 (France); BMBF, DFG, and HGF (Germany); GSRI (Greece); NKFIH (Hungary); DAE and DST (India); IPM (Iran); SFI (Ireland); INFN (Italy); MSIP and NRF (Republic of Korea); MES (Latvia); LAS (Lithuania); MOE and UM (Malaysia); BUAP, CINVESTAV, CONACYT, LNS, SEP, and UASLP-FAI (Mexico); MOS (Montenegro); MBIE (New Zealand); PAEC (Pakistan); MES and NSC (Poland); FCT (Portugal); MESTD (Serbia); MCIN/AEI and PCTI (Spain); MOSTR (Sri Lanka); Swiss Funding Agencies (Switzerland); MST (Taipei); MHESI and NSTDA (Thailand); TUBITAK and TENMAK (Turkey); NASU (Ukraine); STFC (United Kingdom); DOE and NSF (USA).
   
  \hyphenation{Rachada-pisek} Individuals have received support from the Marie-Curie program and the European Research Council and Horizon 2020 Grant, contract Nos.\ 675440, 724704, 752730, 758316, 765710, 824093, 884104, and COST Action CA16108 (European Union); the Leventis Foundation; the Alfred P.\ Sloan Foundation; the Alexander von Humboldt Foundation; the Science Committee, project no. 22rl-037 (Armenia); the Belgian Federal Science Policy Office; the Fonds pour la Formation \`a la Recherche dans l'Industrie et dans l'Agriculture (FRIA-Belgium); the Agentschap voor Innovatie door Wetenschap en Technologie (IWT-Belgium); the F.R.S.-FNRS and FWO (Belgium) under the ``Excellence of Science -- EOS" -- be.h project n.\ 30820817; the Beijing Municipal Science \& Technology Commission, No. Z191100007219010; the Ministry of Education, Youth and Sports (MEYS) of the Czech Republic; the Shota Rustaveli National Science Foundation, grant FR-22-985 (Georgia); the Deutsche Forschungsgemeinschaft (DFG), under Germany's Excellence Strategy -- EXC 2121 ``Quantum Universe" -- 390833306, and under project number 400140256 - GRK2497; the Hellenic Foundation for Research and Innovation (HFRI), Project Number 2288 (Greece); the Hungarian Academy of Sciences, the New National Excellence Program - \'UNKP, the NKFIH research grants K 124845, K 124850, K 128713, K 128786, K 129058, K 131991, K 133046, K 138136, K 143460, K 143477, 2020-2.2.1-ED-2021-00181, and TKP2021-NKTA-64 (Hungary); the Council of Science and Industrial Research, India; the Latvian Council of Science; the Ministry of Education and Science, project no. 2022/WK/14, and the National Science Center, contracts Opus 2021/41/B/ST2/01369 and 2021/43/B/ST2/01552 (Poland); the Funda\c{c}\~ao para a Ci\^encia e a Tecnologia, grant CEECIND/01334/2018 (Portugal); the National Priorities Research Program by Qatar National Research Fund; MCIN/AEI/10.13039/501100011033, ERDF ``a way of making Europe", and the Programa Estatal de Fomento de la Investigaci{\'o}n Cient{\'i}fica y T{\'e}cnica de Excelencia Mar\'{\i}a de Maeztu, grant MDM-2017-0765 and Programa Severo Ochoa del Principado de Asturias (Spain); the Chulalongkorn Academic into Its 2nd Century Project Advancement Project, and the National Science, Research and Innovation Fund via the Program Management Unit for Human Resources \& Institutional Development, Research and Innovation, grant B05F650021 (Thailand); the Kavli Foundation; the Nvidia Corporation; the SuperMicro Corporation; the Welch Foundation, contract C-1845; and the Weston Havens Foundation (USA).
\end{acknowledgments}

\bibliography{auto_generated} 

\providecommand{\href}[2]{#2}\begingroup\raggedright\begin{thebibliography}{10}%
\makeatletter
\providecommand{\hrefCMSnoop }[0]{\@secondoftwo}%
\makeatother
\providecommand{\doi}{\texttt{doi:}\begingroup \urlstyle{tt}\Url}

\bibitem{ARSENE20051}
\hrefCMSnoop {}{{BRAHMS} Collaboration, ``Quark gluon plasma and color glass
  condensate at {RHIC}? {The} perspective from the {BRAHMS} experiment'',}
  \textit{ Nucl. Phys. A} \textbf{ 757} (2005) 1,
  \href{http://dx.doi.org/10.1016/j.nuclphysa.2005.02.130}{\doi{10.1016/j.nuclphysa.2005.02.130}},
\href{http://www.arXiv.org/abs/nucl-ex/0410020}{\texttt{arXiv:nucl-ex/0410020}}.

\bibitem{BACK200528}
\hrefCMSnoop {}{{PHOBOS} Collaboration, ``The {PHOBOS} perspective on
  discoveries at {RHIC}'',} \textit{ Nucl. Phys. A} \textbf{ 757} (2005) 28,
  \href{http://dx.doi.org/10.1016/j.nuclphysa.2005.03.084}{\doi{10.1016/j.nuclphysa.2005.03.084}},
\href{http://www.arXiv.org/abs/nucl-ex/0410022}{\texttt{arXiv:nucl-ex/0410022}}.

\bibitem{ADAMS2005102}
\hrefCMSnoop {}{{STAR} Collaboration, ``Experimental and theoretical challenges
  in the search for the quark gluon plasma: The {STAR} {Collaboration}'s
  critical assessment of the evidence from {RHIC} collisions'',} \textit{ Nucl.
  Phys. A} \textbf{ 757} (2005) 102,
  \href{http://dx.doi.org/10.1016/j.nuclphysa.2005.03.085}{\doi{10.1016/j.nuclphysa.2005.03.085}},
\href{http://www.arXiv.org/abs/nucl-ex/0501009}{\texttt{arXiv:nucl-ex/0501009}}.

\bibitem{ADCOX2005184}
\hrefCMSnoop {}{{PHENIX} Collaboration, ``Formation of dense partonic matter in
  relativistic {nucleus-nucleus} collisions at {RHIC:} {Experimental}
  evaluation by the {PHENIX} {Collaboration}'',} \textit{ Nucl. Phys. A}
  \textbf{ 757} (2005) 184,
  \href{http://dx.doi.org/10.1016/j.nuclphysa.2005.03.086}{\doi{10.1016/j.nuclphysa.2005.03.086}},
\href{http://www.arXiv.org/abs/nucl-ex/0410003}{\texttt{arXiv:nucl-ex/0410003}}.

\bibitem{M_ller_2012}
\hrefCMSnoop {}{B.~M{\"u}ller, J.~Schukraft, and B.~Wys{\l{}}ouch, ``First
  results from {\PbPb} collisions at the {LHC}'',} \textit{ Annu. Rev. Nucl.
  Part. Sci.} \textbf{ 62} (2012) 361,
  \href{http://dx.doi.org/10.1146/annurev-nucl-102711-094910}{\doi{10.1146/annurev-nucl-102711-094910}},
\href{http://www.arXiv.org/abs/1202.3233}{\texttt{arXiv:1202.3233}}.

\bibitem{Armesto_2016}
\hrefCMSnoop {}{N.~Armesto and E.~Scomparin, ``Heavy-ion collisions at the
  {Large Hadron Collider}: {A} review of the results from {Run 1}'',} \textit{
  Eur. Phys. J. Plus} \textbf{ 131} (2016) 3,
  \href{http://dx.doi.org/10.1140/epjp/i2016-16052-4}{\doi{10.1140/epjp/i2016-16052-4}},
\href{http://www.arXiv.org/abs/1511.02151}{\texttt{arXiv:1511.02151}}.

\bibitem{qgpcms_2pc}
\hrefCMSnoop {}{{CMS Collaboration}, ``Measurement of long-range near-side
  two-particle angular correlations in {pp} collisions at $\sqrt{s} =$ 13
  tev'',} \textit{ Phys. Rev. Lett.} \textbf{ 116} (2016) 172302,
  \href{http://dx.doi.org/10.1103/PhysRevLett.116.172302}{\doi{10.1103/PhysRevLett.116.172302}},
\href{http://www.arXiv.org/abs/1510.03068}{\texttt{arXiv:1510.03068}}.

\bibitem{cmshin12015}
\hrefCMSnoop {}{{CMS Collaboration}, ``Observation of long-range, near-side
  angular correlations in {\pPb} collisions at the {LHC}'',} \textit{ Phys.
  Lett. B} \textbf{ 718} (2013) 795,
  \href{http://dx.doi.org/10.1016/j.physletb.2012.11.025}{\doi{10.1016/j.physletb.2012.11.025}},
\href{http://www.arXiv.org/abs/1210.5482}{\texttt{arXiv:1210.5482}}.

\bibitem{PhysRevC.76.024914}
\hrefCMSnoop {}{{NA49} Collaboration, ``Rapidity and energy dependence of the
  electric charge correlations in {A+A} collisions at the {SPS} energies'',}
  \textit{ Phys. Rev. C} \textbf{ 76} (2007) 024914,
  \href{http://dx.doi.org/10.1103/PhysRevC.76.024914}{\doi{10.1103/PhysRevC.76.024914}},
\href{http://www.arXiv.org/abs/0705.1122}{\texttt{arXiv:0705.1122}}.

\bibitem{Dusling_2016}
\hrefCMSnoop {}{B.~S. Kevin~Dusling, Wei~Li, ``Novel collective phenomena in
  high-energy proton{\textendash}proton and proton{\textendash}nucleus
  collisions'',} \textit{ Int. J. Mod. Phys. E} \textbf{ 25} (2016) 1630002,
  \href{http://dx.doi.org/10.1142/s0218301316300022}{\doi{10.1142/s0218301316300022}},
\href{http://www.arXiv.org/abs/1509.07939}{\texttt{arXiv:1509.07939}}.

\bibitem{gqp_hin11001}
\hrefCMSnoop {}{{CMS Collaboration}, ``Long-range and short-range dihadron
  angular correlations in central {\PbPb} collisions at {$\sqrtsNN =
  2.76\TeV$}'',} \textit{ JHEP} \textbf{ 07} (2011) 076,
  \href{http://dx.doi.org/10.1007/jhep07(2011)076}{\doi{10.1007/jhep07(2011)076}},
\href{http://www.arXiv.org/abs/1105.2438}{\texttt{arXiv:1105.2438}}.

\bibitem{hin11006}
\hrefCMSnoop {}{{CMS Collaboration}, ``Centrality dependence of dihadron
  correlations and azimuthal anisotropy harmonics in {\PbPb} collisions at
  {$\sqrtsNN = 2.76\TeV$}'',} \textit{ Eur. J. Phys. C} \textbf{ 72} (2012) 5,
  \href{http://dx.doi.org/10.1140/epjc/s10052-012-2012-3}{\doi{10.1140/epjc/s10052-012-2012-3}},
\href{http://www.arXiv.org/abs/1201.3158}{\texttt{arXiv:1201.3158}}.

\bibitem{cmsppridge}
\hrefCMSnoop {}{{CMS Collaboration}, ``Observation of long-range, near-side
  angular correlations in proton-proton collisions at the {LHC}'',} \textit{
  JHEP} \textbf{ 09} (2010) 091,
  \href{http://dx.doi.org/10.1007/jhep09(2010)091}{\doi{10.1007/jhep09(2010)091}},
\href{http://www.arXiv.org/abs/1009.4122}{\texttt{arXiv:1009.4122}}.

\bibitem{PhysRevC.76.047901}
\hrefCMSnoop {}{E.~Shuryak, ``Origin of the ``ridge'' phenomenon induced by
  jets in heavy ion collisions'',} \textit{ Phys. Rev. C} \textbf{ 76} (2007)
  047901,
  \href{http://dx.doi.org/10.1103/PhysRevC.76.047901}{\doi{10.1103/PhysRevC.76.047901}},
\href{http://www.arXiv.org/abs/0706.3531}{\texttt{arXiv:0706.3531}}.

\bibitem{scott_chemqgp}
\hrefCMSnoop {}{S.~Pratt, ``General charge balance functions: A tool for
  studying the chemical evolution of the quark-gluon plasma'',} \textit{ Phys.
  Rev. C} \textbf{ 85} (2012) 014904,
  \href{http://dx.doi.org/10.1103/PhysRevC.85.014904}{\doi{10.1103/PhysRevC.85.014904}},
\href{http://www.arXiv.org/abs/1109.3647}{\texttt{arXiv:1109.3647}}.

\bibitem{ALICE:2021hjb}
\hrefCMSnoop {}{{ALICE Collaboration}, ``General balance functions of
  identified charged hadron pairs of (\ensuremath{\pi},k,p) in {PbPb}
  collisions at {$\sqrtsNN = 2.76\TeV$}'',} \textit{ Phys. Lett. B} \textbf{
  833} (2022) 137338,
  \href{http://dx.doi.org/10.1016/j.physletb.2022.137338}{\doi{10.1016/j.physletb.2022.137338}},
  \href{http://www.arXiv.org/abs/2110.06566}{\texttt{arXiv:2110.06566}}.

\bibitem{PhysRevLett.85.2689}
\hrefCMSnoop {}{S.~P. Steffen~Bass, Pawel~Danielewicz, ``Clocking hadronization
  in relativistic heavy ion collisions with balance functions'',} \textit{
  Phys. Rev. Lett.} \textbf{ 85} (2000) 2689,
  \href{http://dx.doi.org/10.1103/physrevlett.85.2689}{\doi{10.1103/physrevlett.85.2689}},
\href{http://www.arXiv.org/abs/nucl-th/0005044}{\texttt{arXiv:nucl-th/0005044}}.

\bibitem{Khachatryan:2016odn}
\hrefCMSnoop {}{{CMS Collaboration}, ``Measurement of prompt $\mathrm{D}^0$ and
  {$\overline{\mathrm{D}^0}$} meson azimuthal asymmetry and search for strong
  electric fields in {\PbPb} collisions at ${\sqrtsNN=5.02\TeV}$'',} \textit{
  Phys. Lett. B} \textbf{ 816} (2021) 136253,
  \href{http://dx.doi.org/10.1016/j.physletb.2021.136253}{\doi{10.1016/j.physletb.2021.136253}},
\href{http://www.arXiv.org/abs/2009.12628}{\texttt{arXiv:2009.12628}}.

\bibitem{icpaqgp}
\hrefCMSnoop {}{S.~A. Voloshin, ``Heavy ion collisions: {Correlations and
  Fluctuations} in particle production'',} \textit{ J. Phys.: Conf. Ser.}
  \textbf{ 50} (2006) 111,
  \href{http://dx.doi.org/10.1088/1742-6596/50/1/013}{\doi{10.1088/1742-6596/50/1/013}},
\href{http://www.arXiv.org/abs/nucl-ex/0505003}{\texttt{arXiv:nucl-ex/0505003}}.

\bibitem{ACCDHW:1979mjc}
\hrefCMSnoop {}{{ACCDHW} Collaboration, ``Quantum number effects in events with
  a charged particle of large transverse momentum: {II. C}harge correlations in
  jets'',} \textit{ Nucl. Phys. B} \textbf{ 166} (1980) 233,
  \href{http://dx.doi.org/10.1016/0550-3213(80)90226-6}{\doi{10.1016/0550-3213(80)90226-6}}.

\bibitem{Voloshin:2003ud}
\hrefCMSnoop {}{S.~A. Voloshin, ``{Transverse radial expansion in nuclear
  collisions and two particle correlations}'',} \textit{ Phys. Lett. B}
  \textbf{ 632} (2006) 490,
  \href{http://dx.doi.org/10.1016/j.physletb.2005.11.024}{\doi{10.1016/j.physletb.2005.11.024}},
  \href{http://www.arXiv.org/abs/nucl-th/0312065}{\texttt{arXiv:nucl-th/0312065}}.

\bibitem{BOZEK2005247}
\hrefCMSnoop {}{P.~Bo{\.z}ek, ``The balance function in azimuthal angle is a
  measure of the transverse flow'',} \textit{ Phys. Lett. B} \textbf{ 609}
  (2005) 247,
  \href{http://dx.doi.org/10.1016/j.physletb.2005.01.072}{\doi{10.1016/j.physletb.2005.01.072}},
\href{http://www.arXiv.org/abs/nucl-th/0412076}{\texttt{arXiv:nucl-th/0412076}}.

\bibitem{Pratt:2021xvg}
\hrefCMSnoop {}{S.~Pratt and C.~Plumberg, ``{Charge balance functions for
  heavy-ion collisions at energies available at the CERN Large Hadron
  Collider}'',} \textit{ Phys. Rev. C} \textbf{ 104} (2021) 014906,
  \href{http://dx.doi.org/10.1103/PhysRevC.104.014906}{\doi{10.1103/PhysRevC.104.014906}},
  \href{http://www.arXiv.org/abs/2104.00628}{\texttt{arXiv:2104.00628}}.

\bibitem{Pratt:2018ebf}
\hrefCMSnoop {}{S.~Pratt and C.~Plumberg, ``Evolving charge correlations in a
  hybrid model with both hydrodynamics and hadronic {B}oltzmann
  descriptions'',} \textit{ Phys. Rev. C} \textbf{ 99} (2019) 044916,
  \href{http://dx.doi.org/10.1103/PhysRevC.99.044916}{\doi{10.1103/PhysRevC.99.044916}},
  \href{http://www.arXiv.org/abs/1812.05649}{\texttt{arXiv:1812.05649}}.

\bibitem{Jeon:2001ue}
\hrefCMSnoop {}{S.~Jeon and S.~Pratt, ``Balance functions, correlations, charge
  fluctuations and interferometry'',} \textit{ Phys. Rev. C} \textbf{ 65}
  (2002) 044902,
  \href{http://dx.doi.org/10.1103/PhysRevC.65.044902}{\doi{10.1103/PhysRevC.65.044902}},
  \href{http://www.arXiv.org/abs/hep-ph/0110043}{\texttt{arXiv:hep-ph/0110043}}.

\bibitem{Pratt:2015jsa}
\hrefCMSnoop {}{S.~Pratt, W.~P. McCormack, and C.~Ratti, ``Production of charge
  in heavy ion collisions'',} \textit{ Phys. Rev. C} \textbf{ 92} (2015)
  064905,
  \href{http://dx.doi.org/10.1103/PhysRevC.92.064905}{\doi{10.1103/PhysRevC.92.064905}},
  \href{http://www.arXiv.org/abs/1508.07031}{\texttt{arXiv:1508.07031}}.

\bibitem{Pratt:2019pnd}
\hrefCMSnoop {}{S.~Pratt and C.~Plumberg, ``Determining the diffusivity for
  light quarks from experiment'',} \textit{ Phys. Rev. C} \textbf{ 102} (2020)
  044909,
  \href{http://dx.doi.org/10.1103/PhysRevC.102.044909}{\doi{10.1103/PhysRevC.102.044909}},
  \href{http://www.arXiv.org/abs/1904.11459}{\texttt{arXiv:1904.11459}}.

\bibitem{Pruneau:2019baa}
\hrefCMSnoop {}{C.~A. Pruneau, ``Role of baryon number conservation in
  measurements of fluctuations'',} \textit{ Phys. Rev. C} \textbf{ 100} (2019)
  034905,
  \href{http://dx.doi.org/10.1103/PhysRevC.100.034905}{\doi{10.1103/PhysRevC.100.034905}},
  \href{http://www.arXiv.org/abs/1903.04591}{\texttt{arXiv:1903.04591}}.

\bibitem{Pratt:2012dz}
\hrefCMSnoop {}{S.~Pratt, ``Identifying the charge carriers of the quark-gluon
  plasma'',} \textit{ Phys. Rev. Lett.} \textbf{ 108} (2012) 212301,
  \href{http://dx.doi.org/10.1103/PhysRevLett.108.212301}{\doi{10.1103/PhysRevLett.108.212301}},
  \href{http://www.arXiv.org/abs/1203.4578}{\texttt{arXiv:1203.4578}}.

\bibitem{Ye:2018jwq}
\hrefCMSnoop {}{Y.~J. Ye, Y.~G. Ma, A.~H. Tang, and G.~Wang, ``Effect of
  magnetic fields on pairs of oppositely charged particles in ultrarelativistic
  heavy-ion collisions'',} \textit{ Phys. Rev. C} \textbf{ 99} (2019) 044901,
  \href{http://dx.doi.org/10.1103/PhysRevC.99.044901}{\doi{10.1103/PhysRevC.99.044901}},
  \href{http://www.arXiv.org/abs/1810.04600}{\texttt{arXiv:1810.04600}}.

\bibitem{rhicbf}
\hrefCMSnoop {}{{STAR} Collaboration, ``Balance functions from {Au+Au}, {d+Au},
  and p+p collisions at {$\sqrtsNN = 200\GeV$}'',} \textit{ Phys. Rev. C}
  \textbf{ 82} (2010) 024905,
  \href{http://dx.doi.org/10.1103/PhysRevC.82.024905}{\doi{10.1103/PhysRevC.82.024905}},
\href{http://www.arXiv.org/abs/1005.2307}{\texttt{arXiv:1005.2307}}.

\bibitem{ccubf}
\hrefCMSnoop {}{{ALICE Collaboration}, ``Charge correlations using the balance
  function in {\PbPb} collisions at {$\sqrtsNN = 2.76\TeV$}'',} \textit{ Phys.
  Lett. B} \textbf{ 723} (2013) 267,
  \href{http://dx.doi.org/10.1016/j.physletb.2013.05.039}{\doi{10.1016/j.physletb.2013.05.039}},
\href{http://www.arXiv.org/abs/1301.3756}{\texttt{arXiv:1301.3756}}.

\bibitem{CMS:2008xjf}
\hrefCMSnoop {}{{CMS Collaboration}, ``The {CMS} experiment at the {CERN}
  {LHC}'',} \textit{ JINST} \textbf{ 3} (2008) S08004,
\href{http://dx.doi.org/10.1088/1748-0221/3/08/S08004}{\doi{10.1088/1748-0221/3/08/S08004}}.

\bibitem{HEPData}
\hrefCMSnoop {}{``{HEPD}ata record for this analysis'',} 2022.
\newblock
  \href{http://dx.doi.org/10.17182/hepdata.135972}{\doi{10.17182/hepdata.135972}}.

\bibitem{CMS:2014pgm}
\hrefCMSnoop {}{{CMS Collaboration}, ``Description and performance of track and
  primary-vertex reconstruction with the cms tracker'',} \textit{ JINST}
  \textbf{ 9} (2014) P10009,
  \href{http://dx.doi.org/10.1088/1748-0221/9/10/P10009}{\doi{10.1088/1748-0221/9/10/P10009}},
  \href{http://www.arXiv.org/abs/1405.6569}{\texttt{arXiv:1405.6569}}.

\bibitem{DP-2020-049}
\href {https://cds.cern.ch/record/2743740}{{CMS Collaboration}, ``Track impact
  parameter resolution for the full pseudo rapidity coverage in the 2017
  dataset with the {CMS} phase-1 pixel detector'',} CMS Detector Performance
  Note CMS-DP-2020-049, 2020.

\bibitem{Phase1Pixel}
\hrefCMSnoop {}{{CMS Tracker Group} Collaboration, ``The {CMS} {Phase-1} pixel
  detector upgrade'',} \textit{ JINST} \textbf{ 16} (2021) P02027,
  \href{http://dx.doi.org/10.1088/1748-0221/16/02/P02027}{\doi{10.1088/1748-0221/16/02/P02027}},
  \href{http://www.arXiv.org/abs/2012.14304}{\texttt{arXiv:2012.14304}}.

\bibitem{CMS:LUM-18-001}
\href {https://cds.cern.ch/record/2809613}{{CMS Collaboration}, ``{CMS}
  luminosity measurement using nucleus-nucleus collisions at
  ${\sqrtsNN=5.02\TeV}$ in 2018'',} CMS Physics Analysis Summary
  CMS-PAS-LUM-18-001, 2022.

\bibitem{CMS:LUM-17-003}
\hrefCMSnoop {}{{CMS Collaboration}, ``Precision luminosity measurement in
  proton-proton collisions at $\sqrt{s}={13\TeV}$ in 2015 and 2016 at {CMS}'',}
  \textit{ Eur. Phys. J. C} \textbf{ 81} (2021) 800,
  \href{http://dx.doi.org/10.1140/epjc/s10052-021-09538-2}{\doi{10.1140/epjc/s10052-021-09538-2}},
  \href{http://www.arXiv.org/abs/2104.01927}{\texttt{arXiv:2104.01927}}.

\bibitem{CMS:2020cmk}
\hrefCMSnoop {}{{CMS Collaboration}, ``Performance of the {CMS Level-1} trigger
  in proton-proton collisions at $\sqrt{s} = $ 13 {TeV}'',} \textit{ JINST}
  \textbf{ 15} (2020) P10017,
  \href{http://dx.doi.org/10.1088/1748-0221/15/10/P10017}{\doi{10.1088/1748-0221/15/10/P10017}},
  \href{http://www.arXiv.org/abs/2006.10165}{\texttt{arXiv:2006.10165}}.

\bibitem{CMS:2016ngn}
\hrefCMSnoop {}{{CMS Collaboration}, ``The {CMS} trigger system'',} \textit{
  JINST} \textbf{ 12} (2017) P01020,
  \href{http://dx.doi.org/10.1088/1748-0221/12/01/P01020}{\doi{10.1088/1748-0221/12/01/P01020}},
\href{http://www.arXiv.org/abs/1609.02366}{\texttt{arXiv:1609.02366}}.

\bibitem{strangehadron}
\hrefCMSnoop {}{{CMS Collaboration}, ``Strange hadron collectivity in {\pPb}
  and {\PbPb} collisions'',} \textit{ JHEP} \textbf{ 05} (2023) 007,
  \href{http://dx.doi.org/10.1007/JHEP05(2023)007}{\doi{10.1007/JHEP05(2023)007}},
  \href{http://www.arXiv.org/abs/2205.00080}{\texttt{arXiv:2205.00080}}.

\bibitem{CMS-TDR-15-02}
\href {http://cds.cern.ch/record/2020886}{{CMS Collaboration}, ``Technical
  proposal for the {Phase-II} upgrade of the {Compact Muon Solenoid}'',} CMS
  Technical Proposal CERN-LHCC-2015-010, CMS-TDR-15-02, 2015.

\bibitem{CMS:LUM-17-002}
\href {https://cds.cern.ch/record/2628652}{{CMS Collaboration}, ``{CMS}
  luminosity measurement using 2016 proton-nucleus collisions at
  ${\sqrtsNN=8.16\TeV}$'',} CMS Physics Analysis Summary CMS-PAS-LUM-17-002,
  2018.

\bibitem{symmcum}
\hrefCMSnoop {}{{CMS Collaboration}, ``Observation of correlated azimuthal
  anisotropy fourier harmonics in {pp} and {p+Pb} collisions at the {LHC}'',}
  \textit{ Phys. Rev. Lett.} \textbf{ 120} (2018) 092301,
  \href{http://dx.doi.org/10.1103/PhysRevLett.120.092301}{\doi{10.1103/PhysRevLett.120.092301}},
\href{http://www.arXiv.org/abs/1709.09189}{\texttt{arXiv:1709.09189}}.

\bibitem{2013213}
\hrefCMSnoop {}{{CMS Collaboration}, ``Multiplicity and transverse momentum
  dependence of two- and four-particle correlations in {\pPb} and {\PbPb}
  collisions'',} \textit{ Phys. Lett. B} \textbf{ 724} (2013) 213,
  \href{http://dx.doi.org/10.1016/j.physletb.2013.06.028}{\doi{10.1016/j.physletb.2013.06.028}},
\href{http://www.arXiv.org/abs/1305.0609}{\texttt{arXiv:1305.0609}}.

\bibitem{Hydjet}
Lokhtin\hrefCMSnoop {}{ { et~al.}, ``Heavy ion event generator {HYDJET++}
  ({HYDrodynamics} plus {JETs})'',} \textit{ Comput. Phys. Comm.} \textbf{ 180}
  (2009) 779,
  \href{http://dx.doi.org/10.1016/j.cpc.2008.11.015}{\doi{10.1016/j.cpc.2008.11.015}},
\href{http://www.arXiv.org/abs/0809.2708}{\texttt{arXiv:0809.2708}}.

\bibitem{ampt}
\hrefCMSnoop {}{H.-j. Xu, Z.~Li, and H.~Song, ``High-order flow harmonics of
  identified hadrons in $2.76${A} {TeV} {\PbPb} collisions'',} \textit{ Phys.
  Rev. C} \textbf{ 93} (2016) 064905,
  \href{http://dx.doi.org/10.1103/PhysRevC.93.064905}{\doi{10.1103/PhysRevC.93.064905}},
\href{http://www.arXiv.org/abs/1602.02029}{\texttt{arXiv:1602.02029}}.

\bibitem{HIJING}
\hrefCMSnoop {}{Z.~Moravcova, K.~Gulbrandsen, and Y.~Zhou, ``Generic algorithm
  for multiparticle cumulants of azimuthal correlations in high energy nucleus
  collisions'',} \textit{ Phys. Rev. C} \textbf{ 103} (2021) 024913,
  \href{http://dx.doi.org/10.1103/PhysRevC.103.024913}{\doi{10.1103/PhysRevC.103.024913}},
\href{http://www.arXiv.org/abs/2005.07974}{\texttt{arXiv:2005.07974}}.

\bibitem{HIJING2}
G.~B{\'i}r{\'o}\hrefCMSnoop {}{ { et~al.}, ``Introducing {HIJING++}: the heavy
  ion {Monte Carlo} generator for the high-luminosity {LHC} era'',} in \textit{
  Proceedings of International Conference on Hard and Electromagnetic Probes of
  High-Energy Nuclear Collis\ ions --- PoS(HardProbes2018)}, volume 345,
  p.~045.
\newblock 2019.
\newblock
  \href{http://www.arXiv.org/abs/1901.04220}{\texttt{arXiv:1901.04220}}.
\newblock
\href{http://dx.doi.org/10.22323/1.345.0045}{\doi{10.22323/1.345.0045}}.

\bibitem{CMS:2010mua}
\href {https://cds.cern.ch/record/1279139}{{CMS Collaboration}, ``Measurement
  of tracking efficiency'',} 2010.

\bibitem{GEANT4}
\hrefCMSnoop {}{{GEANT4} Collaboration, ``{\GEANTfour} --- a simulation
  toolkit'',} \textit{ Nucl. Instrum. Meth. A} \textbf{ 506} (2003) 250,
\href{http://dx.doi.org/10.1016/S0168-9002(03)01368-8}{\doi{10.1016/S0168-9002(03)01368-8}}.

\bibitem{CMS:2011iwn}
\hrefCMSnoop {}{{CMS Collaboration}, ``Observation and studies of jet quenching
  in {\PbPb} collisions at {$\sqrtsNN=2.76\TeV$}'',} \textit{ Phys. Rev. C}
  \textbf{ 84} (2011) 024906,
  \href{http://dx.doi.org/10.1103/PhysRevC.84.024906}{\doi{10.1103/PhysRevC.84.024906}},
  \href{http://www.arXiv.org/abs/1102.1957}{\texttt{arXiv:1102.1957}}.

\bibitem{jet1}
\hrefCMSnoop {}{{STAR} Collaboration, ``Long range rapidity correlations and
  jet production in high energy nuclear collisions'',} \textit{ Phys. Rev. C}
  \textbf{ 80} (2009) 064912,
  \href{http://dx.doi.org/10.1103/PhysRevC.80.064912}{\doi{10.1103/PhysRevC.80.064912}},
\href{http://www.arXiv.org/abs/0909.0191}{\texttt{arXiv:0909.0191}}.

\bibitem{jet2}
\hrefCMSnoop {}{{PHOBOS} Collaboration, ``High transverse momentum triggered
  correlations over a large pseudorapidity acceptance in
  $\mathrm{Au}+\mathrm{Au}$ collisions at ${\sqrtsNN=200\GeV}$'',} \textit{
  Phys. Rev. Lett.} \textbf{ 104} (2010) 062301,
  \href{http://dx.doi.org/10.1103/PhysRevLett.104.062301}{\doi{10.1103/PhysRevLett.104.062301}},
\href{http://www.arXiv.org/abs/0903.2811}{\texttt{arXiv:0903.2811}}.

\bibitem{ALICE:2018jco}
\hrefCMSnoop {}{{ALICE Collaboration}, ``{Two particle differential transverse
  momentum and number density correlations in p-Pb and Pb-Pb at the LHC}'',}
  \textit{ Phys. Rev. C} \textbf{ 100} (2019) 044903,
  \href{http://dx.doi.org/10.1103/PhysRevC.100.044903}{\doi{10.1103/PhysRevC.100.044903}},
  \href{http://www.arXiv.org/abs/1805.04422}{\texttt{arXiv:1805.04422}}.

\bibitem{CMS:2014und}
\hrefCMSnoop {}{{CMS Collaboration}, ``{Long-range two-particle correlations of
  strange hadrons with charged particles in {pPb} and {PbPb} collisions at
  {LHC} energies}'',} \textit{ Phys. Lett. B} \textbf{ 742} (2015) 200,
  \href{http://dx.doi.org/10.1016/j.physletb.2015.01.034}{\doi{10.1016/j.physletb.2015.01.034}},
  \href{http://www.arXiv.org/abs/1409.3392}{\texttt{arXiv:1409.3392}}.

\bibitem{PhysRevC.78.054906}
\hrefCMSnoop {}{S.~Pratt and J.~Vredevoogd, ``Femtoscopy in relativistic heavy
  ion collisions and its relation to bulk properties of {QCD} matter'',}
  \textit{ Phys. Rev. C} \textbf{ 78} (2008) 054906,
  \href{http://dx.doi.org/10.1103/PhysRevC.78.054906}{\doi{10.1103/PhysRevC.78.054906}},
\href{http://www.arXiv.org/abs/0809.0516}{\texttt{arXiv:0809.0516}}.

\bibitem{hbt}
\hrefCMSnoop {}{R.~Hanbury~Brown and R.~Q. Twiss, ``Correlation between photons
  in two coherent beams of light'',} \textit{ Nature} \textbf{ 177} (1956) 27,
  \href{http://dx.doi.org/10.1038/177027a0}{\doi{10.1038/177027a0}}.

\bibitem{PhysRevD.84.112004}
\hrefCMSnoop {}{{ALICE Collaboration}, ``Femtoscopy of {pp} collisions at
  $\sqrt{s}=0.9$ and 7 {TeV} at the {LHC} with two-pion {Bose-Einstein}
  correlations'',} \textit{ Phys. Rev. D} \textbf{ 84} (2011) 112004,
  \href{http://dx.doi.org/10.1103/PhysRevD.84.112004}{\doi{10.1103/PhysRevD.84.112004}},
\href{http://www.arXiv.org/abs/1101.3665}{\texttt{arXiv:1101.3665}}.

\bibitem{alicebf}
\hrefCMSnoop {}{{ALICE} Collaboration, ``Multiplicity and transverse momentum
  evolution of charge-dependent correlations in pp, {p-Pb}, and {Pb-Pb}
  collisions at the {LHC}'',} \textit{ Eur. Phys. J. C} \textbf{ 76} (2016) 86,
  \href{http://dx.doi.org/10.1140/epjc/s10052-016-3915-1}{\doi{10.1140/epjc/s10052-016-3915-1}},
\href{http://www.arXiv.org/abs/1509.07255}{\texttt{arXiv:1509.07255}}.

\bibitem{deplection}
\hrefCMSnoop {}{{ALICE Collaboration}, ``Centrality dependence of pion
  freeze-out radii in {PbPb} collisions at {$\sqrtsNN = 2.76\TeV$}'',} \textit{
  Phys. Rev. C} \textbf{ 93} (2016) 024905,
  \href{http://dx.doi.org/10.1103/PhysRevC.93.024905}{\doi{10.1103/PhysRevC.93.024905}},
\href{http://www.arXiv.org/abs/1507.06842}{\texttt{arXiv:1507.06842}}.

\bibitem{STAR:2008szd}
\hrefCMSnoop {}{{STAR} Collaboration, ``Beam-energy and system-size dependence
  of dynamical net charge fluctuations'',} \textit{ Phys. Rev. C} \textbf{ 79}
  (2009) 024906,
  \href{http://dx.doi.org/10.1103/PhysRevC.79.024906}{\doi{10.1103/PhysRevC.79.024906}},
  \href{http://www.arXiv.org/abs/0807.3269}{\texttt{arXiv:0807.3269}}.

\bibitem{EHS:2005lzp}
\hrefCMSnoop {}{{EHS, NA22} Collaboration, ``Boost invariance and multiplicity
  dependence of the charge balance function in $\pi^{+}p$ and $k^{+}p$
  collisions at $\sqrt{s}$ = 22--{GeV/c}'',} \textit{ Phys. Lett. B} \textbf{
  637} (2006) 39,
  \href{http://dx.doi.org/10.1016/j.physletb.2006.04.027}{\doi{10.1016/j.physletb.2006.04.027}},
  \href{http://www.arXiv.org/abs/hep-ex/0506027}{\texttt{arXiv:hep-ex/0506027}}.

\bibitem{Lokhtin:2005px}
\hrefCMSnoop {}{I.~P. Lokhtin and A.~M. Snigirev, ``A model of jet quenching in
  ultrarelativistic heavy ion collisions and high-$\pt$ hadron spectra at
  {RHIC}'',} \textit{ Eur. Phys. J. C} \textbf{ 45} (2006) 211,
  \href{http://dx.doi.org/10.1140/epjc/s2005-02426-3}{\doi{10.1140/epjc/s2005-02426-3}},
\href{http://www.arXiv.org/abs/hep-ph/0506189}{\texttt{arXiv:hep-ph/0506189}}.

\bibitem{pval}
\hrefCMSnoop {}{E.~Gross and O.~Vitells, ``Trial factors for the look elsewhere
  effect in high energy physics'',} \textit{ Eur. Phys. J. C} \textbf{ 70}
  (2010) 525,
  \href{http://dx.doi.org/10.1140/epjc/s10052-010-1470-8}{\doi{10.1140/epjc/s10052-010-1470-8}},
\href{http://www.arXiv.org/abs/1005.1891}{\texttt{arXiv:1005.1891}}.

\bibitem{jetmed}
\hrefCMSnoop {}{{CMS Collaboration}, ``Decomposing transverse momentum balance
  contributions for quenched jets in {\PbPb} collisions at {$\sqrtsNN =
  2.76\TeV$}'',} \textit{ JHEP} \textbf{ 11} (2016) 055,
  \href{http://dx.doi.org/10.1007/jhep11(2016)055}{\doi{10.1007/jhep11(2016)055}},
\href{http://www.arXiv.org/abs/1609.02466}{\texttt{arXiv:1609.02466}}.

\bibitem{jetpbpb}
\hrefCMSnoop {}{{CMS Collaboration}, ``Correlations between jets and charged
  particles in {\PbPb} and pp collisions at {$\sqrtsNN = 2.76\TeV$}'',}
  \textit{ JHEP} \textbf{ 02} (2016) 156,
  \href{http://dx.doi.org/10.1007/jhep02(2016)156}{\doi{10.1007/jhep02(2016)156}},
\href{http://www.arXiv.org/abs/1601.00079}{\texttt{arXiv:1601.00079}}.

\bibitem{cmshinppridge}
\hrefCMSnoop {}{{CMS Collaboration}, ``Ridge correlation structure in high
  multiplicity {pp} collisions with {CMS}'',} \textit{ J. Phys. G} \textbf{ 38}
  (2011) 124051,
  \href{http://dx.doi.org/10.1088/0954-3899/38/12/124051}{\doi{10.1088/0954-3899/38/12/124051}},
\href{http://www.arXiv.org/abs/1107.2196}{\texttt{arXiv:1107.2196}}.

\bibitem{ATLASpPb}
\hrefCMSnoop {}{{ATLAS Collaboration}, ``Measurement of long-range
  pseudorapidity correlations and azimuthal harmonics in {$\sqrtsNN =
  5.02\TeV$} {\pPb} collisions with the {ATLAS} detector'',} \textit{ Phys.
  Rev. C} \textbf{ 90} (2014) 044906,
  \href{http://dx.doi.org/10.1103/PhysRevC.90.044906}{\doi{10.1103/PhysRevC.90.044906}},
\href{http://www.arXiv.org/abs/1409.1792}{\texttt{arXiv:1409.1792}}.

\bibitem{CMS:2015yux}
\hrefCMSnoop {}{{CMS Collaboration}, ``Evidence for collective multiparticle
  correlations in {pPb} collisions'',} \textit{ Phys. Rev. Lett.} \textbf{ 115}
  (2015) 012301,
  \href{http://dx.doi.org/10.1103/PhysRevLett.115.012301}{\doi{10.1103/PhysRevLett.115.012301}},
  \href{http://www.arXiv.org/abs/1502.05382}{\texttt{arXiv:1502.05382}}.

\end{thebibliography}\endgroup
\cleardoublepage \appendix\section{The CMS Collaboration \label{app:collab}}\begin{sloppypar}\hyphenpenalty=5000\widowpenalty=500\clubpenalty=5000
\cmsinstitute{Yerevan Physics Institute, Yerevan, Armenia}
{\tolerance=6000
A.~Tumasyan\cmsAuthorMark{1}\cmsorcid{0009-0000-0684-6742}
\par}
\cmsinstitute{Institut f\"{u}r Hochenergiephysik, Vienna, Austria}
{\tolerance=6000
W.~Adam\cmsorcid{0000-0001-9099-4341}, J.W.~Andrejkovic, T.~Bergauer\cmsorcid{0000-0002-5786-0293}, S.~Chatterjee\cmsorcid{0000-0003-2660-0349}, K.~Damanakis\cmsorcid{0000-0001-5389-2872}, M.~Dragicevic\cmsorcid{0000-0003-1967-6783}, A.~Escalante~Del~Valle\cmsorcid{0000-0002-9702-6359}, P.S.~Hussain\cmsorcid{0000-0002-4825-5278}, M.~Jeitler\cmsAuthorMark{2}\cmsorcid{0000-0002-5141-9560}, N.~Krammer\cmsorcid{0000-0002-0548-0985}, L.~Lechner\cmsorcid{0000-0002-3065-1141}, D.~Liko\cmsorcid{0000-0002-3380-473X}, I.~Mikulec\cmsorcid{0000-0003-0385-2746}, J.~Schieck\cmsAuthorMark{2}\cmsorcid{0000-0002-1058-8093}, R.~Sch\"{o}fbeck\cmsorcid{0000-0002-2332-8784}, D.~Schwarz\cmsorcid{0000-0002-3821-7331}, M.~Sonawane\cmsorcid{0000-0003-0510-7010}, S.~Templ\cmsorcid{0000-0003-3137-5692}, W.~Waltenberger\cmsorcid{0000-0002-6215-7228}, C.-E.~Wulz\cmsAuthorMark{2}\cmsorcid{0000-0001-9226-5812}
% ...
\par}
\vskip\cmsinstskip
\dag:~Deceased\\
$^{1}$Also at Yerevan State University, Yerevan, Armenia\\
$^{2}$Also at TU Wien, Vienna, Austria\\
% ...
\end{sloppypar}
\end{document}